\documentclass[sigconf, nonacm]{acmart}

\newcommand\vldbdoi{XX.XX/XXX.XX}
\newcommand\vldbpages{XXX-XXX}
\newcommand\vldbvolume{14}
\newcommand\vldbissue{1}
\newcommand\vldbyear{2026}
\newcommand\vldbauthors{\authors}
\newcommand\vldbtitle{\shorttitle} 
\newcommand\vldbavailabilityurl{https://gitlab.lip6.fr/benali1/refutationalnormalizationforjsonschema}
\newcommand\vldbpagestyle{plain}

\usepackage{marginnote}
\usepackage{amsmath,amsthm,amsfonts}
\usepackage{graphicx}
\usepackage{csquotes}
\usepackage{float}

\usepackage{ragged2e}
\usepackage{listings}

\usepackage[noend,linesnumbered]{algorithm2e}

\usepackage{graphicx}
\usepackage{bcprules}

\usepackage{subcaption}

\usepackage{todonotes} 

\usepackage{csquotes} 

\usepackage{setspace} 

\usepackage{libertine}

\usepackage{tikz}
\usepackage{ulem}
\usepackage{fancyvrb}
\usepackage{xcolor}

\newif\iflon
\lontrue                        
\iflon
\newcommand{\iflong}[1]{#1}
\newcommand{\ifshort}[1]{}
\else
\newcommand{\iflong}[1]{}
\newcommand{\ifshort}[1]{#1}
\fi

\newif\ifacm

\acmtrue    
\ifacm
\newtheorem{remark}{Remark}
\newtheorem{example}{Example}
\fi

\iflon
\usepackage[appendix=inline]{apxproof} 
\else
\usepackage[appendix=append]{apxproof}
\fi

  {\small \em \tt ccc\begin{table} aaa bbb}%
  {\end{table} ddd}

\newcommand{\json}{JSON}
\newcommand{\kw}[1]{\textbf{#1}}
\renewcommand{\kw}[1]{\ensuremath{\mathtt{#1}}}
\newcommand{\qkw}[1]{\ensuremath{\mathtt{\QQ{#1}\QQ}}}

\renewcommand{\kw}[1]{\akw{#1}}
\renewcommand{\qkw}[1]{\qakw{#1}}

\newcommand{\key}[1]{\ensuremath{\mathit{#1}}}

\newcommand{\akey}[1]{\ensuremath{\mathsf{#1}}}

\newcommand{\rkw}[1]{\ensuremath{\mathsf{#1}}}
\renewcommand{\rkw}[1]{\ensuremath{\mbox{\sf{\small #1}}}}
\newcommand{\akw}[1]{\ensuremath{\mbox{\tt{\small #1}}}}
\newcommand{\tkw}[1]{\ensuremath{\mbox{\tt{\normalsize #1}}}}

\newcommand{\qakw}[1]{\QQ\akw{#1}\QQ}

\newcommand{\qnot}{\qkw{not}}

\newcommand{\xone}{\kw{oneOf}}
\newcommand{\qone}{\qkw{oneOf}}
\newcommand{\xany}{\kw{anyOf}}
\newcommand{\qany}{\qkw{anyOf}}

\newcommand{\qall}{\qkw{allOf}}

\newcommand{\qreq}{\qkw{required}}

\newcommand{\qpattProps}{\qkw{patternProperties}}

\newcommand{\qaddProps}{\qkw{additionalProperties}}

\newcommand{\qunProps}{\qkw{unevaluatedProperties}}

\newcommand{\qmof}{\qkw{multipleOf}}

\newcommand{\xpatt}{\kw{pattern}}
\newcommand{\qpatt}{\qkw{pattern}}

\newcommand{\quniqIts}{\qkw{uniqueItems}}

\newcommand{\qaddIts}{\qkw{additionalItems}}

\newcommand{\qunIts}{\qkw{unevaluatedItems}}

\newcommand{\qunStar}{\qkw{unevaluated*}}

\newcommand{\qaddStar}{\qkw{additional*}}

\newcommand{\eqdef}{\ensuremath{\stackrel{\vartriangle}{=}}}

\newcommand{\gcomment}[1]{}
\newcommand{\oldversion}[1]{}
\newcommand{\hideforspace}[1]{}
\newcommand{\hide}[1]{}
\newcommand{\save}[1]{}
\newcommand{\code}[1]{}

\newcommand{\Implies}{\Rightarrow}

\renewcommand{\comment}[1]{}

\renewcommand{\And}{\wedge}

\newcommand{\Not}{\neg}

\newcommand{\Num}{\akey{Num}}
\newcommand{\Str}{\akey{Str}}
\newcommand{\Nat}{\akey{Nat}}

\newcommand{\Bool}{\akey{Bool}}

\newcommand{\Uri}{\akey{Uri}}

\newcommand{\Uni}{\akey{uniqueItems}}

\newcommand{\keykey}[1]{\key{\underline{#1}}}
\renewcommand{\keykey}[1]{\lfloor{#1}\rfloor}

\newcommand{\Inf}{\infty}

\newcommand{\mJS}{Modern JSON Sche\-ma}

\newcommand{\cJS}{Classical JSON Sche\-ma}

\newcommand{\custcom}[2]{\marginpar{\tiny #1: {#2}}}

\newcommand{\GG}[1]{\custcom{giorgio}{#1}}

\newcommand{\M}{\ |\ }

\newlength{\NL}
\newlength{\SaveNL}

\newcommand{\EmptySet}{\emptyset}

\newcommand{\QQ}{\textnormal{\textquotedbl}}
\newcommand{\Set}[1]{\{\,{#1}\,\}}

\newcommand{\SetOpen}{\{\!|}
\newcommand{\SetClose}{|\!\}}
\renewcommand{\Set}[1]{\SetOpen{#1}\SetClose}
\newcommand{\SSet}[1]{\Set{#1}^{\key{set}}}
\newcommand{\SetST}[2]{\SetOpen{#1}\,\mid\,{#2}\SetClose}

\newcommand{\SetTo}[1]{\Set{1\ldots{#1}}}

\newcommand{\SetFromTo}[2]{\Set{{#1}\ldots{#2}}}

\newcommand{\rlan}[1]{L(#1)}

\newcommand{\jsonsch}{JSON Schema} 

\renewcommand{\SetOpen}{\{\,}
\renewcommand{\SetClose}{\,\}}
\renewcommand{\Set}[1]{\SetOpen{#1}\SetClose}
\renewcommand{\SSet}[1]{\{{#1}\}}
\renewcommand{\SetST}[2]{\SetOpen{#1}\,\mid\,{#2}\SetClose}

\newcommand{\StackDash}[1]{\stackrel{#1}{\vdash}}
\renewcommand{\StackDash}[1]{\mathrel{\vdash\raisebox{.90ex}{\tt\kern-0.5em {\tiny {#1}\kern0.5em}}}}

\newcommand{\Ret}[1]{\,\,\stackrel{#1}{\rightarrow}\,\,}
\newcommand{\RetL}[1]{\,\,{\rightarrow}\,\,}
\renewcommand{\RetL}[1]{\,\,\stackrel{#1}{\rightarrow}\,\,}

\newcommand{\anot}{\qakw{not}}
\newcommand{\atrue}{\akw{true}}
\newcommand{\afalse}{\akw{false}}
\newcommand{\anull}{\akw{null}}
\newcommand{\aone}{\qakw{oneOf}}
\newcommand{\aany}{\qakw{anyOf}}
\newcommand{\aall}{\qakw{allOf}}
\newcommand{\amin}{\qakw{minimum}}
\newcommand{\amax}{\qakw{maximum}}

\newcommand{\atype}{\qakw{type}}

\newcommand{\aaddIts}{\qakw{additionalItems}}

\newcommand{\amof}{\qakw{multipleOf}}
\newcommand{\anotMof}{\qakw{notMultipleOf}}

\newcommand{\apatt}{\qakw{pattern}}
\newcommand{\anotPatt}{\qakw{notPattern}}

\newcommand{\acontAft}{\qakw{containsAfter}}

\newcommand{\aits}{\qakw{items}}

\newcommand{\aconst}{\qakw{const}}

\newcommand{\astr}{\qakw{string}}
\newcommand{\aarray}{\qakw{array}}
\newcommand{\aboolean}{\qakw{boolean}}

\newcommand{\rone}{\rkw{oneOf}}
\newcommand{\rany}{\rkw{anyOf}}

\newcommand{\JObjOpen}{\{}
\newcommand{\JObjClose}{\}}
\newcommand{\JObj}[1]{\JObjOpen\,{#1}\,\JObjClose}

\newcommand{\RAsA}[1]{R\mbox{\ as\ }{A}}

\newcommand{\bshort}{\noindent\begin{minipage}{0.5\textwidth}}
\newcommand{\eshort}{ \end{minipage}}

\newcommand{\ES}{\EmptySet}

\renewcommand{\Ret}[1]{\rightarrow}

\newcommand{\xSub}{<:}

\newcommand{\dFalse}{\akw{dFalse}}

\newcommand{\cTrue}{\akw{cTrue}}
\newcommand{\xFalse}{\akw{xFalse}}
\newcommand{\xTrue}{\akw{xTrue}}

\newcommand{\funkw}[1]{\key{#1}}

\newcommand{\allDS}{\funkw{allDS}}
\newcommand{\allCS}{\funkw{allCS}}
\newcommand{\allCK}{\funkw{allCK}}
\newcommand{\allXX}{\funkw{allXX}}
\newcommand{\anyDD}{\funkw{anyDD}}
\newcommand{\notPush}{\funkw{notPush}}

\newcommand{\CDNF}{\mbox{DNF}}
\newcommand{\aDNF}{\funkw{DNF}}

\newcommand{\DNF}{\akw{DNF}}

\newcommand{\fMap}{\key{fMap}}

\newcommand{\reqList}{\key{reqList}}

\newcommand{\InAppendix}[1]{}

\renewcommand{\anot}{\akw{not}}
\renewcommand{\atrue}{\akw{true}}
\renewcommand{\afalse}{\akw{false}}
\renewcommand{\aone}{\akw{one}}
\renewcommand{\aany}{\akw{any}}
\renewcommand{\aall}{\akw{all}}

\renewcommand{\atype}{\akw{type}}
\newcommand{\atypeSet}{\akw{typeSet}}

\renewcommand{\amin}{\akw{min}}
\renewcommand{\amax}{\akw{max}}
\newcommand{\aexMin}{\akw{exMin}}
\newcommand{\aexMax}{\akw{exMax}}
\renewcommand{\amof}{\akw{mof}}
\renewcommand{\anotMof}{\akw{notMof}}

\newcommand{\amaxLen}{\akw{maxLen}}
\newcommand{\aminLen}{\akw{minLen}}
\renewcommand{\apatt}{\akw{patt}}
\renewcommand{\anotPatt}{\akw{notPatt}}

\newcommand{\apReq}{\akw{pReq}}

\newcommand{\apProp}{\akw{pProp}}

\newcommand{\aminProps}{\akw{minProps}}
\newcommand{\amaxProps}{\akw{maxProps}}

\newcommand{\aitem}{\akw{item}}
\renewcommand{\aaddIts}{\akw{addIts}}

\renewcommand{\acontAft}{\akw{contAft}}
\newcommand{\aminIts}{\akw{minIts}}
\newcommand{\amaxIts}{\akw{maxIts}}

\newcommand{\auniqIts}{\akw{uniqueIts}}
\newcommand{\anotUniqIts}{\akw{notUniqueIts}}

\renewcommand{\aconst}{\akw{const}}
\newcommand{\anotConst}{\akw{notConst}}

\newcommand{\aref}{\akw{ref}}

\newcommand{\aobj}{\akw{object}}
\newcommand{\anum}{\akw{number}}
\renewcommand{\astr}{\akw{string}}
\renewcommand{\aarray}{\akw{array}}
\newcommand{\abool}{\akw{boolean}}
\renewcommand{\anull}{\akw{null}}

\newcommand{\allp}{\akw{allp}}
\newcommand{\anyp}{\akw{anyp}}
\newcommand{\notp}{\akw{notp}}

\newcommand{\NV}[1]{\akw{N}\langle{#1}\rangle}
\renewcommand{\NV}[1]{\akw{N}({#1})}
\newcommand{\RNV}[1]{\aref(N\langle{#1}\rangle)}
\newcommand{\Rf}[1]{\aref({#1})}

\newcommand{\ER}{\ensuremath{\diamond}}

\newcommand{\eeq}{\equiv}

\newcommand{\NE}[1]{{#1}\neeq\ES}

\newcommand{\PI}[2]{{#1}\!\cap\!{#2}}
\newcommand{\PIE}[2]{\PI{#1}{#2}\eeq\ES}
\newcommand{\PINE}[2]{\PI{#1}{#2}\neeq\ES}
\newcommand{\PM}[2]{{#1}\shortminus{#2}}
\renewcommand{\PM}[2]{\PI{#1}{\NP{#2}}}                 
\newcommand{\PME}[2]{\PM{#1}{#2}\eeq\ES}              
\newcommand{\PMNE}[2]{\PM{#1}{#2}\neeq\ES}
\newcommand{\KK}[1]{\,\hat{}\,{#1}\$}
\newcommand{\NP}[1]{\overline{#1}}        

\renewcommand{\PI}[2]{\allp[{#1},{#2}]}
\renewcommand{\NP}[1]{\notp({#1})}

\newcommand{\subt}{<:}

\newcommand{\RNAlg}{Refutational Normalization}
\newcommand{\RN}{RefN}
\newcommand{\RB}{RulB}
\newcommand{\BaseWG}{Base-WG}
\newcommand{\WGBI}{\WG}
\newcommand{\WG}{WitG}

\renewcommand{\RN}{RWG}
\renewcommand{\RB}{RB}
\renewcommand{\BaseWG}{general-purpose witness generation}
\renewcommand{\WG}{WG}

\newcommand{\arrS}[1]{{#1}^l}
\newcommand{\aS}{\arrS{S}}
\newcommand{\aU}{\arrS{U}}
\newcommand{\aC}{\arrS{C}}

\newcommand{\aK}{\arrS{K}}

\newcommand{\aY}{\arrS{Y}}

\newcommand{\SETFN}[1]{\textnormal{\textsf{#1}}}
\renewcommand{\SETFN}[1]{\text{\bf{#1}}}
\renewcommand{\SETFN}[1]{\text{\it{#1}}}
\newcommand{\fAllDS}{\SETFN{allDS}}
\newcommand{\fAllCK}{\SETFN{allCK}}
\newcommand{\fAllCS}{\SETFN{allCS}}
\newcommand{\fAnyDD}{\SETFN{anyDD}}
\newcommand{\fFastFailAllCS}{\SETFN{fastFailAllCS}}

\newcommand{\fMergeFragProp}{\SETFN{mergeFragProp}}
\newcommand{\fEmpty}{\SETFN{empty}}

\newcommand{\fAllXX}{\SETFN{allXX}}

\newcommand{\kPredef}[1]{\ensuremath{\mathtt{#1}}}
\newcommand{\kFMap}{\kPredef{FMap}}

\newcommand{\kxFalse}{\kPredef{xFalse}}
\newcommand{\kpProp}{\kPredef{pProp}}
\newcommand{\kList}{\kPredef{List}}
\newcommand{\kany}{\kPredef{any}}
\newcommand{\kfalse}{\kPredef{false}}
\newcommand{\kdFalse}{\kPredef{dFalse}}
\newcommand{\kall}{\kPredef{all}}
\newcommand{\kref}{\kPredef{ref}}

\newcommand{\kId}[1]{\ensuremath{\mathit{#1}}}
\newcommand{\ke}{\kId{e}}
\newcommand{\kpRef}{\kId{pRef}}

\newcommand{\kreqList}{\kId{reqList}}
\newcommand{\kp}{\kId{p}}
\newcommand{\kx}{\kId{X}}
\newcommand{\kE}{\kId{E}}

\newcommand{\krefinedReqList}{\kId{refinedRL}}
\newcommand{\krefinedPRef}{\kId{refinedPRef}}

\newcommand{\AlgoSetup}{%
\DontPrintSemicolon
\SetKwProg{Fn}{Function}{:}{}
\SetKw{KwReturn}{return}
\SetKw{KwIn}{in}
\SetKw{KwLet}{let}
\SetKw{KwContinue}{continue}
\SetKw{KwRaise}{raise}
\SetKw{KwTry}{try}
\SetKw{KwCatch}{iffails}
\SetKw{KwContinue}{continue}
\SetKw{match}{match}
\SetKw{KwOf}{}
}

\usepackage{booktabs}
\usepackage{multirow}
\usepackage{geometry}

\usepackage[most]{tcolorbox}
\usepackage{listings, color}
\usepackage{paralist}
\usepackage[inline]{enumitem} 
\usepackage{fancyvrb}
\usepackage{hyperref}

\usepackage{xcolor}

\usepackage{multirow}

\usepackage[noend,linesnumbered]{algorithm2e}

\usepackage{mathtools}

\usepackage[english]{babel} 

\newif{\ifMarginalComments}
 \MarginalCommentstrue   

 \definecolor{orange}{rgb}{1,0.45,0}
\definecolor{darkblue}{rgb}{0,0.2,0.8}
\definecolor{darkgreen}{rgb}{0.1,0.4,0.1}
\definecolor{darkviolet}{rgb}{0.45,0,0.65}
\definecolor{darkred}{rgb}{0.8,0,0}
\definecolor{lightblue}{rgb}{0.9,0.9,1}
\definecolor{lightgreen}{rgb}{0.9,1,0.9}

\newcommand{\mrevone}[2]{\textcolor{darkred}{#1}\marginpar{\textcolor{darkred}{#2}}}
\newcommand{\mrevtwo}[2]{\textcolor{orange}{#1}\marginpar{\textcolor{orange}{#2}}}

\newcommand{\mrevfour}[2]{\textcolor{darkviolet}{#1}\marginpar{\textcolor{darkviolet}{#2}}}

\newcommand{\crevone}[1]{\textcolor{darkred}{#1}}

\newcommand{\crevfour}[1]{\textcolor{darkviolet}{#1}}

\newcommand{\bcolorone}{\color{darkred}}
\newcommand{\bcolortwo}{\color{orange}}

\newcommand{\bcolorfour}{\color{darkviolet}}

\newcommand{\ecolor}{\color{black}}

\iflon
\renewcommand{\bcolorone}{\color{black}}
\renewcommand{\bcolortwo}{\color{black}}

\renewcommand{\bcolorfour}{\color{black}}

\renewcommand{\ecolor}{\color{black}}
\renewcommand{\mrevone}[2]{#1}
\renewcommand{\mrevtwo}[2]{#1}

\renewcommand{\mrevfour}[2]{#1}

\renewcommand{\crevone}[1]{{#1}}

\renewcommand{\crevfour}[1]{{#1}}

\fi

\newcommand{\nop}[1]{{}} 

\makeatletter
\newcommand\querysize{\@setfontsize\querysize\@vipt\@viipt}
\makeatother

\lstdefinestyle{query}{
  stepnumber=1,
  numbersep=3pt, 
  tabsize=4,
  showspaces=false,
  showstringspaces=false,
  basicstyle=\linespread{1}\fontfamily{lmtt}\selectfont\querysize,
  keywordstyle=\color{blue},
  stringstyle=\color{purple},
  upquote=true,
  breaklines=true,
  commentstyle=\color{CadetBlue}
}

\definecolor{mygray}{rgb}{0.643,0.643,0.643}

\newtcolorbox{querybox}[2][]{%
  sidebyside align=top,
  enhanced,
  boxsep=0pt,
  arc=0pt,
  top=-3pt, bottom=-3pt,
  left=2pt, right=0pt,
  colback=white,
  colframe=mygray,
  boxrule=0.5pt,
  leftrule=12pt,
  overlay unbroken and first ={%
    \node[rotate=90,
          minimum width=0.5cm,
          anchor=south,
          font=\small\rmfamily,
          yshift=-13pt,
          white]
    at (frame.west) {#2};
  }
} 
\newtcolorbox{querybox2l}[2][]{%
  sidebyside align=top,
  enhanced,
  boxsep=0pt,
  arc=0pt,
  top=-7pt, bottom=-10.5pt,
  left=2pt, right=0pt,
  colback=white,
  colframe=mygray,
  boxrule=0.5pt,
  leftrule=10pt,
  overlay unbroken and first ={%
    \node[rotate=90,
          align=center,
          minimum width=0.5cm,
          anchor=south,
          font=\small\rmfamily,
          yshift=-12pt,
          white]
    at (frame.west) {#2};
  }
}

\makeatletter
\AtBeginDocument{\def\libertine@figurealign{}\libertineOsF}
\newcommand\libertineTabular{\def\libertine@figurealign{T}\libertineLF}
\makeatother

\hypersetup{draft} 

\begin{document}
\title{JSON Schema Inclusion through Refutational Normalization: Reconciling Efficiency and Completeness}


\author{Mohamed-Amine Baazizi}
\affiliation{%
  \institution{Sorbonne Universit\'e, LIP6 UMR 7606}
  \city{Paris}
  \country{France}
}
\email{mohamed-amine.baazizi@lip6.fr}

\author{Nour El Houda Ben Ali}
\affiliation{%
  \institution{Universit\'e Paris-Dauphine -- PSL}
  \city{Paris}
  \country{France}
}
\email{nour-el-houda.ben-ali@dauphine.eu}

\author{Dario Colazzo}
\affiliation{%
  \institution{Universit\'e Paris-Dauphine -- PSL}
  \city{Paris}
  \country{France}
}
\email{dario.colazzo@lamsade.dauphine.fr}

\author{Giorgio Ghelli}
\affiliation{%
  \institution{Universit\`a di Pisa}
  \city{Pisa}
  \country{Italy}
}
\email{giorgio.ghelli@unipi.it}

\author{Stefan Klessinger}
\affiliation{%
  \institution{Universit{\"a}t Passau}
  \city{Passau}
  \country{Germany}
}
\email{stefan.klessinger@uni-passau.de}

\author{Carlo Sartiani}
\affiliation{%
  \institution{Universit\`a della Basilicata}
  \city{Potenza}
  \country{Italy}
}
\email{carlo.sartiani@unibas.it}

\author{Stefanie Scherzinger}
\affiliation{%
  \institution{Universit{\"a}t Passau}
  \city{Passau}
  \country{Germany}
}
\email{stefanie.scherzinger@uni-passau.de}

\begin{abstract}

\mrevone{JSON Schema is the de facto standard for describing the structure of JSON documents. 
Deciding JSON Schema inclusion --- whether every instance of a schema $S$
is also an instance of a schema $S^{\prime}$ --- 
is a building block for version and API compatibility checks, schema refactoring, and large-scale corpus 
analysis. The problem is EXPTIME-complete, and the two existing families of algorithms fail in practice in complementary ways: rule-based algorithms are fast but incomplete by design, while witness-generation algorithms are complete but so slow that, under any reasonable timeout, they are incomplete in practice 
as well.
We show that this specific problem admits a way out. We redesign the normalization core of the complete algorithm, so that, on the easy cases, it mimics the behavior of the fast rule-based 
algorithm, without losing completeness.
We prove that it is exactly as efficient as the rule-based approach on every inclusion that approach 
can prove, while remaining complete on all others.
We validate this on a vast collection of real-world schemas and hand-crafted schemas. 
We show that our algorithm advances the 
state of the art, making tractable a range of use cases that were out of reach for existing tools.
}{1.1,4.1}

\end{abstract}

\maketitle

\pagestyle{\vldbpagestyle}
\begingroup\small\noindent\raggedright\textbf{PVLDB Reference Format:}\\
\vldbauthors. \vldbtitle. PVLDB, \vldbvolume(\vldbissue): \vldbpages, \vldbyear.\\
\href{https://doi.org/\vldbdoi}{doi:\vldbdoi}
\endgroup
\begingroup
\renewcommand\thefootnote{}\footnote{\noindent
This work is licensed under the Creative Commons BY-NC-ND 4.0 International License. Visit \url{https://creativecommons.org/licenses/by-nc-nd/4.0/} to view a copy of this license. For any use beyond those covered by this license, obtain permission by emailing \href{mailto:info@vldb.org}{info@vldb.org}. Copyright is held by the owner/author(s). Publication rights licensed to the VLDB Endowment. \\
\raggedright Proceedings of the VLDB Endowment, Vol. \vldbvolume, No. \vldbissue\ %
ISSN 2150-8097. \\
\href{https://doi.org/\vldbdoi}{doi:\vldbdoi} \\
}\addtocounter{footnote}{-1}\endgroup

\ifdefempty{\vldbavailabilityurl}{}{
\vspace{.3cm}
\begingroup\small\noindent\raggedright\textbf{PVLDB Artifact Availability:}\\
The source code, data, and/or other artifacts have been made available at \url{\vldbavailabilityurl}.
\endgroup
}
\renewcommand{\subt}{\ensuremath{\preceq}}
\renewcommand{\subt}{\ensuremath{\subseteq}}
\newcommand{\subtr}{\subt}

\section{Introduction}\label{sec:intro}

JSON Schema has become a \emph{de facto} standard for describing and validating JSON data structures across diverse application domains, from web APIs to database systems.\footnote{We target here {\cJS}\ifshort{, as opposed to {\mJS}~\cite{DBLP:journals/pacmpl/AttoucheBCGSS24}}.
\iflong{Boolean inclusion for {\mJS} can be decided by  mapping a schema $S$ in {\mJS} into an equivalent schema $S^{\prime}$ in {\cJS} through the rewriting rules described by Attouche et al.~\cite{DBLP:journals/tcs/AttoucheBCGKSS26}.}}
As schema-driven development practices become increasingly prevalent, the ability to formally reason about relationships between schemas is essential.

Schema inclusion --- determining whether all instances that are valid against one schema are also valid against another --- is a fundamental operation that underpins critical tasks including schema evolution, data migration, query optimization, access control enforcement, schema 
refactoring, and schema analysis.
\mrevone{Specific use cases documented in the literature include static type checking in ML pipe\-lines~\cite{DBLP:conf/issta/HabibSHP21} and a case study with schemas describing a cloud monitoring service~\cite{DBLP:conf/models/ColantoniGBWB21}.}{1.13}
\mrevone{We next present a synthetic example.}{}

\begin{example} \label{ex:schema} \em
Below, we show a small JSON Schema~$S_o$. It declares that any valid instance must be a string that must match exactly one (\rone) of two regular expressions, or patterns (\xpatt).
The strings ``Pizza dell'alleanza'' and ``Antica margherita'' are valid instances, while ``Pizza margherita'' is not: it matches both patterns.

\begin{Verbatim}[fontsize=\small]
 {"oneOf":[{"pattern":"^Pizza"},{"pattern":"margherita$"}]}
\end{Verbatim}

For schema~$S_a$ with \rany\ in place of \rone, ``Pizza margherita'' is valid, since both patterns may be matched.
Every instance that is valid for $S_o$ is also valid for $S_a$, hence $S_o$ is \emph{included} in~$S_a$: $S_o \subt S_a $.
\end{example}

\hide{
Existing approaches to schema inclusion checking often lack formal guarantees, relying on heuristics or incomplete algorithms that may produce incorrect results or fail to recognize valid inclusion relationships. 
However, the practical applications of schema inclusion demand both soundness and completeness: soundness ensures that reported inclusions are genuine, preventing runtime errors and security vulnerabilities, while completeness guarantees that all valid inclusions are recognized, enabling maximum automation and avoiding unnecessary manual verification. Without these guarantees, developers cannot trust automated tooling for schema refactoring, API evolution, or data integration tasks.}

\hide{\subsection{Approaches to JSON Schema inclusion}}

\hide{
{
    "anyOf": [
        {
            "minProperties": 2
        },
        {
            "patternProperties": {
                "a": {
                    "multipleOf": 3
                }
            },
            "maxProperties": 2
        },
        {
            "patternProperties": {
                "a": {
                    "multipleOf": 5
                }
            },
            "maxProperties": 20
        }
    ]
}

{
    "patternProperties": {
        "a": {
            "anyOf" : [
                {
                    "pattern": "B"
                },
                {
                    "pattern": "C"
                }
            ]
        }
    }
}
}

Two different approaches have been proposed in the past to decide inclusion between JSON Schema schemas:
the \emph{rule-based} approach~\cite{DBLP:conf/issta/HabibSHP21} and the \emph{witness-generation} approach~\cite{DBLP:journals/pvldb/AttoucheBCGSS22}.

With the rule-based approach, a \mrevtwo{schema inclusion}{2.5} problem is reduced to a collection of subproblems by employing a set of deduction rules, until all subproblems have been proven or until a subproblem is found that cannot be proven.
In practice, a small set of rules is sufficient to cover a substantial portion of real-world schemas, making this approach well-suited for tools that prioritize efficiency, simplicity, and practicality.
However, the rule system is \emph{incomplete}, meaning that there are schema pairs
$S_1$, $S_2$ where inclusion holds but the rules are unable to prove this fact.
\iflong{This is not just a problem of the only currently available formulation \cite{DBLP:conf/issta/HabibSHP21},
but is a general problem of the approach, as we discuss in Section \ref{sec:rules}.}

With witness generation\iflong{ proposed by Attouche et al.~\cite{DBLP:journals/pvldb/AttoucheBCGSS22}}, the problem $S_{1} \subt  S_{2}$ is first reduced to unsatisfiability of the schema
$\qall : [ S_{1}, \{ \qnot: S_{2} \} ]$, a schema that is satisfied by all and only  the counterexamples of 
$S_{1} \subt  S_{2}$.
This problem is then solved using a complete witness-generation algorithm,
that is, an algorithm that either generates an instance of its input schema, i.e., a \emph{witness}, or reports its
unsatisfiability.
This approach is complete in principle, but may take too long in practice, since   witness-generation is based on the computation of a Disjunctive Normal Form (DNF) for the input schema,
and the computation of a DNF for a schema such as  $\qall : [ S_{1}, \{ \qnot: S_{2} \} ]$, which combines 
conjunction and negation, can have exponential cost even when the schemas~$S_1$ and~$S_2$ are quite simple.

While these approaches are satisfactory for some use cases~\cite{DBLP:journals/pvldb/AttoucheBCGSS22,10.1145/3799416},
there are cases where
the incompleteness of the first approach and the inefficiency of the second  constitute real obstacles for
their adoption.
Rather than having to choose between speed and completeness, one would like to reconcile both,
achieving both completeness and efficiency.
To this aim, we propose here to completely
redesign the core  of the witness-generation ap\-proach, which is the normalization algorithm,
in a way that takes inspiration from the inclusion 
rules, 
so to reconcile the efficiency of the rule-based
approach with the completeness of witness-generation.
\iflong{
The rule-based approach is designed to optimize the process of proving $S \subt S'$ when 
inclusion holds.

This happens when $\qall : [ S_{1}, \{ \qnot: S_{2} \} ]$ is not satisfiable, which means that
the DNF of $\qall : [ S_{1}, \{ \qnot: S_{2} \} ]$ is equivalent to $\afalse$. Hence, a normalization
algorithm that mirrors the rules is optimized with the aim of proving that $\aDNF(S)$ is equivalent to
$\afalse$. For this reason, we call it a \emph{Refutational Normalization} algorithm.
Refutational Normalization can be described as a technique to embed the speed advantages of the
rule-based approach inside the normalization algorithm.}
\ifshort{
This algorithm is optimized for cases when 
$\aDNF(S)$ is equivalent to
$\afalse$; for this reason, we call it a \emph{Refutational Normalization} algorithm.}

\paragraph*{Contributions}
\mrevone{We present the following contributions.}{1.1,4.1}
\bcolorone
\begin{enumerate}
\item An \mrevtwo{inclusion}{2.5}  algorithm that combines the efficiency of the rule-based approach with the completeness of witness generation, supporting recursion and the unrestricted use of negation. \mrevone{This algorithm supports all operators of Classical JSON Schema (from Draft4 to Draft7), except {\quniqIts}.}{1.4}

\item A notion of \emph{refutational normalization}, an approach 
to normalization  that internalizes the behavior of the inclusion rules and is optimized for the fast detection of the cases where the normalized term is unsatisfiable, without sacrificing the complementary cases.

\item A proof that, on \emph{rule-provable} judgments --- those where inclusion can be 
proved by the rule-based approach --- our complete algorithm is as efficient as the incomplete 
rule-based one. On judgments that are not rule-provable our approach may be slower, since 
it performs a complete proof search, but on the rule-provable cases it adds no extra cost.

\item A targeted experimental evaluation, 
showing that the approach attains the completeness of witness generation together with the 
efficiency of the rule-based approach, making it possible to treat use cases that were out of reach for the previous state of the art.
\end{enumerate}
\ecolor

\mrevone{Beyond the algorithm itself, the paper carries a methodological lesson. The normalization of 
$\qall : [ S_1, \{ \qnot : S_2 \} ]$
and  the rule-based comparison of $S_1$ with $S_2$
are commonly regarded as two unrelated processes, with complementary strengths. 
We show in this paper that the former can in fact be implemented so as to mirror the latter,
combining the advantages of the two approaches.
This is the main idea behind our algorithm, and it may have applicability beyond JSON Schema inclusion.}{1.1,1.14, 4.1}

\ecolor


\emph{Paper Outline.}
\iflong{The paper is structured as follows.}
In Section \ref{sec:relworks} we discuss related work.
In Section \ref{sec:json} we provide a quick overview of JSON Schema.
In Sections \ref{sec:WGinclusion} and ~\ref{sec:rulebased}, we describe the wit\-ness-generation-based approach, and the rule-based inclusion approach\hide{ and focus, in particular, on that of Habib et al.~\cite{DBLP:conf/issta/HabibSHP21}}. In Section \ref{sec:refutational} we present our inclusion algorithm.
In Section~\ref{sec:efficiency} 
we show that, in the rule-provable cases, it is as efficient as the 
rule-based approach.
 In Section~\ref{sec:expeval} we present an experimental evaluation validating our claims of efficiency and completeness. In Section~\ref{sec:concl}, we draw our conclusions.
%

\section{Related Work}\label{sec:relworks}

JSON Schema has been repeatedly studied in academic research from various angles, such as sche\-ma discovery~\cite{DBLP:journals/vldb/BaaziziCGS19,DBLP:conf/edbt/BaaziziBCGS20,DBLP:conf/sigmod/SpothKLHL21,DBLP:conf/edbt/KlessingerFGKSS23,DBLP:conf/btw/KlettkeSS15},
and validation~\cite{DBLP:journals/pacmpl/AttoucheBCGSS24}. \ifshort{Pezoa et al.~\cite{DBLP:conf/www/PezoaRSUV16} formalized
{\jsonsch}, proving polynomial-time validation and EXPTIME-hard satisfiability.
Bourhis et al.~\cite{DBLP:conf/pods/BourhisRSV17} later mapped {\jsonsch} to recursive
JSL and showed that satisfiability and inclusion are EXPTIME-complete for schemas without
{\Uni}, and in 2EXPTIME with {\Uni}.}

\iflong{Overviews of schema languages for {\json} have been presented
 by Pezoa et al.~\cite{DBLP:conf/www/PezoaRSUV16} and Bourhis et al.~\cite{DBLP:conf/pods/BourhisRSV17}. Pezoa et al.~\cite{DBLP:conf/www/PezoaRSUV16}  introduced the first formalization of {\jsonsch} and showed that it cannot be captured by MSO or tree automata because of the {\Uni} constraints. They 
focused on validation and proved that it can be decided in  polynomial time. They also showed that {\jsonsch} can simulate tree automata; hence,  {\jsonsch} satisfiability is EXPTIME-hard. (See  Su{\'a}rez Barr{\'i}a   \cite{suarez2016thesis} for a detailed proof of how a quantified alternating tree automaton can be encoded in polynomial time into an equivalent {\jsonsch}.)

Bourhis et al.~\cite{DBLP:conf/pods/BourhisRSV17} refined the analysis of Pezoa et al.\ by mapping {\jsonsch} onto an equivalent modal logic, called recursive~JSL. They proved that satisfiability is
EXPTIME-complete for recursive schemas without {\Uni}, and it is in {2}EXPTIME for recursive schemas with {\Uni}. 
Since satisfiability and inclusion for {\cJS} are equivalent, these results trivially extend to inclusion and equivalence checking.}

\ifshort{ A tool for checking schema inclusion, called \emph{jsonsubschema},
has been described by Habib et al.~\cite{DBLP:conf/issta/HabibSHP21}\iflong{; it
was developed in the context of IBM's open source AutoML framework LALE~\cite{DBLP:journals/corr/abs-2007-01977}}.
This is the rule-based approach that we present in greater detail in Section~\ref{sec:rulebased}. }

\iflong{A tool for checking schema inclusion has been described by Habib et al.~\cite{DBLP:conf/issta/HabibSHP21}. Their \emph{jsonsubschema} tool, written in Python and available through {\tt{pip}} and GitHub\footnote{\url{https://github.com/IBM/jsonsubschema}},%
was developed in the context of IBM's open source AutoML framework LALE~\cite{DBLP:journals/corr/abs-2007-01977}.  \emph{jsonsubschema}  has been tailored for this use case, which explains the presence of severe restrictions, such as lack of support for recursion and negation in complex schemas.
The tool, which supports {\cJS} (Draft 4) without recursion and generalized negation, takes as input two schemas $S_{1}$ and $S_{2}$, and returns three possible results: (i) $false$, if $S_{1}$ is not included in $S_{2}$; (ii) $true$, if $S_{1}$ is included in $S_{2}$; or (iii) $unknown$, if the tool was not able to take a decision. The $unknown$ result is motivated by the fact that \textsc{jsonsubschema} exploits a traditional rule-based approach by relying on an incomplete set of rules; therefore, it is possible that no further rule could be applied when comparing two schemas.

Despite the lack of support for recursion and generalized negation, the tool has been successfully used to identify 38 real bugs in the LALE framework.
This is the rule-based approach that we present in greater detail in Section~\ref{sec:rulebased}.}

An alternative approach 
based on witness generation has been defined by Attouche et al.~\cite{DBLP:journals/pvldb/AttoucheBCGSS22} and is presented in Section~\ref{sec:WGinclusion}.

The reduction of inclusion to unsatisfiability of
\(S_1 \And \Not S_2\) has a long history across 
communities.
In description logics, concept subsumption \(C \sqsubseteq D\)
is standardly reduced to unsatisfiability of \(C \sqcap \neg D\)
and solved by tableau algorithms~\cite{DLHandbook2010};
a rich optimization literature addresses the resulting blowup through techniques such as absorption,
lazy unfolding, and early clash detection~\cite{DBLP:journals/logcom/HorrocksP99,DBLP:journals/sLogica/BaaderS01},
with conceptual parallels with our refutational normalization.
\iflong{For XML, Genev{\`e}s et al.~\cite{DBLP:conf/pldi/GenevesLS07}
reduce XPath containment and type-checking problems to satisfiability of \(\mu\)-calculus formulas over finite trees, solved using BDD-based symbolic techniques.
XDuce~\cite{DBLP:conf/popl/HosoyaP01,Hosoya2001,DBLP:journals/toit/HosoyaP03,DBLP:journals/toplas/HosoyaVP05}
checks inclusion of regular-expression types for XML using tree-automata techniques; subsequent work~\cite{DBLP:conf/wia/HosoyaM03,DBLP:conf/wia/SudaH05} addresses the complementation blowup through
divide-and-conquer and non-back\-tracking top-down containment algorithms.}

\hide{In the semantic subtyping tradition, CDuce~\cite{DBLP:conf/lics/FrischCB02,Frisch2004,DBLP:conf/ppdp/CastagnaF05,DBLP:journals/jacm/FrischCB08},
addresses the blowup through compact representations (decision trees) and specialized emptiness procedures.
More broadly, automata-theo\-retic language inclusion and related automata problems have been extensively studied through antichain, simulation, and coinductive techniques \cite{DBLP:conf/cav/WulfDHR06,DBLP:conf/wia/BouajjaniHHTV08,DBLP:conf/popl/BonchiP13}.}

\mrevone{In the semantic subtyping tradition, CDuce~\cite{DBLP:conf/lics/FrischCB02,Frisch2004,DBLP:conf/ppdp/CastagnaF05,DBLP:journals/jacm/FrischCB08} handles the blowup through compact representations and specialized emptiness procedures, and automata-theoretic inclusion has been studied via antichain, simulation, and coinductive techniques \cite{DBLP:conf/cav/WulfDHR06,DBLP:conf/wia/BouajjaniHHTV08,DBLP:conf/popl/BonchiP13}.
Our reduction has a natural counterpart there \cite{tata2008}: a schema
can be understood as an alternating tree automaton with references as states and
$\qall$/$\qany$ as conjunction/disjunction.
The DNF computation corresponds to the classical translation to non-deterministic automata, 
so the blowup of $\DNF(\qall : [ S_{1}, \{ \qnot: S_{2} \} ])$ corresponds to the blowup of this
construction. 
The classical construction is eager and corresponds to the witness-generation algorithm we start from,
although JSON Schema adds a lot of technical complexities, having to do with the combination of unordered
objects with field names specified as patterns, ordered arrays with constraints applied only after a given position,
and the possibility to impose limits on the number of fields and elements.}{1.2}

\mrevone{These works differ from ours in three respects.
First, prior work studies the optimization of satisfiability 
procedures or Boolean representations, whereas we make the complete normalization itself 
mirror the rule-based structural comparison of the two schemas: 
it interleaves lazy expansion with early contradiction detection, matching rule-based efficiency 
without inheriting its incompleteness. 
This combination, though related to several of the lines above, does not appear to have been 
evaluated before.
Second, those lines target description logics, XML types, or automata, whereas we address 
Classical JSON Schema, where the negation-induced blowup interacts with schema-specific 
features such as pattern-defined field names and numeric bounds on the number of fields. 
Third, 
as the problem is EXPTIME-hard and the 
exponential explosion unavoidable, we focus on the empirical criterion of the success rate 
on substantial  collections of real-world schemas,  
where existing approaches provide neither satisfactory coverage nor performance.}{1.1,4.1}



\section{JSON Schema}\label{sec:json}


We say that a logical formalism is \emph{algebraic} if, whenever you substitute a subterm $S_1$ with an equivalent
subterm $S_2$ in any context~$C$, the resulting term is equivalent to $S_1$. 
We say that a logical formalism that includes negation ($\anot$) is \emph{closed under negation} when
$\anot$ can be pushed through all operators with De Morgan style rules and can thus be eliminated.

As discussed in \cite{DBLP:journals/tcs/BaaziziCGSS23}, JSON Schema is \emph{almost} algebraic, but not completely, 
and is \emph{almost} closed under negation but not completely.
Hence, in order to simplify the formal treatment, we adopt the same approach as \cite{DBLP:journals/tcs/BaaziziCGSS23}:
in our formalization, and in our implementation, we use a syntax for JSON Schema that is algebraic and 
closed under negation. This algebraic syntax facilitates formal treatment and has the same expressive 
power as the JSON schema,
since the translation from JSON schema to this syntax and vice versa takes linear time and space.
The translation is detailed 
in \cite{DBLP:journals/tcs/BaaziziCGSS23}.

\ifshort{\mrevone{This is an excerpt of the syntax, sufficient to illustrate the approach,
where the only structural operators are those on objects and strings;
the full syntax, formalizing the entirety of Classical JSON Schema, is in the full paper
\cite{full}.}{1.3}
\mrevone{$\Uri$ is the set of all URI-refs.}{}\bcolorone}
\iflong{This is the syntax we adopt.
The set $\Num$ contains the decimal numbers, that is, those that admit a finite representation
in decimal notation.
$\Uri$ is the set of all URI-refs, that we assume to be a well-defined subset of $\Str$.
}
\ifshort{$$
\begin{array}{llllll}
\multicolumn{3}{l}{
x\in\Uri, k\in\Str, n\in\Nat, m\in\Nat,b\in\Bool,  r\in \mbox{\em reg-exps}
}\\[\NL]
S & := &  \atrue \M \afalse \M \aall\ \aS  \M \aany\ \aS \M \aone\ \aS \M \anot(S)  \M \aref(X)  \\
             &&  \M \atype(U) \M \atypeSet\ \aU   \M \apatt(e) \\ 
             && \M \apProp(e,S) \M \apReq(e,S) \M \aminProps(n) \M \amaxProps(n\hide{|\Inf}) \\
\aS \!\!\!& := & [\, S_1,\ldots,S_n \,] \\
e & :=  & r \M \keykey{k} 
                \iflong{\M  \amaxLen(n) \M \aminLen(n) \\ & & }
                 \M \notp(e) \M \allp[e_1,\ldots,e_n] \M \anyp[e_1,\ldots,e_n]  \\
U & := & \aobj \M \aarray \M \astr \M \anum \M \abool \M \anull  \\
\aU \!\!\!\!\!\!& := & [\, U_1,\ldots,U_n \,] \\
E & := & [ (X_1: S_1), \ldots, (X_n: S_n) \,] \ \\
X & := & x \M \NV{x} \\
P & := & (S,E) 
\end{array}
$$\ecolor}
\iflong{$$
\begin{array}{llllll}
\multicolumn{3}{l}{
x\in\Uri, k\in\Str, n\in\Nat, m\in\Nat,b\in\Bool, q\in\Num}\\
\multicolumn{3}{l}{A \in (\Num\cup\Bool),  r\in \mbox{\em reg-exps}
}\\[\NL]
S & := &\atype(U) \M \atypeSet\ \aU \M \aconst(A) \M \anotConst(A)    
        \M \aref(X)  \\
             &&  \M \atrue \M \afalse \M \aall\ \aS  \M \aany\ \aS \M \aone\ \aS \M \anot(S)  \\
             && \M \apProp(e,S) \M \apReq(e,S) \M \aminProps(n) \M \amaxProps(n\hide{|\Inf}) \\
 \iflong{           && \M \aitem(n,S) \M \aaddIts(n,S) \M \acontAft(n,S) \M \aminIts(n)  \\
            &&  \M \amaxIts(n\hide{|\Inf})  \M \auniqIts \M \anotUniqIts \\ }
            && \M \amin(\hide{-\Inf|}q) \M \aexMin(q) \M  \amax(q\hide{|\Inf}) \M \aexMax(q) 
              \\&& 
             \M \amof(q) \M \anotMof(q)  \\ 
             \iflong{ && \M \apatt(e) \\ }
\aS \!\!\!& := & [\, S_1,\ldots,S_n \,] \\
e & :=  & r \M \keykey{k} 
                \iflong{\M  \amaxLen(n) \M \aminLen(n) \\ & & }
                 \M \notp(e) \M \allp[e_1,\ldots,e_n] \M \anyp[e_1,\ldots,e_n]  \\
U & := & \aobj \M \aarray \M \astr \M \anum \M \abool \M \anull  \\
\aU \!\!\!\!\!\!& := & [\, U_1,\ldots,U_n \,] \\
E & := & [ (X_1: S_1), \ldots, (X_n: S_n) \,] \ \\
X & := & x \M \NV{x} \\
P & := & (S,E) 
\end{array}
$$}

An algebraic JSON Schema document $P$ contains two components, a schema $S$ and an environment $E$, which is a function
that maps each reference $X$ onto a schema.

The function $E$ may be ``recursive'', meaning that it may map $X$ to a term that depends on $X$ directly,
as in $(X : \apProp(\qkw{a}, \aref(X)))$, or indirectly, that is, through a chain of references $X_{0}, X_{1}, \ldots , X_{n}$,  where each $X_i$ depends on $X_{i-1}$ and $X_{0} = X_{n} = X$.
\iflong{We say that a chain of dependencies is \emph{guarded} when there exists at least one reference $X_i$
such that every occurrence of $X_{i-1}$ in the definition
of~$X_i$ is \emph{guarded}, that is, is in the scope of an operator that is different from
{\aany}, {\aall}, {\aone}, and {\anot} (that is, an operator that is chosen among
$\apProp, \apReq, \aitem, \aaddIts, \acontAft$).}

The function $E$ is defined only on a finite set of URIs; we say that 
URI $x$ is \emph{reachable} from a 
document $P=(S,E)$ when either $\aref(x)/\aref(\NV{x})$ appears in $S$, or it appears in $E(y)/E(\NV{y})$ for some 
$y$ that is
reachable from~$P$; of course, this is a recursive definition.
\ifshort{The URIs that appear in a real JSON Schema document can either be local 
or remote; in this study, we assume that they are all local, i.e., that $E$ defines all  
URIs that are reachable 
from $S$.}

\iflong{We say that a document $P=(S,E)$ is well-formed when:
\begin{enumerate}
\item for every URI $x$ that is reachable from $P$, the function $E(x)$ is defined;
\item for every URI $x$, any chain of dependencies going from $x$ to $x$ is guarded.
\end{enumerate}
}

\iflong{The URIs that appear in a real JSON Schema document can either be local (those that start with $\#$) or remote. 
$E(x)$ is defined inside the document when the reference is local, and somewhere on the web otherwise.
This distinction is irrelevant for our study, hence, in any algebraic JSON Schema document 
$(S,E)$, we assume that $E$ contains all the local and the remote schemas that are reachable from $S$.}

We can now specify when  an instance $J$ satisfies a schema $S$ in the context of an environment $E$.

\noindent\textbf{Boolean operators.}
Every $J$ satisfies $\atrue$, no $J$ satisfies $\afalse$.
$J$ satisfies $\aall[S_1,\ldots,S_n]$ if it satisfies every $S_i$; every $J$ satisfies $\aall[\ ]$.
$J$ satisfies $\aany[S_1,\ldots,S_n]$ if it satisfies any $S_i$; no $J$ satisfies $\aany[\ ]$.
$J$ satisfies $\aone[S_1,\ldots,S_n]$ if it satisfies one $S_i$, and only one.
$J$ satisfies $\anot(S)$ if it does not satisfy $S$.

\noindent\textbf{References.} $J$ satisfies  $\aref(X)$ if it satisfies $E(X)$.
$\aref(X)$ is called the \emph{reference} operator, and 
$X$ is called ``a reference''. 
We assume that every reference ($X$) is either a 
plain URI ($x$) or a ``negated'' URI ($\NV{x}$), with the property that
whenever $E$ contains $x:S$ it also contains $\NV{x}:\anot(S)$; this assumption is 
easy to implement
and simplifies not-elimination; $\NV{\NV{x}}$ is defined to be $x$ itself.

\noindent\textbf{Structural (typed) operators.}
\mrevone{$J$ satisfies $\atype(U)$ if it has type~$U$;
it satisfies $\atypeSet\ \aU$ if its type belongs to the set $\aU$.}{1.3}
\iflong{It  satisfies $\aconst(A)$ if it is equal to $A$, and it satisfies $\anotConst(A)$ if it is different 
from $A$.}

$J$ satisfies  $\apProp(e,S)$ iff, if $J$ is an object, then, every field of $J$ whose name matches $e$ has a value that satisfies $S$.
Here, ``if $J$ is an object'' means that every $J$ that is not an object \emph{satisfies} $\apProp(e,S)$. This is true for all structural
operators that we describe next --- which we call the \emph{analytical structural operators}: if $J$ does not belong to the type that is being analyzed, 
then $J$ satisfies the operator.

$J$ satisfies  $\apReq(e,S)$ iff, if $J$ is an object, then it has at least one field whose name matches $e$ and whose value satisfies $S$.

$J$ satisfies  $\aminProps(n)$ iff, if $J$ is an object, then it has at least~$n$ fields.
$J$ satisfies  $\amaxProps(n)$ iff, if $J$ is an object, then it has at most~$n$ fields.
\iflong{Every $J$ satisfies  $\amaxProps(\Inf)$.}

\iflong{$J$ satisfies  $\aitem(n,S)$ iff, if $J$ is an array, then, if it has an item at position $n$, then its value satisfies $S$; positions are numbers 
starting from $0$. Observe that an array with no item at position $n$ satisfies $\aitem(n,S)$.

$J$ satisfies  $\aaddIts(n,S)$ iff, if $J$ is an array, then, if it has any item at a position $i\geq n$, then its value satisfies $S$.
For example, both $\aall[   \aaddIts(0,\Rf{X})]$
and $\aall[  \aitem(0,\Rf{X}), \aaddIts(1,\Rf{X})]$ force all items to satisfy $\Rf{X}$.

$J$ satisfies  $\acontAft(n,S)$ iff, if $J$ is an array, then it has at least one item at a position $i\geq n$ whose value satisfies $S$.
For example, $\acontAft(0,S)$ is satisfied by an array with an item in any position that satisfies $S$,
while $\acontAft(1,S)$ requires an item that satisfies $S$ in a position different from the first (whose index is $0$).

$J$ satisfies  $\aminIts(n+1)$ if it satisfies $\acontAft(n,\atrue)$; every $J$ satisfies $\aminIts(0)$. 

$J$ satisfies  $\amaxIts(n)$ if it satisfies 
$\aitem(n,\afalse)$\iflong{; every $J$ satisfies $\amaxIts(\Inf)$}. 

$J$ satisfies  $\auniqIts$  iff, if $J$ is an array that contains at least two items;
for example, all of $[\ ]$, $[ 1 ]$, and $[1,2,3]$ satisfy $\auniqIts$,
while $[1,2,1]$ violates it. 

$J$ satisfies  $\anotUniqIts$  iff, if $J$ is an array, then it contains at least two items with equal values.
For example, all of $[\ ]$, $[ 1 ]$, and $[1,2,3]$ violate $\anotUniqIts$,
while $[1,2,1]$ satisfies it.
}

\iflong{$J$ satisfies  $\amin(q) $ iff, if $J$ is a number, then $J\geq q$\iflong{; any $J$ satisfies $\amin(-\Inf)$}.
$J$ satisfies  $\aexMin(q) $ iff, if $J$ is a number, then $J> q$.
$J$ satisfies  $\amax(q) $ iff, if $J$ is a number, then $J\leq q$\iflong{; any $J$ satisfies $\amax(\Inf)$}.
$J$ satisfies  $\aexMax(q) $ iff, if $J$ is a number, then~$J< q$.

$J$ satisfies  $\amof(q)$ iff, if $J$ is a number, then there exists an integer~$l$ such that $J=l\times q$; here, $q$ can be any decimal number. 
$J$ satisfies  $\anotMof(q)$ iff, if $J$ is a number, then there exists no integer~$l$ such that $J=l\times q$.}
\iflong{

\noindent\textbf{Regular expressions.}}
\mrevone{$J$ satisfies  $\apatt(e) $ iff, if $J$ is a string, then it belongs to the set $\rlan{e}$
of the
strings that match the pattern $e$; $\rlan{e}$ is defined as follows:
if  $e=r$ with $r\in \mbox{\em reg-exps}$, then $\rlan{e}$ contains the strings matched by 
$r$.}{1.3}

$\rlan{\keykey{k}}$ only contains~$k$. $\rlan{\notp(e)}$ is the complement of $\rlan{e}$.
\iflong{$\rlan{\amaxLen(n)}$ contains the strings whose length is less or
equal than $n$. Likewise, $\rlan{\aminLen(n)}$ contains the strings whose length is greater or
equal than $n$. }

$\rlan{\allp[e_1,\ldots,e_n]}$ is the intersection of $\rlan{e_1}\ldots\rlan{e_n}$.

$\rlan{\anyp[e_1,\ldots,e_n]}$ is  their union.


Unlike JSON Schema, this algebra is \emph{not-complete}, meaning that for 
every
operator, such as $\amof(q)$, we have a dual operator, in this case $\anotMof(q)$,
which allows $\anot(\amof(q))$
to be expressed without negation.
\iflong{

Most of JSON Schema operators actually have such a dual; the five exceptions are
$\apProp(e,S)$, $\aaddIts(n,S)$, $\amof(q)$, $\auniqIts$, and $\apatt(e)$;
these operators have the following duals in the algebra, respectively:
$\apReq(e,S)$, $\acontAft(n,S)$, $\anotMof(q)$, $\anotUniqIts$, and $\apatt(\anotPatt(e))$.
These operators do not add expressive power, but they greatly simplify the formal treatment.}
\ifshort{These dual operators do not add expressive power w.r.t.\ JSON Schema, but they greatly simplify the formal treatment.}

\section{The \emph{Witness-Generation} inclusion algorithm}\label{sec:WGinclusion}

The \emph{Witness-Generation Inclusion} algorithm ({\WG}) \cite{DBLP:journals/pvldb/AttoucheBCGSS22}, in order to check whether $S_1 \subt S_2$, applies a 
{\BaseWG} 
algorithm
to the schema $\aall[S_1,\anot(S_2)]$\iflong{: if a
witness $J$ is produced, then there exists an element of $S_1$ that is not in $S_2$,
hence the inclusion does not hold};
 if the result is \key{unsatisfiable}, no such element exists, hence the inclusion holds.

We are now going to describe the role of normalization in this algorithm,
and why normalization is problematic.

\subsection{Witness Generation algorithm and {\CDNF}}\label{sec:DNF}

\renewcommand{\KK}{K}


A schema $D$ is in \emph{DNF} (\emph{Disjunctive Normal Form}) if it is 
a disjunction ($D$) of conjunctions ($C$) of structural operators  ($K$):
\[
\begin{array}{lll@{\quad}lll}
D  := & \aany[\,(C,)^* \ C\, ] \ \M\ \aany [\, ]
  \\[\NL] C  := & \aall[ \, (K,)^* \ K\, ] \ \M\  \aall [\, ] \\[\NL]
K  := & \atype(U) \M \apatt(e) \ldots
\end{array}
\]
When the 
{\BaseWG} 
algorithm is applied to any schema $S$, it
first  computes a \emph{DNF} of $S$,
that satisfies some extra properties:
\begin{enumerate}
\item it is \emph{canonical}: for every conjunction $C_i=\aall[\KK^i_{1},\ldots,\KK^i_{m_{i}}]$,
the first argument $\KK^i_{1}$ is exactly $\atype(T)$, for some type $T$, and all the 
other $\KK^i_j$ are structural operators that analyze type $T$:
every conjunction is \emph{type-homogeneous};
\item it is \emph{stratified}: when any $\KK^i_j$ has a schema $S'$ as argument,
   as in $\KK^i_j=\apProp(e,S')$, that schema $S'$ is a reference $\aref(x)$:
   $\KK^i_j=\apProp(e,\aref(x))$ \mrevtwo{(nesting is mediated by references);}{2.6}
\item every object disjunct  $\aall[\atype(\aobj),\KK^i_1,\ldots,\KK^i_n]$ is \emph{prepared}, 
meaning that:
\begin{enumerate}
\item  is partitioned: for any pair $(\KK^i_{j_1},\KK^i_{j_2})$,
      where
      $$\begin{array}{lllll}
                  \KK^i_{j_1}&=&(\apProp\ |\ \apReq)(e_1,\aref(x_1)) \\
                  \KK^i_{j_2}&=&(\apProp\ |\ \apReq)(e_2,\aref(x_2))
\end{array}$$
        we have that either
       $e_1=e_2$ or $\rlan{e_1}\cap\rlan{e_2}=\emptyset$;
       this applies to homogeneous pairs (\apProp-\apProp, \apReq-\apReq)
       and to heterogeneous pairs (\apProp-\apReq, \apReq-\apProp)
\item is \emph{and-completed}: 
          for any pair of operators $(\KK^i_{j_1},\KK^i_{j_2})$ defined as in the previous condition,
          if we are in the case $e_1=e_2$, then the environment $E$ contains a reference that
          is equivalent to $\aall[\aref(x_1),\aref(x_2)]$.
\end{enumerate}
\ifshort{\item every array type is also \emph{prepared}; preparation for an array disjunct
   is described in the full \mrevone{paper \cite{full}}{1.10}.
}
\iflong{
\item every array type is \emph{prepared}, where preparation for an array disjunct
     $C_i=\aall[\KK^i_j,\ldots,\KK^i_{m(j)}$ can be described as follows:
\begin{enumerate}
\item is partitioned: exactly one of the $\KK^i_j$ schemas is \\
        $\aaddIts(n_a,\aref(x_a))$;
        for every $l\in\SetFromTo{0}{n_a-1}$, exactly one of the $\KK^i_l$ schemas is $\aitem(l,\aref(x_l))$;
        for every $j$ such that $\KK^i_{j}=\acontAft(n_j,\aref(x_j))$, we have that $n_j \geq n_a$;
\item is and-completed: 
          for any $j$ such that \\
          $\KK^i_{j}=\acontAft(n_j,\aref(x_j))$, then the environment $E$ contains a 
          reference that is equivalent to $\aall[\aref(x_a),\allowbreak\aref(x_j)]$;
          for any $j_1$ and $j_2$ such that $\KK^i_{j_1}=\acontAft(n_{j_1},\allowbreak\aref(x_{j_1}))$,
          and $\KK^i_{j_2}=\acontAft(n_{j_2},\aref(x_{j_2}))$, $E$ contains a 
          reference that is equivalent to $\aall[\aref(x_{j_1}),\aref(x_{j_2})]$;
\end{enumerate}
}
 \end{enumerate}
 
 \ifshort{
Given a document $(S,E)$, once $S$ and all schemas $E(x)$  are in a canonical and prepared DNF, 
the algorithm performs a bottom-up  generation phase, where it generates a witness for every reference~$x$, until a witness for the root schema is generated, or until it determines that $S$ is unsatisfiable; stratification, partitioning, and and-completion
ensure that the generation of a witness for an object or for an array schema can be reduced to generation 
of a witness for all the references that are reachable from that schema.
 }
 
\iflong{
 The partitioning property for objects ensures that whenever two operators, which may be
 both $\apProp$, both $\apReq$, or one of each, may both affect 
 the same field, since its name matches the patterns of both, then the two operators have exactly 
 the same pattern, so that their interaction can be analyzed; and-completion ensures that, in this case, the compatibility of the respective schemas can be analyzed by checking the satisfiability of
 a specific reference, the one that is equivalent to
 $\aall[\aref(x_1),\aref(x_2)]$.
 Similarly, for arrays, partitioning ensures that all $\aaddIts$ and all $\aitem$ affect independent
 positions of the array, and that every $\acontAft$ only interacts with the only $\aaddIts$
 and with the other $\acontAft$; and-completion ensures that, for any interacting operators, the compatibility of the respective schemas can be analyzed by checking the satisfiability of
 a reference in $E$.

 This computation of the DNF is then followed by a phase of \emph{generation} where the DNF is used to
 determine whether $S$ was unsatisfiable, or to generate a witness;
 we do not describe it here.
 }


\subsection{Normalizing $\tkw{all}[S_1,\tkw{not}(S_2)]$}\label{sec:DNFNotObj}

\newcommand{\AndA}{\wedge}
\newcommand{\OrO}{\vee}

\newcommand{\AndDD}{\,_{\akw{D}}\!\!\AndA_{\akw{D}}}
\newcommand{\OrDD}{\,_{\akw{D}}\!\!\OrO_{\akw{D}}}
\newcommand{\AndCC}{\,_{\akw{C}}\!\AndA_{\akw{C}}}

\newcommand{\cpar}[3]{%
  \mspace{10mu}\mathllap{{}_{\!#1}\mkern-2mu}{#2}_{#3}\,%
}

\renewcommand{\AndDD}{\cpar{\akw{D}}{\wedge}{\akw{D}}}
\renewcommand{\AndCC}{\cpar{\akw{C}}{\wedge}{\akw{C}}}
\renewcommand{\OrDD}{\cpar{\akw{D}}{\vee}{\akw{D}}}

The {\WG}  inclusion algorithm, when applied to $S_1\subt S_2$, applies 
the 
{\BaseWG} algorithm
 to $\aall[S_1,\anot(S_2)]$, hence it computes a {\CDNF}
of $\aall[S_1,\anot(S_2)]$.
\iflong{This step 
is potentially very expensive.}

The recursive algorithm used to compute this {\CDNF} 
can be described as follows,
where we assume that every $\KK_i$ has already been rewritten in a form that is stratified and has no negation.

$$
\begin{array}{llllll}
\!\!\aDNF(\aref(x)) &\!\!=\!\!& \aDNF(E(x)) \\
\!\!\aDNF(\aany[S_1,\ldots,S_n]) &\!\!=\!\!& \aDNF(S_1)\OrDD  \ldots\OrDD   \aDNF(S_n) \\
\!\!\aDNF(\aall[S_1,\ldots,S_n]) &\!\!=\!\!& \aDNF(S_1)\AndDD  \ldots\AndDD   \aDNF(S_n) \\
\!\!\aDNF(S) &\!\!=\!\!& \aany[\aall[S]]   \ \ \ \ \qquad\qquad \mbox{\em otherwise} & 
\end{array}
$$

{\small
$$
\begin{array}{llllll}
\multicolumn{3}{l}{
\!\!\aany[C^1_1,\ldots,C^1_n] \OrDD   \aany[C^2_1,\ldots,C^2_m]  } \\
\qquad\qquad\qquad\qquad\qquad\quad
\!\!&\!\!=\!\!& 
\!\!\aany[\Set{C^1_1,\ldots,C^1_n}\cup \Set{C^2_1,\ldots,C^2_m}] \\[2\NL]
\multicolumn{3}{l}{
\!\!\aany[C^1_1,\ldots,C^1_n] \AndDD   \aany[C^2_1,\ldots,C^2_m] 
}\\
\!\!&\!\!=\!\!& 
\!\!\aany[\SetST{C^1_i\AndCC    C^2_j}{i\!\in\! 1..n, \ j\!\in\! 1..m}] \\[2\NL]
\multicolumn{3}{l}{
\!\!\aall[\KK^1_1,\ldots,\KK^1_n] \AndCC    \aall[\KK^2_1,\ldots,\KK^2_m] 
}\\
\!\!&\!\!=\!\!& 
\!\!\aall[\Set{\KK^1_1,\ldots,\KK^1_n}\cup \Set{\KK^2_1,\ldots,\KK^2_m}] \\
\end{array}
$$
}

If $D_1$ and $D_2$ are schemas in  {\CDNF}, then $D_1\OrDD  D_2$ computes their disjunction, and
 $D_1\AndDD  D_2$ computes their conjunction, by taking a pairwise conjunction $C^1_i\AndCC    C^2_j$ 
 of each disjunct $C^1_i$ of $D_1$ with each  disjunct $C^2_j$ of $D_2$. 
Hence, if we have a set of schemas $D_i$ each with $m$ disjuncts, then 
 $D_1 \AndDD \ldots \AndDD D_n$, has $m^n$ disjuncts.

\hide{
$$
\begin{array}{llllll}
\aany\  \aC_1 \cup \aany\ \aC_2 = \aany\ (\aC_1 \cup \aC_2) \\[\NL]
\aany [\aall\ \aS^1_1, \ldots,  \aall\ \aS^1_n]\times \aany [\aall\ \aS^2_1, \ldots,  \aall\ \aS^2_m] &\!\!=\!\! 
\\
\qquad\qquad \aany \SetST{\aall (\ \aS^1_i \cup \aS^2_j)}{  i \!\in\! 1..n, j \!\in\! 1..m }\\
\end{array}
$$
}

\hide{
In the rest of this section, and in other occasions in the paper, 
we will write that 
``the negation of $\apProp(e,S)$ is $\apReq(e,\anot(S))$'', which is wrong,
since the negation is actually
$$\aany[\atype(\anull,\astr,\ldots),\apReq(e,\anot(S))];$$
we make this imprecise statement in situations where the extra context
$\aany[\atype(\ldots),\_]$ complicates the description but does not weaken the
general observation that we are stating.
}

The {\CDNF} of a schema $S$ is, in the worst case, exponentially bigger than $S$.
However, the vast majority of schemas that are found on GitHub 
are just object or array schemas that either contain
no boolean operator or contain only disjunctions of object schemas\iflong{, as happens in the
specification of an interface that may accept requests whose possible shapes are listed as ``either
$S_1$ or $S_2$ or...'' where each $S_i$ is an object schema}.
Hence, as observed in~\cite{DBLP:journals/pvldb/AttoucheBCGSS22}, the {\CDNF} of a real world schema
is, typically, not much bigger than the original schema\ifshort{.}\iflong{, with the important exception of schemas using the
$\aone$ operator, that are discussed in the next subsection}\mrevtwo{\footnote{\bcolortwo
Of course, this observation may be biased by the specific dataset that was considered. \ecolor}}{2.4}

\newcommand{\nx}{\NV{x}}
\newcommand{\nxt}[2]{\NV{x^{#1}_{#2}}}
\newcommand{\pushNot}{\anot}

The situation is totally different when you consider the schema
$\aDNF(\aall[S_1,\anot(S_2)])$. Assume $S_2=\aany[O_1,\ldots,O_n]$, 
where every~$O_i$ is an object schema.
For simplicity, assume that each $O_i$ is a schema that contains $m$
assertions with shape $\apProp(\keykey{k^i_j},\aref(x^i_j)).$
Then, when we apply not-elimination to $\anot(S_2)$, we get:
\iflong{$$
\begin{array}{llllll}
\pushNot(S_2) = \pushNot(\aany[\ldots,O_i,\ldots]) \\
= \pushNot(\aany[\ldots,\aall [ \ldots,\apProp(\keykey{k^i_j},\aref(x^i_j)),\ldots],\ldots]) \\
\to \aall[\ldots,\pushNot(\aall [ \ldots,\apProp(\keykey{k^i_j},\aref(x^i_j)),\ldots]),\ldots] \\
\to \aall[\ldots,\aany [ \ldots,\pushNot(\apProp(\keykey{k^i_j},\aref(x^i_j))),\ldots],\ldots] \\
\to \aall[\ldots,\aany [ \ldots,\apReq(\keykey{k^i_j},\aref(\nxt{i}{j})),\ldots],\ldots] \\
\end{array}
$$
To sum up, $\anot(S_2)$ is rewritten as}
$$
\begin{array}{lllll}
\aall [ \!\!\!&\aany [\apReq(\keykey{k^1_1},\aref(\nxt{1}{1})), \ldots, \apReq(\keykey{k^1_m},\aref(\nxt{1}{m}))],\\
&\ldots, \\
   &  \aany [\apReq(\keykey{k^n_1},\aref(\nxt{n}{1})), \ldots, \apReq(\keykey{k^n_m},\aref(\nxt{n}{m}))]
       ]
\end{array}
$$

When we apply $\aDNF$ to this term, we multiply $n$ disjunctions of~$m$ subschemas each,
hence we end up with a {\CDNF} term with~$m^n$ different disjuncts.
\iflong{More precisely, for each of the $m^n$  finite functions $f \in (1..n) \to (1..m)$,
the DNF contains the corresponding conjunction:
\begin{equation*}
	\begin{split}
		C_f = \aall[ &\apReq(\keykey{k^1_{f(1)}},\aref(\nx^1_{f(1)})), \ldots, \\
		&\apReq(\keykey{k^n_{f(n)}},\aref(\nx^n_{f(n)}))]
	\end{split}
\end{equation*}
}

In practice, $n$ may be \mrevone{large enough}{1.6} to make a term of $m^n$ size unmanageable.
\hide{

Finally, we compute $\aDNF(\aall[S_1,\anot(S_2)])$ as $\aDNF(S_1) \times D_2$.

if $S_1=S_2$, we combine the $m$ disjuncts of $S_1$ with the $m^n$ disjuncts of $D_2$,
and we get $m^{n+1}$ disjuncts with the following shapes:
for $i\in 1..n, f\in (1..n) \to (1..m)$:
$$
\begin{array}{lllll}
C_{i,f} = \aall [ O_i, C_f ]
\end{array}
$$

$C_{i,f}$ is hence a conjunction that contains the $m$ $\apProp(\keykey{k^i_j},\aref(x^i_j))$ of 
$O_i$ and the $n$ $\apReq(\keykey{k^{i'}_{f(i')}},\aref(\nx^{i'}_{f(i')}))$ of $C_f$.
Hence, for any $i\in 1..n, f\in (1..n) \to (1..m)$, $C_{i,f}$
contains both $\apProp(\keykey{k^i_{f(i)}},\aref(x^i_{f(i)}))$
and 
$\apReq(\keykey{k^i_{f(i)}},\aref(\nx^i_{f(i)}))$.
These two schemas are incompatible, since the second requires the presence of one field with name
$k^i_{f(i)}$ and schema $\aref(\nx^i_{f(i)})$, but the first requires every  field with name
$k^i_{f(i)}$ to satisfy $\aref(x^i_{f(i)})$, and $\aref(x^i_{f(i)})$ is incompatible with $\aref(\nx^i_{f(i)})$.
}
\hide{

Finally, we compute $\aDNF(\aall[S_1,\anot(S_2)])$ as $\aDNF(S_1) \AndDD   D_2$,
generating a DNF with $m^{n+1}$
disjuncts each of size $(m+n)$. If the inclusion holds, all of these disjuncts  are unsatisfiable;
hence, if we enrich the base algorithm with an immediate
test of compatibility between the elements of a conjunctions, then every of these $m^{n+1}$ disjuncts
is immediately eliminated from the DNF as still as it is generated, but still they have to be all generated.

}
Exponential explosion during normalization of $\anot(S_2)$ 
is the \mrevone{core issue}{1.6} that makes the {\WG}  algorithm problematic when used to test inclusion of large schemas.

\iflong{
\subsection{The problem with $\tkw{one}$}

$J$ satisfies $\aone[S_1,\ldots,S_n]$ holds when $J$ satisfies one and only one of the $S_i$
schemas, hence it is translated as follows:

$$
\begin{array}{lllll}	
\aone[S_1,\ldots,S_n] = \aany[
    \!\!\!&\aall[S_1,\anot(S_2),\ldots,\anot(S_n)],
    \ \ldots,\\[\NL]
        &\aall[\anot(S_1),\ldots,\anot(S_{n-1}),S_n]
        &\!\!\!]
\end{array}
$$

Similarly to in the previous case, the conjunction of a positive
object and $n-1$ negated object schemas each of size 
$m$ results in a DNF with $m^{n-1}$ disjuncts of size $m+n-1$. In this case,
the exponentiality is not generated by the $\anot$ due to the inclusion, but 
depends on the negation that is implicit in the $\aone$ operator.
We do not discuss this issue in this paper for reasons of space, but also
because the problem presents itself in the rule-based algorithm and in the 
{\WG}  algorithm in the same way, but {\RNAlg} can help in this case.
}

\hide{
$$
\begin{array}{lllll}	
\aone[S_1,\ldots,S_n] \\
= \aany[\ldots,
    \aall[\anot(S_1),\ldots,\anot(S_{i-1}),S_i,\anot(S_{i+1}),\ldots,\anot(S_n)],
    \ldots]
\end{array}
$$
}

\hide{

 more difficult to reason about and to validate than 
$\aany$, since, while $\aany$ just accepts the union of its arguments, while $\aone$ requires a 
negative test 

In \cite{usage} we measured the use of logical operators in real-world schemas, and we discovered
that the $\aone$ operators is much more common that $\aany$.
We were surprise, since $\aone$ is more complex to test and to understand than $\aany$, despite
the fact that is implicitly used negation, since $\aone[S_1,S_2]$ 
We analyzed the schema using $\aone$ and we discovered that, in the vast majority of the cases $\aone$ was equivalent to $\aany$ since the different branches 
}


\section{Rule-based Inclusion}\label{sec:rulebased}


Rule-based \mrevtwo{schema inclusion}{2.5} checking, as defined in \cite{DBLP:conf/issta/HabibSHP21}, consists of two steps.
First, the compared schemas are  reduced to a normal form that is almost a DNF
(canonicalization and simplification).

\iflong{After canonicalization/simplification (Section \ref{sec:canonic}), the algorithm compares two schemas 
$\aany[C^1_1,\ldots,C^1_n]$
and $\aany[C^2_1,\ldots,C^2_m]$ that are \emph{almost} in DNF, by verifying, through 
the set of rules described in 
Section \ref{sec:rules}, that
for every conjunction $C^1_i$ there exists a conjunction $C^2_j$ such that $C^1_i \xSub C^2_j$.

The algorithm may return either ``included'', ``not included'', or ``do not know''.
}

\ifshort{After canonicalization/simplification (Section \ref{sec:canonic}), the algorithm compares two schemas 
$\aany[C^1_1,\ldots,C^1_n]$
and $\aany[C^2_1,\ldots,C^2_m]$ that are \emph{almost} in DNF, 
using a set of structural rules (Section \ref{sec:rules}).}

We now discuss these two phases and explain why this algorithm is, by design, 
sound but not complete.

\subsection{Canonicalization and simplification} \label{sec:canonic}

\newcommand{\ADNF}{almost-DNF}

Canonicalization and simplification transform any schema into a
Normal Form that we call here
\emph{Almost Disjunctive Normal Form}\iflong{,
to emphasize the fact that it is similar in spirit to the DNF, but is not,
technically speaking, a DNF}.

\ifshort{\mrevtwo{The main difference between this {\ADNF} and a complete DNF as described in Section \ref{sec:DNF}
comes from the fact that 
the algebra used in \cite{DBLP:conf/issta/HabibSHP21} is not closed under negation, because of the operators 
{\qmof}, {\qpattProps}, and {\qaddIts}, which makes it impossible to push negation
through these operators (see~\cite{DBLP:journals/tcs/BaaziziCGSS23}). This is just a technical 
detail: if the RB adopted our algebra and hence a full DNF, this would enlarge the amount
of schemas that it can treat, but would not address the real reasons of its intrinsic incompleteness,
which we are going to describe next.
}{2.2}}

\iflong{The main difference between this {\ADNF} and a complete DNF as described in Section \ref{sec:DNF}
comes from the fact that 
the algebra used in \cite{DBLP:conf/issta/HabibSHP21} is not closed under negation, because of the operators 
{\qmof}, {\qpattProps}, and {\qaddIts}, which makes it impossible to push negation
through these operators (see~\cite{DBLP:journals/tcs/BaaziziCGSS23})\iflong{; 
moreover, the authors do not introduce any technique to push negation 
through references}.
\iflong{For this reason,  the {\ADNF} does not completely eliminate negation, and is
not able to completely eliminate the presence of any boolean operator inside conjunction.}

The fact that {\ADNF} is not a real DNF is the first reason why this approach,
as implemented in \cite{DBLP:conf/issta/HabibSHP21}, is not complete,
but we do not consider this as a fundamental limitation of the approach: if one adopts
our algebra to represent JSON Schema, then this limit can be overcome.
The real reasons why the approach is incomplete  are introduced in the next section.
}



\subsection{Rule-based analysis} \label{sec:rules}

\paragraph{Rule (anyOf)}
The rule-based phase compares two schemas that are in {\ADNF} using the two (\textsc{anyOf-*})
rules that we report here using our notation.\footnote{
In \cite{DBLP:conf/issta/HabibSHP21}, the two rules are actually merged into one, and they add
a \emph{nonOverlapping} condition, which is irrelevant for our discussion.} 

\hide{
{\small
\begin{verbatim}
(anyOf)
%\kw{nonOverlapping}([S1,..Sm])
\kw{forall}\ i \in 1..n. \kw{exists}\ j \in 1..m. C'i < Cj
----------------------------------------------------------------------------------------
{ anyOf : [S'1,..,S'n] }  \subtr { anyOf : [S1,..,Sm] }
\end{verbatim}
}}

\infrule[anyOf-l]
{
\forall i \in \{1, \ldots , n\}.\ \ C^{\prime}_{i} \subtr D
}
{ \aany [C^{\prime}_{1}, \ldots ,C^{\prime}_{n}]  \subtr D
}

\infrule[anyOf-r]
{
\exists j \in \{1, \ldots , m\}.\ C^{\prime} \subtr C_{j}
}
{ C^{\prime} \subtr \aany  [C_{1}, \ldots ,C_{m}] 
}

Rule (\textsc{anyOf-l}) is very natural and is complete: $\aany [C^{\prime}_{1}, \ldots ,C^{\prime}_{n}]$ is
\mrevone{included in}{1.7} $D$ if, and only if, every $C^{\prime}_{i}$ is \mrevone{included in}{1.7} $D$.

Rule (\textsc{anyOf-r}) is applied after rule (\textsc{anyOf-l}) has eliminated $\aany$
from the LHS, and it 
requires that the LHS is completely \mrevone{included in}{1.7} just one disjunct of the RHS.
\iflong{Hence, the combination of these two rules reduces the problem
of comparing two schemas in DNF with $m$ and $n$ disjuncts 
to that of comparing at most $m\times n$ pairs of 
canonical conjunctive schemas.}

The (\textsc{anyOf-r}) rule is sound, but is not complete. Consider, for instance,  the following comparison.
$$
\begin{array}{llllllllll}
\apProp(\keykey{a},\aany[S_1,S_2]) \subt \aany[\apProp(\keykey{a},S_1),\apProp(\keykey{a},S_2)]
\end{array}
$$

The subschema is satisfied by any 
\iflong{JSON value that is not an object, and by any }
JSON object $J$ such 
that, for every field of $J$ whose name matches $\keykey{a}$,
its value matches $\aany[S_{1},S_{2}]$. 
Every such object has at most one field that matches $\keykey{a}$,
hence the value of this only field either satisfies $S_1$ or it satisfies $S_2$, 
hence every object that satisfies the only branch of the subschema also satisfies one of the 
two branches of the superschema.

However,
rule (\textsc{anyOf-r}) reduces this judgment to the disjunction of the following two 
 subproblems:

$$
\begin{array}{llllllllll}
\apProp(\keykey{a},\aany[S_1,S_2]) \ \subt\ \apProp(\keykey{a},S_1)\\[\NL]
\apProp(\keykey{a},\aany[S_1,S_2]) \ \subt\ \apProp(\keykey{a},S_2)
\end{array}
$$

None of the two separate judgments above holds, hence the rule (\textsc{anyOf-r})
cannot prove the original judgment, hence that rule is not complete.
\iflong{This is a second source of incompleteness of the rule-based approach, and this
one seems
very difficult to solve in this context.}
\ifshort{This is a fundamental source of incompleteness of the rule-based approach.}

\hide{
The (\textsc{anyOf-r}) rule is sound, but is not complete; consider the following comparison.\GG{Translate
using our notation}

{\begin{verbatim}
{ pattProps { a+ : (S1 or S2) } } 
  \subtr  { minProps(2) or { pattProps(a+:S1), maxProps(1)}
            or { pattProps(a+:S2), maxProps(1) }}
\end{verbatim}}

The subschema is satisfied by any JSON value that is not an object, and by any JSON object such that, for every object field whose name matches \qkw{a+}
its value matches $\aany[S_{1},S_{2}]$. 
Every such object either has at least two fields, or it has at most one field; in this second case, this field, if it is present and if it matches \qkw{a+}, either
has a value that satisfies $S_{1}$ or has a value that satisfies $S_{2}$, hence every object that satisfies the subschema also satisfies one of the 
three branches of the superschema.

Rule (\textsc{non-overlapping anyOf}) of \cite{DBLP:conf/issta/HabibSHP21} reduces this judgment to the disjunction of the following three subproblems:

{\begin{verbatim}
{ pattProps { a+ : (S1 or S2) } } \subt   minProps(2) 
{ pattProps { a+ : (S1 or S2) } } \subt   { pattProps(a:S1), maxProps(1) } 
{ pattProps { a+ : (S1 or S2) } } \subt   { pattProps(a:S2), maxProps(1) }
\end{verbatim}}

However, none of the three judgments above holds, hence the rule (anyOf)
cannot prove the original judgment.
}

\paragraph{Rule (uninhabited)}
The incomplete (\textsc{anyOf-r}) rule reduces the problem
of comparing two schemas in DNF to that of comparing pairs of 
canonical conjunctive schemas.

A canonical schema $C'$ is \mrevone{included in}{1.7} a canonical schema $C$
in two cases: (1) when
 $C'$ is unsatisfiable, and (2) when
$C'$ and $C$ \mrevone{describe values of the same primitive JSON type}{1.8, 2.7}, and the constraints that are encoded by 
$C'$  imply those that are encoded by $C$.

This first possibility is encoded by rule (\textsc{uninhabited}).

\infrule[uninhabited]
{
\neg \ \kw{satisfiable}(C_{1})
}
{C_{1} \subtr C_{2}
}

Rule (\textsc{uninhabited}) can be  implemented either using a complete 
satisfiability test, or an incomplete one.
However, the only complete satisfiability
algorithm currently published is  the {\BaseWG} algorithm 
by Attouche et al.~\cite{DBLP:journals/pvldb/AttoucheBCGSS22},  and a systematic use of such algorithm for every step of the rule-based
algorithm would make the entire
algorithm very slow; once one decides to include the entire
{\BaseWG} algorithm inside the rule-based
approach, maintaining
both algorithms in the same system makes little sense.
The only reasonable choice is the one adopted by Habib et al.~\cite{DBLP:conf/issta/HabibSHP21}, 
which use an efficient but incomplete satisfiability algorithm.
Hence, this rule is another intrinsic cause of incompleteness
of the rule-based approach. 

\paragraph{The structural rules}
\iflong{
When the (\textsc{uninhabited}) rule fails since~$C_1$ is not proved to be unsatisfiable,
either the two schemas regard different types (for example, {\aobj} and {\aarray}), 
hence  $C_1$ is not included in $C_2$,
or they regard the same type, and we apply one of the six structural rules of 
Figure \ref{fig:rules}.
In these rules, $C_1$ and $C_2$ are canonical conjunctions, and,
following the notation of  \cite{DBLP:conf/issta/HabibSHP21},
we use~$C.\akw{x}$ to extract the parameter of the $\akw{x}$ operator
that appears in $C$.

We do not describe the details of the algorithm, but just the general approach.
}

\ifshort{
When the (\textsc{uninhabited}) rule fails since~$C_1$ is satisfiable,
either the two schemas \mrevone{apply to different types}{1.8, 2.7} (for example, {\aobj} and {\astr}), 
hence  $C_1$ is not included in $C_2$,
or they apply to the same basic JSON type (for example, $C_{1}$ applies to {\aobj} and $C_{2}$ to {\aobj} as well), and we apply the structural rule of 
that type.

For example, the rule for objects is the following one; we adopt  the notation of  \cite{DBLP:conf/issta/HabibSHP21},
and use $C.\akw{x}$ to extract the parameter of the $\akw{x}$ operator
that appears in $C$.
}

\infrule[object]
{
C_{1}.\atype = \aobj \qquad C_{2}.\atype = \aobj\\
C_{1}.\aminProps \geq C_{2}.\aminProps \qquad C_{1}.\amaxProps \leq C_{2}.\amaxProps\\
\forall (e_2,S_{2}) \in C_{2}.\apProp:\ \exists  (e_1,S_{1}) \in C_{1}.\apProp:\ 
                                                  \ e_1 \supseteq e_2 \ \mbox{\ and\ } \ S_{1} \subtr S_{2} \\
\forall (e_2,S_{2}) \in C_{2}.\apReq:
\exists\ (e_1,S_{1}) \in C_{1}.\apReq: \ \ e_1 \subseteq e_2 \ \mbox{\ and\ } \ S_{1} \subtr S_{2} \\
}
{
C_{1} \subtr C_{2}
}

Here, the condition
$C_{1}.\aminProps \geq C_{2}.\aminProps$ indicates that the $\aminProps$
constraint of $C_1$ implies that of $C_2$, and similarly for $\amaxProps$.
The condition
$$\forall (e_2,S_{2}) \in C_{2}.\apProp:\ \exists  (e_1,S_{1}) \in C_{1}.\apProp:\ 
                                                  \ e_1 \supseteq e_2 \ \mbox{\ and\ } \ S_{1} \subtr S_{2} $$
requires that, for every field in the object that is constrained by a 
$\apProp (e_2,S_{2}) \in C_{2}$, there exists $\apProp (e_1,S_{1}) \in C_{1}$
that matches the same field and imposes a schema $S_1$ that is stronger than $S_2$.
The condition for $\apReq$ is analogous. 

This rule captures most common inclusion cases, but it is not complete, since 
there are situations where some operator of $C_2$
is implicitly enforced by the combined effect of different operators in~$C_1$.
For example, the following inclusion holds when $S_{u}$ is unsatisfiable,
since, in that case, the only name available for a field of an object that satisfies the 
schema in the  left hand side is \qkw{a}; this inclusion is not provable in the rule-based system:
$$
 \apProp(\anotPatt(\keykey{a}) : S_{u}) \ \ \subtr\ \  \amaxProps(1)
$$


The (\textsc{array}) rule has similar issues; hence, we can conclude that the incompleteness
of the array and object rules, which cannot be solved without a complete satisfiability test,
is another inherent cause of incompleteness for the rule-based approach.

\renewcommand{\akw}[1]{\ensuremath{\mbox{\tt{\scriptsize #1}}}}

\iflong{
\begin{figure}[htbp]
{\footnotesize
\infrule[null]
{
C_{1}.type = null \qquad C_{2}.type = null
}
{C_{1} \subtr C_{2}
}

\infrule[boolean]
{
C_{1}.type = boolean \qquad C_{2}.type = boolean  \qquad s{1}.enum \subseteq C_{2}.enum
}
{C_{1} \subtr C_{2}
}

\infrule[string]
{
C_{1}.type = string \qquad C_{2}.type = string  \qquad s{1}.pattern \subseteq C_{2}.pattern
}
{C_{1} \subtr C_{2}
}

\infrule[number]
{
\forall i  \in  \{1, \ldots ,k\}, not \not\in dom(C_{i}) \wedge  C_{i}.type = number \\
\forall i  \in  \{k + 1, \ldots , n\}, not \in dom(C_{i}) \wedge C_{i}.not.type = number \\
\forall i  \in  \{1, \ldots , l \}, not \not\in dom(t_{i}) \wedge t_{i}.type = number \\ 
\forall i  \in  \{l + 1, \ldots , n \}, not \in dom(t_{i}) \wedge t_{i}.not.type = number \\
subNumber(([C_{1}, \ldots , C_{k} ], [C_{k+1}, \ldots , C_{n}]), ([t_{1}, \ldots , t_{l} ], [t_{l+1}, \ldots , t_{m}]))
}
{\{allOf:[C_{1}, \ldots , C_{k} , C_{k+1}, \ldots , C_{n}]\} \subtr \{allOf:[t_{1}, \ldots , t_{l} , t_{l+1}, \ldots , t_{m}]\}
}

\infrule[array]
{
C_{1}.\atype = \aarray \qquad C_{2}.\atype = \aarray\\
C_{1}.\aminIts \geq C_{2}.\aminIts \qquad C_{1}.\amaxIts \leq C_{2}.\amaxIts\\
C_{1}.items = [C_{11} , \ldots , C_{1k} ] \qquad  C_{2}.items = [C_{21} , \ldots , C_{2m}]\\
\forall i  \in  \{0, \ldots , max(k) + 1 \}, (C_{1i} \| C_{1}.\aits)  \subtr (C_{2i} \| C_{2}.\aits)\\
C_{2}.\auniqIts \Implies  (C_{1}.\auniqIts \vee \mathit{allDisjointItems}(C_{1})) 
}
{
C_{1} \subtr C_{2}
}

\infrule[object]
{
C_{1}.\atype = \aobj \qquad C_{2}.\atype = \aobj\\
C_{1}.\aminProps \geq C_{2}.\aminProps \qquad C_{1}.\amaxProps \leq C_{2}.\amaxProps\\
\forall (e,S_{2}) \in C_{2}.\apReq:
\exists\ (e,S_{1}) \in C_{1}.\apReq: \ \ S_{1} \subtr S_{2} \\
\forall (e_2,S_{2}) \in C_{2}.\apProp:\ \exists  (e_1,S_{1}) \in C_{1}.\apProp:\ 
                                                  \ e_1 > e_2 \ \mbox{\ and\ } \ S_{1} \subtr S_{2} \\
}
{
C_{1} \subtr C_{2}
}

%
}
\caption{Structural rules.}
\label{fig:rules}
\end{figure}
}

\renewcommand{\akw}[1]{\ensuremath{\mbox{\tt{\small #1}}}}




\hide{
The \textsc{(\textsc{object})} rule that is defined in \cite{DBLP:conf/issta/HabibSHP21} is actually different
--- it is the following one.

\infrule[object]
{
s_{1}.type = object \qquad s_{2}.type = object\\
s_{1}.minProperties \geq s_{2}.minProperties \qquad s_{1}.maxProperties \leq s_{2}.maxProperties\\
s_{1}.required \supseteq s2.required \\
\forall p_{1}: s_{p1} \in s_{1}.patternProperties, p_{2} : s_{p2} \in s_{2}.patternProperties, p_{1} \cap p_{2} \neq \emptyset \Longrightarrow s_{p1} \subtr s_{p2} 
}
{
\aall[\  \atype(\aobj)    (,\apProp(e,\aref(X)))^*\ (,\apReq(e,\aref(X)))^*\ ( ,\aminProps(n))^?\ (,\amaxProps(n))^? \ ]   \subtr s_{2}
}

The above rule is not sound, since it allows one to prove that
$\{ props : \{ \qkw{a} : Int \} \} \subt \{props : \{ \qkw{b} : Bool \} \}$
which is not correct, since the two types above are clearly unrelated.

However, we tested the code published by the authors, and it seems that the code implements the rule that we
reported in Figure \ref{fig:rules}, which is a sound rule.
\ER
}

\iflong{
Thus, this approach is based on four sets of rules: (1) those that compute an almost-DNF in the first 
phase\iflong{, which are essentially
as complex as our DNF rules - they are only lacking some of the not-elimination cases};
(2) the incomplete (\textsc{anyOf-*}) rules to compare two DNF's (3) the (\textsc{uninhabited}) rule (4) the structural rules.
}

When the schemas to be compared are simple and contain neither negation nor recursion, the
rule-based approach is efficient and is very often able to provide a definite answer.
However, when the schemas are more complex, incompleteness becomes a problem,
especially because these rules, at least in the version of~\cite{DBLP:conf/issta/HabibSHP21},
do not support recursive schemas. 


\section{Refutational normalization}\label{sec:refutational}

In Section \ref{sec:WGinclusion} we have seen that the {\WG}  algorithm, when applied to inclusion-checking, is complete;  \mrevone{however, it  is too slow in practice. For example, in Section \ref{sec:expeval} we 
present several datasets derived from real-world schemas where it hits a 10-minute timeout
in a fraction of cases that ranges from $13\%$ to $90\%$.}{1.9}
\mrevone{Since inclusion for {\cJS} is EXPTIME-hard, we cannot hope to have an algorithm that is 
efficient in any possible 
situation; our aim is to make it ``fast enough'' in most cases of practical interest,
and, 
at least, as fast as the rule-based approach in the rule-provable cases.}{1.1}

%

We achieve this by proposing a different approach for the computation of the DNF of 
$\aall[S_1,\anot(S_2)]$: instead of adopting the standard and general-purpose DNF algorithm, we adopt a \emph{\RNAlg} algorithm, an algorithm that is  optimized for the case when $\aall[S_1,\anot(S_2)]$
reduces to $\afalse$, 
\mrevone{by adopting a combination of techniques, all inspired by the basic idea
of mirroring the behavior of the rules during the process of normalization:}{1.1}
\begin{enumerate}
\item \emph{lazy normalization}: we have seen in Section \ref{sec:DNFNotObj} that it is common to find
  a schema $S$ such that the {\CDNF} of $\anot(S)$ may have an exponential size; for this reason  we 
  normalize $\aall[S_1,\anot(S_2)]$ using a \emph{lazy} approach that does not completely compute 
   $\aDNF(\anot(S_2))$ in the rule-provable cases;
\item \emph{fast complement-absorption}: in the rule-based approach, the reflexive case
   $S\subt S$ is proved in linear time,
   hence we deploy techniques to be able to reduce $\aall[S,\anot(S)]$ to $\afalse$ in linear time;
    we call this reduction ``complement-absorption'', since $\anot(S)$ is the complement of $S$;
\item \emph{eager reference-evaluation}: in the DNF algorithm of \cite{DBLP:journals/pvldb/AttoucheBCGSS22}, when we meet
      a typed operator that guards a reference, such as $K=\apReq(\keykey{a},\aref(y))$, we just return $K$
      (case \emph{otherwise} in the definition of $\aDNF$ in Section \ref{sec:DNF});
      in Refutational Normalization, we eagerly evaluate the body of $y$ since, if it 
      reduces to $\afalse$, then the entire $K$ 
      can be reduced to
      $\afalse$.

\iflong{\item \emph{rapid component access}: 
    consider the following representation of an object inclusion statement, where the two object types
    are represented by the two inner $\aall[]$ schemas:
     $$
     \begin{array}{lllll}
       \aall[\!\!\!\!&\aall[\apProp(e_1,S_1),\ldots,\apProp(e_{n+l},S_{n+l})],\\[\NL]
                         &\anot(\aall[\apProp(e_1,S'_1),\ldots,\apProp(e_n,S'_n)])\ \ ]
                \end{array}
      $$
This representation is equivalent to
      $$
     \begin{array}{lllll}
       \aall[\!\!\!\!&\aall[\apProp(e_1,S_1),\ldots,\apProp(e_{n+l},S_{n+l})],\\[\NL]
                         &\aany[\anot(\apProp(e_1,S'_1)),\ldots,\anot(\apProp(e_n,S'_n))]\ \  ]
                \end{array}
      $$
      
       To rapidly normalize that expression, we must match every $\anot(\apProp(e_i,S'_i))$ 
        in the RHS
       with the complementary $\apProp(e_i,S_i)$ in the LHS, which requires that we store the 
       entire LHS in just one structure that is hashed on the $e_i$ patterns for rapid component access.
    } 
\item $\anot(\aany[\ldots])$-optimization: in order to achieve the same speed as the incomplete 
    (\textsc{anyOf-r}) rule in the rule-provable cases, when the RHS matches $\aany [\ldots]$,
    we exploit a fast-fail version of the algorithm to check whether the incomplete but
    efficient strategy of the (\textsc{anyOf-r}) rule suffices to prove the inclusion;
    when the incomplete (\textsc{anyOf-r}) strategy is not sufficient, we switch to the complete
    algorithm.
\end{enumerate}

\mrevfour{%
Inclusion problems where schemas contain large disjunctions, recursion and/or negation, as well as big {\qone} operators, are the likely candidates to benefit most from refutational 
normalization.}{4.6}

 We now describe how these principles lead to the definition of a complete
 and efficient {\RNAlg} algorithm.
\iflong{\subsection{The syntax of {\CDNF}'s: disjunctions $D$, canonical conjunctions $C$, c-references $X$}\label{subsec:dnfsyn}}
\ifshort{\subsection{{\CDNF} syntax}\label{subsec:dnfsyn}}

The aim of the normalization algorithm is to transform a document $P=(S,E)$
into an equivalent document $P'=(D,E')$, where the root $D$ is in {\CDNF}, and the body
of every reference in $E'$ is in {\CDNF}.

A {\CDNF} $D$ is defined as a disjunction of canonical conjunctions~$C$, as defined by the  grammar
in Figure \ref{fig:cangrammar}\iflong{,
that refines the grammar presented in Section \ref{sec:DNF}}:
a canonical conjunction $C$ is either a type-set conjunction $CTS$ or a typed conjunction 
\ifshort{\mrevone{($CO$, $CS$, etc.)}{1.3}}
\iflong{($CO$, $CA$, etc.)}
of stratified typed operators.
\hide{Some operators are redundant but are used for efficiency, such as 
 $\aminIts(n)$ and $\amaxIts(n)$.}
 
 \newcommand{\cref}[1]{\kw{cr}{#1}}
\begin{figure}[tb]
\bcolorone
\[
\begin{array}{llllllll}
D & := & \aany [\ ]\ \M\ \aany [\ (C,)^* \ C\ ] \\[\NL]
C & := & CTS  \M  CO  \M  CA  \M  CS  \M  CN  \M  CB  \M  CL  \\[\NL]
CTS \!\!\!& := &  
          \atypeSet\ \aU       \ (\mbox{where\ }\aU\not\eq [])    \\[\NL] 
CO & := & \aall[\  \atype(\aobj)\ \\
& &        \qquad  (,\apProp(e,\aref(X)))^*\ (,\apReq(e,\aref(X)))^*\ \\ 
        && \qquad (,\aminProps(n))^?\ (,\amaxProps(n))^? \ ]   \\[\NL]
\iflong{CA & := & \aall[\  \atype(\aarray)\ 
	     (,\aitem(n,\aref(X)))^*\ \\
	&& \qquad (,\aaddIts(n,\aref(X)))^?\ (,(\acontAft(n,\aref(X))^*\ \\
        && \qquad   (,\aminIts(n))^?\ (,\amaxIts(n))^? \ ]   \\[\NL]
}
\iflong{CN & := & \aall[\  \atype(\anum)\ 
         (,\amof(q))^?\ (,\anotMof(q))^*\  \\
        && \qquad (,\amin(\hide{-\Inf|}q)|\aexMin(q))^?\ (,\amax(q)|\aexMax(q))^?\ ]      \\[\NL] }
CS & := & \aall[\  \atype(\astr),\ (,\apatt(e))^? \ ]  \\[\NL]
\iflong{CB & := & \aall[\  \atype(\abool) \ (,\aconst(b))^? \ ]   \\[\NL]
CL & := & \aall[\  \atype(\anull) \ ]    \\[\NL]}
\ifshort{ 
CA  &::=& \aall[\atype(\aarray)\ldots]\ \ \ 
CN  ::= \aall[\atype(\anum)\ldots]\ \ 
}  \\[\NL]
\ifshort{ 
CL  &::=& \aall[\atype(\anull)\ldots]\ \ \ 
\ \,CB  ::= \aall[\atype(\aboolean)\ldots]\ \  \
}  \\[\NL]
X & := & 
                \cref{[ ((x|\NV{x}),)^*(x|\NV{x})]}
\end{array}
\]
\caption{Grammar of canonical conjunctions.}
\label{fig:cangrammar}
\ecolor
\end{figure}

\hide{
Assume we normalize a document $(S,E)$ to produce a normalized pair $(D,E')$.
As we will see,
the normalization process may require the creation of new reference-body pairs,
starting from the references in $E$;
these new references have a very specific shape: they correspond to the \emph{conjunction}
of original references that where in $E$ or their complements,
and for this reason we call then \emph{c-references}.
The production $X$ defines the syntax of the c-references that may appear in $(D,E')$.
}

\noindent\textbf{C-references.}
The normalization process creates new references when it merges
related operators.
\bcolorfour
\mrevfour{%
For example, assume we normalize the following $(S,E)$ document,
where $S$ is the conjunction of two object schemas:}{4.3, M.1}
$$\begin{array}{llll}
(\ \ \aall[\ \aall[\atype(\aobj),\ \apReq(\keykey{a}:\aref(x))\ ], \\
\qquad\ \ \ \ \aall[\atype(\aobj),\ \apProp(\keykey{a}:\aref(\NV{y})\ ]\ \ ], \\[\NL]
\ \ \ [\ x: S_x,\ \ y: S_y,\ \  \NV{x}: \anot(S_{x}),\ \ \NV{y}: \anot(S_{y})\ ]\ \ )
\end{array}
$$

As we will see, the normalization process merges the $\aref(x)$ schema of 
the $\apReq$ operator with that of the corresponding $\apProp$, creating a new
reference $\cref{[ x,\NV{y}]}$, which is also added
to $E$:
$$\begin{array}{llll}
(\ \aany[\aall[\atype(\aobj), \apReq(\keykey{a}\!:\!\aref(x)),\\
   \qquad\qquad\qquad\qquad\quad\ \ \  
                              \apProp(\keykey{a}\!:\!\aref(\cref{[ x,\NV{y}]})]], \\[\NL]
\ \ \ [\ x: \DNF(S_x),\ \ \ldots,
         \ \  \cref{[ x,\NV{y}]} : \DNF(\aall[S_x,\anot(S_{y})])\ ]\  )
\end{array}
$$
\ecolor
All new references (such as $\cref{[ x,\NV{y}]}$) created during normalization correspond to the \emph{conjunction}
of original references that were in $E$,
and for this reason, we call them \emph{c-references}.

The production $X$ in the grammar
defines the syntax of these c-references. 
Any c-reference is fully defined by a set
$$\Set{x_1,\ldots,x_n,\NV{x^2_1},\ldots,\NV{x^2_{n'}}}$$
 where every $x_i$ and $\NV{x^2_i}$ is in the domain
of the original $E$, hence we will often identify c-references with the corresponding sets, and we will write
operations such as $X_1 = X_2 \cup X_3$, and write
$\aref(\Set{x,\NV{y}})$ instead of $\aref(\cref{[x,\NV{y}]})$.

\noindent\textbf{Abbreviations.}  
We use the following abbreviations, so that 
$\cTrue$ 
belongs to $C$, $\dFalse$ 
belongs to $D$, and $\xFalse$ and $\xTrue$ belong to~$X$;
 $\xFalse$ definition is based on one arbitrary reference $x_1$\iflong{ in the original~$E$}.
%
$$
\begin{array}{l@{\qquad}l@{\qquad}l}
  \multicolumn{3}{l}{\cTrue \ = \ \atypeSet [\abool, \astr, \anull, \anum, \aobj, \aarray]} \\[\NL]
  \dFalse = \aany[\ ] &
  \xTrue = \cref{[\ ]} &
  \xFalse = \cref{[x_1,\NV{x_1}]}
\end{array}
$$
\bcolorfour
\noindent%
\emph{Running example}
\mrevfour{%
We define the following schemas:}{4.3, M.1}
$$
\begin{array}{lcl}
S^1_L & = & \aall[\atype(\aobj),\ \apReq(\keykey{a},\Rf{\SSet{x^1}})] \\[\NL]
S^2_L & = & \aall[\atype(\aobj),\ \apReq(\keykey{a},\Rf{\SSet{x^2}})] \\[\NL]
S_R & = & \aany[S_a,S_{ab}] \\[\NL]
S_a & = & \aall[\atype(\aobj),\ \apReq(\keykey{a},\Rf{\SSet{y}})] \\[\NL]
S_{ab} & = & \aall[\atype(\aobj),\ \apReq(\anyp[\keykey{a},\keykey{b}],\Rf{\SSet{z}})] \\[\NL]
\end{array}
$$
with reference environment:
$$
\begin{array}{l@{\quad}l}
E(\SSet{x^1}) = E(\SSet{y}) = \atype(\anum) \\[\NL]
E(\SSet{x^2}) = \atypeSet[\anum,\astr] \\[\NL]
E(\SSet{z}) = \atype(\astr) \\[\NL]
E(\SSet{\NV{x^1}}) = \anot(\atype(\anum)), \ E(\SSet{\NV{y}}) =  \ldots
\end{array}
$$

We use as running examples the inclusions $S^1_L\subt S_R$ and   $S^2_L\subt S_R$.

The inclusion $S^1_L\subt \aany[S_a,S_{ab}]$ holds because $S^1_L$ is equivalent to $S_{a}$.
The inclusion $S^2_L\subt \aany[S_a,S_{ab}]$ holds because an object of $S^2_L$ has either a shape
$\JObj{\qkw{a}:q,\ldots}$ with $q\in\Num$, which satisfies $S_a$,
or $\JObj{\qkw{a}: k,\ldots}$ with $k\in\Sigma^*$, which satisfies $S_{ab}$. 

\ecolor

\subsection{$\tkw{allDS}(D,S,E)$, $\tkw{allCS}(C,S,E)$: lazy normalization, $\tkw{not}(\tkw{any}[\ldots])$-optimization}
\label{sec:notany}\label{sec:allCS}

Normalization in the {\WG} algorithm 
proceeds bottom up and is defined as follows
(Section~\ref{sec:DNF}):
$\OrDD$ just takes the union of the argument lists of the two disjunctions; $D_1 \AndDD D_2$ merges any argument 
of the disjunction
$D_1$ with any argument of $D_2$ ---  performing in a sense a cartesian product 
---  while $\kw{notPush}(S)$ pushes $\anot$ through all other operators until it is absorbed in the leaves 
of~$S$.
$$
\begin{array}{lllll}
\aDNF(\aany[S_1,S_2]) &=& \aDNF(S_1) \OrDD\aDNF(S_2) \\[\NL]
\aDNF(\aall[S_1,S_2]) &=& \aDNF(S_1) \AndDD \aDNF(S_2) \\[\NL]
\aDNF(\anot(S)) &=& \aDNF(\kw{notPush}(S)) \\[\NL]
\aDNF(\aref(x)) &=& \aDNF(E(x)) \\[\NL]
\aDNF(S) &=& S & \mbox{otherwise} \\[\NL]
\end{array}
$$

\newcommand{\opLR}[3]{\!\ _{\akw{#2}}{#1}_{\akw{#3}}}

\newcommand{\AndDS}{\cpar{\akw{D}}{\wedge}{\akw{S}}}

When we normalize 
$\aall[S_1,\anot(S_2)]$, 
the {\WG} algorithm computes the {\DNF} of $\anot(S_2)$, which may be very big.
The \emph{lazy normalization} approach of Refutational Normalization is based on the
idea 
of normalizing $S_1$ to $D_1$,  
without normalizing $\anot(S_2)$ immediately, but just pushing negation one
level down of $S_2$, obtaining a term $N_2$ --- not normalized ---
and then extracting one piece $N'_2$ at a 
time from
$N_2$ to combine that piece with $D_1$, hoping to reduce the 
$\aall[D_1,N'_2]$ conjunction to $\afalse$, so that the computation may
immediately stop.
Hence, instead of a function $D_1 \AndDD D_2$ that merges two terms in {\CDNF},
we define a function $D \AndDS S$ that merges a {\CDNF} $D$ with a non normalized schema $S$;
from now on, we call it $\allDS(D,S,E)$;
\emph{lazy normalization} refers to the fact that the first argument $D$ is normalized, but the second argument $S$, the one that contains negation, is not.
\mrevfour{In our running example, instead of computing the {\CDNF} of
$\aall[S^i_L,\anot(S_R)]$, we would first compute the 
$\CDNF$ of $S^i_L$, whose shape is $\aany[C^i_L]$, and will
then invoke 
\begin{equation}
\allDS(\aany[C^i_L],\anot(S_R),E)
\label{ex:allDS}
\end{equation}
for the lazy normalization.{ (We number the main steps of the running example to make it easier to follow,
so this is Step 1.)}}{4.3, M.1}

$\allDS(D,S,E)$ is mutually recursive
with $\allCS(C,S,E)$, 
which merges a conjunction $C$ with a schema $S$, with the same ``lazy'' approach,
with $\allCK(C,K,E)$ (Section \ref{sec:allCK}), which merges a conjunction $C$ with a typed operator $K$,
and
with $\allXX(X_1,X_2,E)$ (Section \ref{sec:allXX}), which returns a c-reference equivalent to $\aall[X_1,X_2]$.
The {\CDNF} of a schema is computed by computing
$\allDS(\aany[\cTrue],S, E)$. 
The pseudocode of $\allDS$ is shown in Algorithm \ref{alg:allDS}.

\newcommand{\fastFailAllCS}{\akw{fastFailAllCS}}

$\allDS(D,S,E)$ just applies distributivity to the $D$ 
argument. 
In the empty case, $\aany[]$ is $\dFalse$, hence the computation immediately stops,
without computing the {\CDNF} of $S$.
In the non-empty case, we use $\anyDD(D_1,D_2)$ 
(i.e., $D_1 \opLR{\vee}{D}{D} D_2$ ) to merge the result of 
$\allCS(C_1,S,E)$ and the result of the recursive call on the tail.
\mrevfour{In our running example, $\allDS(\aany[C^i_L],\anot(S_R),E)$
first computes 
\begin{equation}
D_1=\allCS(C^i_L,\anot(S_R),E)
\label{eq:D1}
\end{equation}
and then it trivially
merges the result $D_1$ with an empty $\aany[]$ by invoking $\anyDD(D_1,\allDS(\aany[],\anot(S_R),E))$ 
(lines 4-5 of Algorithm \ref{alg:allDS}).}{4.3, M.1}
%

\RestyleAlgo{ruled}
\begin{algorithm}[t]
\small
\AlgoSetup

\Fn{\fAllDS$(D,S,E)$}{
  \match $D$ \KwOf{
    \uCase(
                )
                          {$\kany[\,]$}{\KwReturn $\kany[\,]$\;
    }
    \uCase(
               ){$\kany[C_1,\;\aC]$}{
      \KwReturn \ $\ \fAnyDD(
                                  \fAllCS(C_1,S,E),\fAllDS(\kany[\aC],S,E))$\;
    }
  }
}
\caption{\fAllDS$(D,S,E)$}
\label{alg:allDS}
\end{algorithm}
 
Function $\allCS(C,S,E)$, whose pseudocode is shown in Algorithm \ref{alg:allCS},  works by cases on $S$. A reference is substituted with its body, retrieved from $E$ (lines 2-3). When $S$ is $\anot(S')$, we push negation just one level below, in agreement with
the lazy normalization principle (lines 4-5).
The $\notPush$ function, defined in detail in~\cite{DBLP:journals/pvldb/AttoucheBCGSS22,DBLP:journals/tcs/BaaziziCGSS23},
uses well known De Morgan dualities:
$$\notPush(\aall[S_1,\ldots,S_n]) = \aany[\anot(S_1),\ldots,\anot(S_n)]$$
or dualities between structural operators, as the one between $\apProp$ and $\apReq$:
$$\notPush(\apProp(e,S)) = \aall[\atype(\aobj), \apReq(e,\anot(S))]
$$
\hide{For references, we assume that every reference 
$x$ has a complementary reference $\NV{x}$ whose
body is just $\anot(E(x))$, and $\notPush$ exchanges 
$\aref(x)$ with $\aref(\anot(x))$ and vice versa.}
\ifshort{(For a complete definition of $\notPush$, see the full \mrevone{paper \cite{full}.)}{1.10}}
\iflong{
We use a not-completed version of JSON Schema, according to the terminology of \cite{DBLP:journals/tcs/BaaziziCGSS23},
so that every  JSON Schema has a De Morgan dual.
Here we report most cases; we assume that every schema is stratified, but it would be very each to generalize to the non-stratified case.
\begin{figure}[htb]
\small
$$
{
\setlength{\arraycolsep}{1pt}
\begin{array}{lllll}
\notPush(\aall[S_1,\ldots,S_n]) & = & \aany[\anot(S_1),\ldots,\anot(S_n)]\\[\NL]
\notPush(\aany[S_1,\ldots,S_n]) & = & \aall[\anot(S_1),\ldots,\anot(S_n)]\\[\NL]
\notPush(\anot(S)) & = & S\\[\NL]
\notPush(\aref(x)) & = & \RNV{x}\\[\NL]
\notPush(\RNV{x}) & = & \aref(x)\\[\NL]
\notPush(\aconst(A)) & = & \anotConst(A)\\[\NL]
\notPush(\anotConst(A)) & = & \aconst(A)\\[\NL]
\notPush(\atypeSet\ \aU) & = & \aany[\atype(U_1),\ldots,\atype(U_n)] \\
 &&  \kw{where\ }\Set{U_1,\ldots,U_n}=types\setminus \aU\\[\NL]
\notPush(\atype(U)) & = & \aany[\atype(U_1),\ldots,\atype(U_5)] \\
&& \kw{where\ }\Set{U_1,\ldots,U_5}=types\setminus \Set{U}\\[\NL]
\notPush(\apProp(p,\aref(x))) & = & \aall[\atype(\aobj), \apReq(p,\RNV{x})]\\[\NL]
\notPush(\apReq(p,\aref(x))) & = & \aall[\atype(\aobj), \apProp(p,\RNV{x})]\\[\NL]
\notPush(\aminProps(0)) & = & \afalse \\[\NL]
\notPush(\aminProps(m+1)) & = & \aall[\atype(\aobj), \amaxProps(m)]\\[\NL]
\notPush(\amaxProps(m)) & = & \aall[\atype(\aobj), \aminProps(m+1)]\\[\NL]
\notPush(\aaddIts(n,\aref(x))) & = & \aall[\atype(\aarray), \acontAft(n,\RNV{x})]\\[\NL]
\notPush(\acontAft(n,\aref(x))) & = & \aall[\atype(\aarray), \aaddIts(n,\RNV{x})]\\[\NL]
\notPush(\aitem(n,\aref(x))) & = &  \aall[\atype(\aarray), \\
&& \quad \;\;\; \aitem(n,\RNV{x}), \aminIts(n+1)]\\[\NL]
\notPush(\auniqIts) & = & \aall[\atype(\aarray), \anotUniqIts]\\[\NL]
\notPush(\anotUniqIts) & = & \aall[\atype(\aarray), \auniqIts]\\[\NL]
\notPush(\apatt(e)) & = & \aall[\atype(\astr), \apatt(\notp(e))]\\[\NL]
\notPush(\amin(q)) & = & \aall[\atype(\anum), \aexMax(q)]\\[\NL]
\notPush(\aexMin(q)) & = & \aall[\atype(\anum), \amax(q)]\\[\NL]
\notPush(\amax(q)) & = & \aall[\atype(\anum), \aexMin(q)]\\[\NL]
\notPush(\aexMax(q)) & = & \aall[\atype(\anum), \amin(q)]\\[\NL]
\notPush(\amof(q)) & = & \aall[\atype(\anum), \anotMof(q)]\\[\NL]
\notPush(\anotMof(q)) & = & \aall[\atype(\anum), \amof(q)]\\[\NL]
\ldots
\end{array}
}
$$
\vspace{-10pt}
\caption{\notPush(S).}\label{fig:notpush}
\end{figure}
}

\mrevfour{In our example, $\allCS(C^i_L,\anot(S_R),E)$,
with $S_R=\aany[S_a,S_{ab}]$ (Step \ref{eq:D1})
invokes (lines 4-5 of Algorithm \ref{alg:allCS})):
\begin{equation}
\allCS(C^i_L,\aall[\anot(S_a),\anot(S_{ab})]),E)
\label{eq:allCS}
\end{equation}}{4.3, M.1}%
\indent Function $\allCS(C,S,E)$ applies distributivity in case 
$S=\aany[\ldots]$ (lines 8 - 11). The interesting case of $\allCS$ is the one where $S=\aall[S_1,\ldots,S_n]$;
in this case, we combine
lazy normalization with the $\anot(\aany[\ldots])$ optimization (lines 12 - 20).

The $\anot(\aany[\ldots])$ optimization is designed to ensure that, in those 
cases where the  (\textsc{anyOf-r}) rule
suffices to verify inclusion,
refutational normalization mimics its efficient behavior, without losing completeness
in cases
where (\textsc{anyOf-r}) does not suffice.

\mrevfour{Assume we compare $S'$ with 
$\aany[S_1,\ldots,S_n]$. 
This eventually 
invokes
$\allCS(C',\anot(\aany[S_1,\ldots,S_n]),E)$
for all 
conjunctions $C'$ of the {\CDNF} of $S'$,
as in the running example,
which invokes (see Step~\ref{eq:allCS}):
\begin{equation*}
\allCS(C',\aall[N_1,\ldots,N_n],E) \mbox{\ \ with\ }N_i=\anot(S_i)
\mbox{\ \ for \ } i\in 1...n
\end{equation*}}{4.3, M.1}
\indent Now, 
we should merge $C'$ with $N_1$, the result with
$N_2$, and so on, until we get $\dFalse$, as we do in lines 17-20
of Algorithm \ref{alg:allCS}.
\iflong{
This approach is already an optimization w.r.t. the base algorithm,
since we do not normalize the entire expression $\aall[N_1,\ldots,N_n]$:
we extract each $N_i$ to merge it with $\akw{result}$ until we arrive at $\akw{result}=\dFalse$; from this 
moment, we will compute $\allDS(\dFalse,N_{i+l},E)$ which returns $\dFalse$ immediately,
without inspecting $N_{i+l}$ --- this is one important instance of the lazy normalization approach.
}

In cases when the inclusion holds, this ``accumulative'' computation, before 
eventually resulting in $\dFalse$,
may produce intermediate results that grow exponentially.
However, we observe that, in most real-world scenarios, the incomplete
rule (\textsc{anyOf-r}) suffices since~$C'$ is completely included in one of the RHS addends~$S_i$, that is, $\allCS(C',\anot(S_{i}),E)$ reduces to $\dFalse$.
Hence, before trying the complete, exponential, approach of merging $C'$ with all the $N_i$'s in sequence, we tentatively merge $C'$ with each $N_i$ separately;
if one of these mergings results in $\dFalse$ we can immediately
return $\dFalse$; hence, in the easy (and common) case
where  (\textsc{anyOf-r}) suffices,  we prove inclusion as efficiently as the 
rule-based approach. 

\bcolorfour
\mrevfour{Let us go back to the running example, Step \ref{eq:allCS}:}{4.3, M.1} 
$$\allCS(C^i_L,\aall[\anot(S_a),\anot(S_{ab})]),E)$$
In this case $C^1_L$ and $C^2_L$ differ.
In both cases, 
 we normalize separately first $\allCS(C^i_L,\anot(S_a),E)$ and then
$\allCS(C^i_L,\anot(S_{ab}),E)$ 
(lines 2-3 of Algorithm \ref{alg:fastCheck}, $\anot(\aany[\ldots])$ optimization).

In the first example, since  $C^1_L$ is equivalent to $S^1_L$, we have 
that $C^1_L\subseteq S_a$ ;  this is a case that rule (\textsc{anyOf-r}) solves,
and indeed
\begin{equation}
\allCS(C^1_L,\anot(S_a),E)
\label{eq:notSa}
\end{equation}
reduces to $\dFalse$ (as we will see 
in Section \ref{sec:allXX}),
hence the optimized code returns $\dFalse$.

In the second example, $C^2_L\subseteq S_R$, 
but neither $C^2_L\subseteq S_a$ nor $C^2_L\subseteq S_{ab}$ holds,
and rule (\textsc{anyOf-r}) would  fail.
Our algorithm first computes $\allCS(C^2_L,\anot(S_a),E)$ and $\allCS(C^2_L,\anot(S_{ab}),E)$,
and indeed none of the two terms is $\dFalse$.
Hence, \akw{fastCheck} fails, and our  algorithm stops mirroring  (\textsc{anyOf-r}) and
switches to the complete approach, by first computing  
\begin{equation}
D_1=\allCS(C^2_L,\anot(S_a),E)
\label{eq:completefirst}
\end{equation}
and then
\begin{equation}
\key{result} \ = \ \allDS(D_1,\anot(S_{ab}),E)
\label{eq:complete}
\end{equation}
(lines 17 and 19 respectively, Algorithm \ref{alg:allCS}).
This computation eventually results in $\dFalse$, as shown in Section \ref{sec:allCK}.
\ecolor
%
%


\begin{algorithm}[t]
\small
\AlgoSetup

\Fn{\fAllCS$(C,S,E)$}{
  \match $S$ \KwOf{
    \uCase(){$\kref(x)$}{
      \KwReturn \fAllCS$(C,\SETFN{getBody}(x,E),E)$\;
    }
    \uCase(){$\anot(S')$}{
      \KwReturn \fAllCS$(C,\SETFN{notPush}(S'),E)$\;
    }
    \uCase(){$\kany[\,]\ \text{or}\ \kfalse$}{
      \KwReturn $\kany[\,]$\;
    }
 \uCase(){$\kany[S_{h},\;\aS]$}{
      $\kId{Head} \leftarrow \fAllCS(C,S_{h},E)$\;
      $\kId{Tail} \leftarrow \fAllCS(C,\kany[\aS],E)$\;
      \KwReturn \SETFN{anyDD}$(\kId{Head},\kId{Tail})$\;
          }
    \uCase(){$\kall[\;]\ \text{or}\ \kPredef{true}$}{
      \KwReturn $\kany[C]$\;
    }
    \uCase(){$\kall[S_1,\ldots,S_n]$}{
      \KwTry{
        \KwReturn \SETFN{fastCheck}$(C,\kall[S_1,\ldots,S_n],E)$\;
      }
      \KwCatch{\;
      \Indp
        $\kId{result} \leftarrow \fAllCS(C,S_1,E)$\;
        \For{$i \leftarrow 2$ \KwTo $n$}{
          $\kId{result} \leftarrow \SETFN{allDS}(\kId{result},S_i,E)$\;
        }
        \KwReturn $\kId{result}$\;
      }
      \Indm
    }
    \uCase(\tcp*[f]{\texttt{K is any structural keyword}}){$K$}{
      \KwReturn \fAllCK$(C,K,E)$\;
    }
  }
}
\caption{$\fAllCS(C,S,E)$}
\label{alg:allCS}
\end{algorithm}

\ifshort{
\begin{algorithm}[t]
\small
\AlgoSetup

\Fn{\SETFN{fastCheck}$(C,\kall[S_1,\ldots,S_n],E)$}{
  \For{$i \leftarrow 1$ \KwTo $n$}{
    \KwTry{
      \If{\SETFN{allCS}$(C,S_i,E)=\kdFalse$}{
        \KwReturn $\kdFalse$\;
      }
    }
    \KwCatch{
      \KwContinue\;
    }
  }
  \KwRaise \kId{exception}\;
}
\caption{\SETFN{fastCheck}$(C,\kall[S_1,\ldots,S_n],E)$}
\label{alg:fastCheck}
\end{algorithm}
}

\iflong{
\begin{algorithm}[t]
\small
\AlgoSetup

\Fn{\SETFN{fastCheck}$(C,\kall[S_1,\ldots,S_n],E)$}{
  \For{$i \leftarrow 1$ \KwTo $n$}{
    \KwTry{
      \If{\SETFN{fastFailAllCS}$(C,S_i,E)=\kdFalse$}{
        \KwReturn $\kdFalse$\;
      }
    }
    \KwCatch{
      \KwContinue\;
    }
  }
  \KwRaise \kId{exception}\;
}
\caption{\SETFN{fastCheck}$(C,\kall[S_1,\ldots,S_n],E)$}
\label{alg:fastCheck}
\end{algorithm}
}

\iflong{This (\textsc{anyOf-r}) optimization is implemented by the
function \akw{fastCheck}.
Observe that \akw{fastCheck} uses \akw{fastFailAllCS} ra\-ther than {\allCS}.
\akw{fastCheck} is intended to be incomplete but fast, as fast as the rule-based system.
If $C$ is included in the complement of $S_i$, for some $i\leq n$,
\akw{fastCheck} invokes the test $\allCS(C,S_{i},E)$ for $i$ times, and the test
returns a non-$\dFalse$ value for the first 
$i-1$ times; hence, these $i-1$ ``unhelpful'' tests must fail rapidly.
For this reason, instead of $\akw{allCS}(C,S_i,E)$,
we use here a function $\akw{fastFailAllCS}(C,\allowbreak S_i,E)$ that 
fails rapidly, as rapidly as the corresponding test in the rule-based approach.

$\akw{fastFailAllCS}$
behaves like $\allCS$
in all cases apart from $\aall[\ldots]$, where it behaves as the rule-based
system: it fails when the optimized test fails.
Hence,
while $\allCS$ is complete but may be slow, 
$\akw{fastFailAllCS}$ is not complete 
--- which is OK since it is only used to implement a non-complete optimization ---
but is fast.
The use of this fast-fail approach makes the $\akw{fastCheck}$ test as efficient as
the rule-based approach in all cases when the judgment is rule-provable.

In principle, $\akw{fastFailAllCS}$ is defined by substituting every call to $\allCS$ by a call to $\fastFailAllCS$,
and similarly by substituting calls to $\allCS$ with $\akw{fastFailAllCS}$ to and $\allCK$  with $\akw{fastFailAllCK}$.
In the code, we just add one extra parameter to all the functions that makes them use either
the ``fast'' or the complete approach.}


\iflong{
\begin{algorithm}[t]
\DontPrintSemicolon
\AlgoSetup

\Fn{\SETFN{fastFailAllCS}$(C,S,E)$}{
  \match $S$ \KwOf{
    \uCase(\tcp*[f]{same as \fAllCS, but invoking \fFastFailAllCS}){$\cdots$}{
      \tcp*[f]{(cases omitted)}
    }
    \uCase(\tcp*[f]{\texttt{all[S1...Sn]}}){$\kall[S_1,\ldots,S_n]$}{
      \KwReturn \SETFN{fastCheck}$(C,\kall[S_1,\ldots,S_n],E)$\;
    }
    \uCase(\tcp*[f]{\texttt{K} is any structural keyword}){$K$}{
      \KwReturn \SETFN{fastFailAllCK}$(C,K,E)$\;
    }
  }
}
\caption{$\fFastFailAllCS(C,S,E)$}
\label{alg:fastFailAllCS}
\end{algorithm}
}
 
\hide{ 
 \begin{remark}
 Actually, it should be like this.
 {\small\begin{verbatim}
fastCheck(C,all[S1...Sn],E): 
       for i in 1..n if fastFailAllCS(C,Si,E)='unsat' then return('unsat')
      return('unknown')  
\end{verbatim}
}
{\small\begin{verbatim}
fastFailAllCS(C,S,E): 
    case S of 
       ref(x): return(fastFailAllCS(C,getBody(x,E),E))
       not(S'): return(fastFailAllCS(C,notPush(S'),E))
       any[] or false: return('unsat')
       any[S,\aS]: if fastFailAllCS(C,S,E)='sat'  then return('sat')
                  else return allCS(C,"any[{\aS}]",E)
       all[] or true: return('sat')   
       all[S1...Sn]:  
          if fastCheck(C,all[S1...Sn],E) = 'unsat' return 'unsat'
          else return 'unknown'                                 
       K: return(allCK(C,K,E))   -- K is any structural keyword???
  \end{verbatim}
 }
 \ER
 \end{remark}
 }
 
This is how we implement the principle of lazy normalization and
the $\tkw{not}(\tkw{any}[\ldots])$-optimization in
functions $\allDS$ and $\allCS$, which deal with
boolean operators.
We now explore how the other principles are implemented by the $\allXX$ and the 
$\allCK$ functions.

\subsection{$\tkw{allXX}(X,Y,E)$:
eager reference-normalization to enable
fast complement-absorption}\label{sec:allXX}


$\allXX$ is the function that merges references, and which is crucial in the
implementation of the ``eager reference-normalization''
and ``fast complement-absorption'' principles.

\bcolorfour
\mrevfour{
In the running example,
\akw{fastCheck}  invokes $\allCS(C^1_L,\anot(S_a),E)$ (Step \ref{eq:notSa}).
}{4.3,M.1}
$\allCS$ pushes negation, distributes over $\aany$, pushes negation again,
and finally the second argument becomes a structural operator
$\apProp$, so that $\allCK$ is invoked:
\begin{align}
&\allCS(C^1_L, \anot(S_a),E) \label{eq:allCSnot}\\
&= \allCS(C^1_L, \anot(\aall[\atype(\aobj),\apReq(\keykey{a},\Rf{\Set{y}})]),E) \notag\\
&\to \allCS(C^1_L, \aany[\ldots,\anot(\apReq(\keykey{a},\Rf{\Set{y}}))],E) \notag\\
&\to \allCS(C^1_L, \anot(\apReq(\keykey{a},\Rf{\Set{y}})),E) \notag\\
&\to \allCK(C^1_L,\apProp(\keykey{a},\Rf{\Set{\NV{y}}}),E) \notag
\end{align}

$C^1_L$ will be described in Section \ref{sec:allCK} - Step \ref{eq:CL}, but it is 
equivalent to $S^1_L$, hence it contains $\atype(\aobj)$,
$\apReq(\keykey{a},\Rf{\Set{x^1}})$, plus some redundant ``normalization scaffolding'':
\begin{equation*}
\begin{split}
&C^1_L=\aall[\ \atype(\aobj),\ \apReq(\keykey{a},\Rf{\Set{x^1}}),\ \ldots] \\ 
\end{split}
\end{equation*}

The  schemas
$\apReq(\keykey{a},\Rf{\SSet{x^1}})$ and $\apProp(\keykey{a},\Rf{\SSet{\NV{y}}})$ are complementary;
in the \mrevone{next section}{1.11} we will see how exactly $\allCK$ works, but it will eventually
merge them and combine the c-refer\-ences ${\SSet{x^1}}$ and ${\SSet{\NV{y}}}$
by invoking ${\allXX}(\SSet{x^1},\SSet{\NV{y}},E)$.
\ecolor

A function $\allXX(X,Y,E)$ that merges two c-references $X$ and~$Y$ into one may just return 
the combined c-reference 
$X\cup Y$; this is sound and is what is done in the {\WG} algorithm~\cite{DBLP:journals/pvldb/AttoucheBCGSS22}, yet it does not discover that  
\mrevfour{$\apReq(\keykey{a},\Rf{\Set{x^1}})$ and $\apProp(\keykey{a},\Rf{\Set{\NV{y}}})$ are complementary,}{4.3, M.1}
hence their combination is not reduced 
to $\dFalse$, hence \emph{fast comple\-ment-absorption} is not obtained.

For this reason, ${\allXX}(X,Y,E)$, before resorting to returning $X\cup Y$, tries to
prove that $X\cup Y$ is actually unsatisfiable.
It first checks whether  $X\cup Y$ contains any contradictory pair $\Set{z,\NV{z}}$, in which case
it returns $\xFalse$; otherwise, it checks whether
$\aDNF(E(X\cup Y))$ has already been computed and stored in $E$; 
if it has, and it is $\dFalse$, it returns $\xFalse$; if it has not been computed yet,
it computes it, memorizes it into $E$, and, if the result is $\dFalse$, it returns $\xFalse$,
\mrevfour{as it happens in our first running example,
where
$$E(\Set{x^1,\NV{y}})=\aall[\atype(\anum),\anot(\atype(\anum))]$$
which reduces to $\dFalse$.
If {\allXX} cannot prove that $\DNF(E(X\cup Y))$
is $\dFalse$, then ${\allXX}(X,Y,E)$ just returns $X\cup Y$;
this happens in the second running examples, where 
$$E(\Set{x^2,\NV{y}})=\aall[\atypeSet[\anum,\astr],\anot(\atype(\anum))]$$
is not $\dFalse$, hence 
\begin{equation}
{\allXX}(\Set{x^2},\Set{\NV{y}},E)\ \ \to \ \ 
\Set{x^2,\NV{y}}.
\label{eq:allXX}
\end{equation}}{4.3, M.1}

\bcolortwo
\begin{remark} {Normalizing recursive schemas:}
\mrevtwo{The {\WG} approach normalizes the bodies $E(X)$ of a system of equations $E$,
and then generates a witness using the normalized system.
Recursion plays no special role during normalization, but the  
witness is generated 
bottom-up, 
similar to 
Datalog 
systems~\cite{DBLP:journals/pvldb/AttoucheBCGSS22, DBLP:books/aw/AbiteboulHV95}, 
to avoid infinite loops.}{2.3}

Our algorithm follows the same approach as {\WG}, with a difference:
because of the eager  reference-normalization described in this section, 
${\allXX}(X,Y,E)$, instead of just returning $X\cup Y$,
invokes $\aDNF(E(X\cup Y))$, which, when the schema is recursive,
may invoke  ${\allXX}(X,Y,E)$ again, producing an infinite loop. 
We solve this problem by keeping track of the active invocations of ${\allXX}$, so that
when ${\allXX}(X,Y,E)$ is invoked recursively while evaluating  
${\allXX}(X,Y,E)$, we just return $X\cup Y$, as happens with the {\WG}
algorithm.
Hence, recursion does not impose any specific overhead on our algorithm.

\iflong{This approach is coherent with our aim of preserving 
the completeness of
the {\WG} algorithm while matching the speed of the {\RB} approach whenever
possible; in this specific case, the second aim is moot, since the {\RB} approach does not
support recursion.}
\ecolor
\end{remark}

\hide{THIS PART WAS IN THE IFLONG SECTION NOW IS JUST HIDDEN
However, this approach does not fully implement the fast complement-absorption.

Consider a schema $\aref(\Set{x})$ in the recursive schema\GG{We may shorten this}
$E=(x : \apProp(\keykey{a},\Set{x}))$, when combined with its complement will not 
reduce to $\dFalse$ but to $E(\Set{x,\NV{x}})$, where $E'$ extends $E$ with the following mapping:
$$\Set{x,\NV{x}} = \aall [ \atype(\aobj),
\apProp(\keykey{a},\aref(\Set{x}))),
\apReq(\keykey{a},\aref(\Set{x,\NV{x}})))]
$$
The recursive reference $\aref(\Set{x,\NV{x}})$ is not satisfiable, since, in order, to build a term of depth
$n$
that satisfies  $\aref(\Set{x,\NV{x}})$ we need a term of depth $n-1$, and so on forever, but any strictly
decreasing sequence of natural numbers must stop at zero.

Hence, while we do not fully implement fast complement-absorption, this reduction is still fast
(that is, linear time), and it is complete: the witness generator algorithm will detect the fact that
this schema is not satisfiable.

The fact that we do not normalize comparison of recursive schemas to $\dFalse$ does not
contradict our claim that we are as fast as the rule-based approach in rule-provable cases,
since the rule-based approach returns \emph{``do not know''} on recursive schemas \cite{DBLP:conf/issta/HabibSHP21}.
}

\hide{THIS WAS IFSHORT
This approach is not able, in general, to immediately reduce a comparison of two recursive 
schemas such that $S_1\subt S_2$ to $\dFalse$; in this case, the fact that the DNF of $\aall[S_1,\anot(S_2)]$ is unsatisfiable
is discovered during the final step of the witness generation algorithm, when the recursive
generation phase that follows normalization is not able to 
generate any witness.
This late discovery  does not
contradict our claim that our approach is as fast as the rule-based approach in rule-provable cases,
since the rule-based approach returns \emph{``do not know''} on recursive schemas \cite{DBLP:conf/issta/HabibSHP21}.
}


\subsection{$\tkw{allCK}(C,K,E)$:
fast complement-absorption}\label{sec:allCK}

The last function to describe is $\allCK(C,K,E)$, that merges a structural operator $K$ into a canonical schema $C$.
\iflong{
The argument $C$ is either a $CTS$  $\atypeSet\,\aU$ (see Section \ref{subsec:dnfsyn})
or it is a typed conjunction $\aall[\atype(U),\ldots]$.
The argument $K$ is any term from the $S$ grammar that is neither boolean not $\aref(X)$ (Section \ref{sec:json}),
hence we have two cases: 
either $K\in\Set{\atype(U_1),\atypeSet\,\aU_1, \aconst(A), \anotConst(A)}$ or $K$ is an analytical structural operator.
Hence, we have four cases:
\begin{enumerate}
\item $C = \atypeSet\ \aU$ - $K\in\Set{\atype(U_1),\atypeSet\ \aU_1}$: we just remove from 
      $\aU$ all types apart from $U'$ or those in $\aU_1$ (when the result is an empty set,
      we return $\dFalse$);
\item $C=[\atype(U),\ldots]$ typed schema - 

$K\in\Set{\atype(U'),\atypeSet\ \aU_1}$: if $U$ 
    is $U_1$, or appears in $\aU_1$, then we return $C$, else we return $\dFalse$;
\item $C = \atypeSet \aU$ - $K$ typed operator; let $U'$ be the type that is analyzed by $K$;
         if $U'\not\in \aU$, then we just return $\aany[C]$, since it is not affected by $K$;
         otherwise we return a disjunction of $\atypeSet ( \aU \setminus \Set{U'})$, which is not affected by
               $K$, and the result of $\allCK(\aall[\atype(U')],K,E])$, computed as described later on.
\item $C=[\atype(U),\ldots]$  - $K$ typed operator: let $U'$ be the type that is analyzed by $K$;
         if $U_1!=U$, then we just return $\aany[C]$, since it is not affected by $K$; otherwise we insert
         $K$ inside $U$, and this is the interesting case.
\end{enumerate}
The specific algorithms that we use in the homogeneous case depend on $U$, but are all based on the Refutational Normalization principles. We describe here the case for $U=\aobj$ since it touches
on all of these principles, 
and it is the most important case in practice.
}
\ifshort{
The crucial case is when $C=\aall[\atype(U),\ldots]$ (see the grammar in Section \ref{subsec:dnfsyn}) 
and 
$K$ is an analytical operator that 
acts on $U$; the other cases are trivial, and they are described in the full paper  \cite{full}.
We describe the case for $U=\aobj$, 
since it is the most complex and the most important case in practice.}
\ifshort{$\allCK$ implements, cooperating with $\allXX$, the fast complement-absorption principle,
ensuring that, when $S$ is merged  with $\anot(S)$, the contradiction is discovered in linear
time.}

\paragraph{Canonical object schemas}

Canonical object schemas contain the following operators: a set of $\apProp(e,\Rf{X})$ operators, 
a set of $\apReq(e,\Rf{X})$ operators, and optional $\aminProps$ and $\amaxProps$:
$$
\begin{array}{llllll}
CO\!: \aall[\  \atype(\aobj)\ 
                        (,\apProp(e,\aref(X)))^*\ (,\apReq(e,\aref(X)))^*\ 
                        \\ 
                       \qquad\qquad 
                       (,\aminProps(n))^?\  (,\ \amaxProps(n\hide{|\Inf}))^? \ ]   \\[\NL]
\end{array}
$$

\ifshort{
\renewcommand{\PIE}[2]{\rlan{\PI{#1}{#2}}{=}\ES}
\renewcommand{\NE}[1]{\rlan{{#1}}{\neq}\ES}
\renewcommand{\PINE}[2]{\rlan{\PI{#1}{#2}}{\neq}\ES}
\renewcommand{\PME}[2]{\rlan{\PM{#1}{#2}}{=}\ES}              
\renewcommand{\PMNE}[2]{\rlan{\PM{#1}{#2}}{\neq}\ES}
}

All object schemas produced by our algorithm satisfy two properties: \emph{partitioning}
and \emph{refinement}.
The partitioning property specifies that, if we enumerate all 
$\apProp(e,\Rf{X})$ assertions of an object as $\apProp(e_1,\Rf{X_1}),\ldots,\apProp(e_n,\Rf{X_n})$,
then $n>0$ and the patterns $[e_1,\ldots,e_n]$ represent a partition of the set of all strings, which means that
they satisfy the following properties\iflong{ (hereafter $e_1\eeq  e_2$  means $\rlan{e_1}=\rlan{e_2}$)}:
\begin{enumerate}
\item disjunction: $\forall i,j\in\SetTo{n}.\ i\neq j\ \Implies\ \PIE{e_i}{e_j}$
\item covering: $\forall k\in \Sigma^*.\ \exists i\in\SetTo{n}.\ k\in\rlan{e_i}$
\item non-emptiness: $\forall i\in\SetTo{n}.\ \NE{e_i}$
\end{enumerate}
Partitioning ensures that all $\apProp$ constraints are mutually independent,
and every field name has exactly one corresponding $\apProp$.

The \emph{refinement} property specifies that each $\apReq(e,\Rf{Y})$ has
a corresponding
$\apProp(e,\Rf{X})$, and  that the $\apReq$ schema $\Rf{Y}$
refines the $\apProp$ schema $\Rf{X}$. 
In detail: 
\begin{enumerate}
\item for any $\apReq(e,\Rf{Y})$, there exists one $\apProp(e',\Rf{X})$ with $e = e'$;
\item for every $\apReq(e,\Rf{Y})$ assertion, if $X$ is the c-reference of the corresponding
$\apProp(e,\Rf{X})$, then $Y\supseteq X$ ($Y$ \emph{refines} $X$),
\iflong{ that is, $Y$ contains more references than $X$, }so that if a field satisfies
$\apReq(e,\Rf{Y})$, then it also satisfies $\apProp(e,\Rf{X})$.
\end{enumerate}
\bcolorfour
\mrevfour{Partitioning and refinement are obtained by (1) splitting the patterns of all $\apProp$ 
and $\apReq$ constraints until they are either equal or disjoint, and by (2) refining the c-reference of each 
$\apReq$ assertion so that it includes the c-reference of the corresponding $\apProp$.}{4.3, M.1}

For the running left-hand schema $S^i_L$, the canonical object is
the following one; observe how the assertion  $\apProp(\qkw{.*},\Rf{\xTrue})$, implicit in $S^i_L$,
was split in order to
match the pattern of $\apReq$:
\begin{equation}
\begin{split}
C^i_L = \aall[&\atype(\aobj), \\
         &\apProp(\keykey{a},\Rf{\xTrue}), \apReq(\keykey{a},\aref(\SSet{x^i})),\\
         &\apProp(\NP{\keykey{a}},\Rf{\xTrue})\ \ \    ]   
\end{split}
\label{eq:CL}
\end{equation}
\ecolor



In our code, we represent a canonical object as
a \emph{fragment map} $[f_1,\ldots,f_n]$ enriched with a pair $\aminProps(m)$, $\amaxProps(M)$:
$$\begin{array}
{llll}
CO & = & \{\ \fMap: [f_1,\ldots,f_n]\ (,\aminProps(m))^?\ (,\amaxProps(M)^? \} \\[\NL]
f_i & = & (e_i,X_i,[Y^i_1,\ldots,Y^i_{m_i}])
\end{array}
$$
A fragment map is a list of \emph{fragments} $f_i$;
each fragment $f_i$ is a triple $(e_i,X_i,[Y^i_1,\ldots,Y^i_{m_i}])$ ---  (pattern, propRef, reqRefList)
--- that represents the conjunction (with $m_i \geq 0$):
$$ \aall [ \apProp(e_i,\Rf{X_i}), \apReq(e_i,\Rf{Y^i_1}), \ldots, \apReq(e_i,\Rf{Y^i_{m_i}})]$$

\bcolorfour
\mrevfour{Hence,}{4.3, M.1} the object $C^i_L$ from the running example is represented as
follows; the $\apReq$ assertion is encoded by the $\SSet{x^i}$ c-reference in the first fragment:
\begin{equation}
\begin{split}
C^i_L &\text{\ (Step 9) is represented as:\ \ }\\
&\{\ \fMap: [(\keykey{a},\xTrue,[\SSet{x^i}]),(\NP{\keykey{a}},\xTrue,[\,])] \ \}
\end{split}
\label{eq:rep}
\end{equation}


\hide{
Finally, as an optimization, we observe that when the reqList $[Y^i_1,\ldots,Y^i_{m_i}]$ contains two c-references such that \GG{We may remove this}
$Y_1 \Implies Y_2$ (for example, because $Y_1\supseteq Y_2$),
then $Y_2$ can be removed from the list, since $\apReq(e_i,\Rf{Y_1})$ implies $\apReq(e_i,\Rf{Y_2})$.
Hence, we use a sound function ${\fastImplies}$ such that  ${\fastImplies}(Y_1,Y_2)$ is a sufficient condition
for $Y_1 \Implies Y_2$, and we ensure that the reqList never contains two c-references such that  ${\fastImplies}(Y_1,Y_2)$.
In our implementation,
${\fastImplies}(Y_1,Y_2)$ returns true when the two c-references are equal and when $Y_1=xFalse$,
so that no c-reference appears twice, and so that, once $xFalse$ is inserted into the list, it removes all other c-references.
}

\mrevfour{In our example, we call
$\allCK(C^i_L,\apProp(\keykey{a},\Rf{\Set{\NV{y}}}),E)$
(Step \ref{eq:allCSnot}).
}{4.3, M.1}
To add a new assertion $\apProp(p,\Rf{X})$ 
to a canonical object $C$, the function $\allCK$ combines the $\apProp(p,\Rf{X})$ with 
all fragments $(e_i,X_i,[\aY])$ of $C$, in a way that depends 
on which of the following cases describes the relation between ~$e_i$ and $p$:\ecolor\\
(1) $e_i$ \emph{disjoint} from $p$ ($\PIE{e_i}{p}$); \\
(2) $e_i$ \emph{included} in $p$ ($\PINE{e_i}{p}$ and  $\PME{e_i}{p}$); \\
(3) $e_i$ \emph{divided} by $p$ ($\PINE{e_i}{p}$ and  $\PMNE{e_i}{p}$).\\

The combination of the new assertion and a single fragment is implemented by the 
\akw{mergeFragProp} operation of 
Algorithm \ref{alg:mergeFragProp};
in the full \mrevone{paper \cite{full}}{1.10} we show how {\allCK} applies \akw{mergeFragProp}
to all the fragments of the fMap of $C$, and how it combines the results
with a form of cartesian product, with a shortcut
that immediately returns $\dFalse$ when a fragment produces the empty lists.

The \akw{mergeFragProp} function, given a fragment \akw{(e,pRef,reqList)} and a 
$\apProp$ parameter
\akw{pProp(p,X)}, returns 
a list of fMaps, which is equivalent to the conjunction of the fragment with
$\apProp(p,\aref(X))$; when the returned list is empty, the conjunction is unsatisfiable.

\hide{
--returns a list of lists of fragments, that is, a list of
--fMaps, whose union is equivalent to the conjunction 
--of the fragment (e,pRef,reqList) with pProp(p,x) 
--when applied to an object field that matches e}




\begin{algorithm}[htb]
\small
\AlgoSetup

\SetAlgoLined


\Fn{\fMergeFragProp$((\ke,\kpRef,\kreqList),\,\kpProp(\kp,\kx),\,\kE): \kList[\kFMap]$}{
  \If{\fEmpty($\ke \cap \kp$)}{
    \KwRet $[\kFMap[(\ke,\kpRef,\kreqList)]]$\;
  }

  \uIf(\tcp*[f]{e included in p}){\fEmpty($\ke \setminus \kp$)}{  
  \krefinedReqList\ $\gets$  
  \lFor{$\kId{ref} \bf{\ in\ } \kreqList$}{
     $\fAllXX(\kId{ref},\kx,\kE)$
  }
    \If{$\kxFalse \in \krefinedReqList$}{
      \KwRet $[]$\;
    }
    $\krefinedPRef \gets \fAllXX(\kpRef,\kx,\kE)$\;
    \KwRet $[\kFMap[(\ke,\krefinedPRef,\krefinedReqList)]]$\;
  }
  \Else(\tcp*[f]{e divided by $\kp$}){
    $\krefinedPRef \gets \fAllXX(\kpRef,\kx,\kE)$\;
    \If{$\kreqList = []$}{
      \KwRet $[\kFMap[(\ke \cap \kp,\krefinedPRef,[]),(\ke \setminus \kp,\kpRef,[])]]$\;
    }
    \Else(\ldots \tcp*[f]{exponential decomposition}){ 
    }
  }

}
\caption{\SETFN{mergeFragProp}\((\kId{frag},\kpProp(\kp,\kx),\kE)\)}
\label{alg:mergeFragProp}
\end{algorithm}

%
%
%
%

When the fragment is disjoint from $p$ (lines 2 - 3),
the fragment is returned as it is (more precisely, it is encapsulated into a unary list that 
contains a unary fMap that contains the fragment).
 
When the fragment is included in $p$ (lines 4 - 9), 
we 
refine all c-references in its \kw{reqList}, as well as  its  \kw{pRef}, using $X$.
In order to refine a c-reference $Y$ with 
$X$, we use the 
function $\allXX(Y,X,E)$, which,
as described in Section \ref{sec:allXX}, computes a c-reference equivalent to $Y\cup X$,
and immediately tries to falsify it, 
returning $\xFalse$ in case the falsification attempt is successful.
This eager falsification is exploited in line 6:
$
        \mbox{\bf if\ } \xFalse \in \krefinedReqList  \mbox{\bf\ return}\ [\, ] 
$.
We return ``$[\,]$'', that means ``unsatisfiable'', since,
when a reqList contains $Y^i_j=\xFalse$,
then the corresponding
$\apReq(e_i:Y^i_j)$  is not satisfiable, hence the entire object schema is
not satisfiable,
and $\allCK(C,\apProp(p,X),E)$ can immediately return $\dFalse$.
This is the mechanism that $\allCK$ 
uses to implement the  ``fast complement-absorption'' principle
in the case when $C$ contains $\apReq(p,Y)$ 
and $K=\apProp(p,X)$ is its 
complement
: we refine $Y$ with $X$ using $\allXX$, the function $\allXX$
tries eagerly to falsify $E(Y\cup X)$, it returns $\dFalse$, and 
$\allCK$ returns $\dFalse$.
\iflong{The dual case that was discussed in Section \ref{sec:allXX},
when $C=\apProp(p,X)$ and $K$ is its complement $\apReq(p,Y)$,
is implemented similarly when
$\apReq(p,Y)$ is inserted into a $C$ that contains a fragment  $(p,X,\ldots)$.}

\bcolorfour
\mrevfour{Consider again the  example.}{4.3, M.1} With
$
K=\apProp(\keykey{a},\Rf{\SSet{\NV{y}}})
$,
$\allCK(C^1_L,K,E)$ (see Step \ref{eq:allCSnot}) invokes
\akw{mergeFragProp($f_1$,K,E)} over the first fragment $f_1$ of $C^1_L$ 
(see Step \ref{eq:rep}),
where:
\begin{equation*}
f_1=(e,\kw{pRef},\kw{reqList})=(\keykey{a},\xTrue,[\SSet{x^1}]).
\end{equation*}
Since $\akw{empty}(e\!\setminus\! p)$, 
we refine $\kw{reqList}$ by invoking (lines 4-5):
\begin{equation}
\allXX(\SSet{x^1},\SSet{\NV{y}},E)
\end{equation}
$\allXX$ eagerly normalizes 
$E(\SSet{x^1,\NV{y}})$: since $x^1$ and $y$ both denote $\atype(\anum)$,
$E(\SSet{x^1,\NV{y}})$ is unsatisfiable and $\allXX$ returns $\xFalse$,
hence $\krefinedReqList$ contains $\xFalse$.
Hence, \akw{mergeFragProp} returns the empty list (line 7), which 
allows $\allCK$ to immediately return $\dFalse$.

In the second example, instead, ${\allXX}(\SSet{x^2},\SSet{\NV{y}},E)$  returns
$\SSet{x^2,\NV{y}}$ (Step \ref{eq:allXX}).
In this case, we have not been able to falsify the conjunction, hence (line 8) we
merge $\key{pRef}=\xTrue$ with the new $\apProp$ c-reference $X=\Set{x^2}$, 
and \akw{mergeFragProp} returns the  refined
fMap:
$\key{fm}=\kFMap[(\keykey{a},\Set{x^2},\Set{x^2,\NV{y}}]$ (line 9),
representing:
\begin{equation*}
S_{\key{fm}} = \aall[\ \apProp(\keykey{a},\Rf{\Set{x^2}}),\ \apReq(\keykey{a},\aref(\Set{x^2,\NV{y}}))\  ]  
\end{equation*}
Hence, the computation of $\allCS(C^2_L,\anot(S_a),E)$ (Step \ref{eq:completefirst}) that invoked
$\allCK$ returns a DNF $D_1$ that includes $S_{\key{fm}}$:
\begin{equation*}
\begin{split}
D_1 = \aany[C_\key{fm}] \qquad C_\key{fm} = \aany[\aall[\atype(\aobj), S_{\key{fm}}, \ldots ]]
\end{split}
\end{equation*}
and now $\allCS$ invokes $\allDS$, as anticipated in Section \ref{sec:allCS}, Step \ref{eq:complete},
and again $\allDS$ eventually invokes $\allCK$, as in the previous case: 
\begin{equation}
\begin{split}
&\allDS(D_1,\anot(S_{ab}),E) \\
&\to^* \allCK(C_\key{fm},\ \apProp(\anyp[\keykey{a},\keykey{b}],\Rf{\SSet{\NV{z}}}),\ E) 
\end{split}
\end{equation}
Since $C_\key{fm}$ contains $S_{\key{fm}}$, hence 
$\apReq(\keykey{a},\aref(\Set{x^2,\NV{y}}))$, this  results in $\allCK$ 
using $\fMergeFragProp$ to merge
$\apReq(\keykey{a},\aref(\Set{\ldots}))$ with $\apProp(\anyp[\keykey{a},\keykey{b}],\Rf{\SSet{\NV{z}}})$.

Since $\keykey{a}\setminus\anyp[\keykey{a},\keykey{b}]$ is empty, we
compute
${\allXX}(\key{pRef},X,E)$
 (line 5), that is,
${\allXX}(\Set{x^2,\NV{y}},\SSet{\NV{z}},E)$.
We have:
$$
\begin{array}{llllll}
\!\!\!E(\Set{x^2,\NV{y},\NV{z}}) = \aall[\!\!\!\!&\!\!\atypeSet[\anum,\astr], \\[\NL]
&\!\!\anot(\atype(\anum)),\anot(\atype(\astr))\ ]\ 
\end{array}$$
so that this time $\allXX$ returns $\xFalse$, hence $\allCK$ returns $\dFalse$,
and $\allCS$ returns $\dFalse$, so that $C^2_L\subseteq S_R$
is proved. Hence, in a situation where 
rule (\textsc{\textsc{anyof-r}}) fails, RWG answers correctly.
\ecolor

\medskip
Finally, in the case of line 10, 
the $e_i$ fragment is divided into two fragments
with patterns $\PI{e_i}{p}$ and $\PM{e_i}{p}$, and the $\akw{pRef}$ of the
first fragment must be refined
with $X$.
When the divided fragment has an empty $\reqList$, nothing else must be done
(lines 12-13).
\iflong{If, however, a divided fragment $(e_i,\Rf{X_i},\reqList)$ has a non-empty 
$\reqList=[Y^i_1,\ldots,Y^i_{m_i}]$, then each of the
$\apReq(e_i,\Rf{Y^i_l})$ assertions may either be satisfied by a field matching $\PI{e_i}{p}$ or by a field matching $\PM{e_i}{p}$;
this means that we must list $2^{m_i}$ alternatives for the divided fragment, 
where every alternative describes how each $\apReq(e_i,\Rf{Y^i_l})$  is satisfied:
by the $\PI{e_i}{p}$ fragment or by the $\PM{e_i}{p}$ fragment.}
\ifshort{If, however, a divided fragment $(e_i,\Rf{X_i},\reqList)$ has a non-empty 
$\reqList=[Y^i_1,\ldots,Y^i_{m_i}]$, then the algorithm is more complex, and
 is described in the full \mrevone{paper \cite{full}}{1.10}.
}

\ifshort{
\medskip
\mrevone{We have shown how function $\allCK$ implements the principle of 
	fast complement-absorption for the $\apReq$-$\apProp$ pair, exploiting eager reference-normalization 
	implemented by $\allXX$. This is one of the most complex, and important, cases.
	The same principle is actually implemented for all pairs of complementary operators; more cases
	are detailed in the full paper \cite{full}.}{1.10}}

\hide{
In the full \mrevone{paper \cite{full}}{1.10} we show how {\allCK} applies \akw{mergeFragProp}
to all the fragments of the fMap of $C$, getting a list of fMaps from any fragments,
how it takes a cartesian product of these lists to get the final result, with a shortcut
that immediately returns $\dFalse$ when a fragment produces the empty lists.
}

\iflong{
$ {\allCK}(C,K,E)$, in the case where $K=\apProp(p,x)$, invokes a 
function $\akw{objectInsert}(C,pProp(p,X),E)$ that uses \akw{recPropInsert} to recursively
apply \akw{mergeFragProp}
to all the fragments of the fMap of $C$, getting a list of fMaps from any fragments,
taking a cartesian product of these lists to get the final result, with a shortcut
that immediately returns $[ ]$, mapped to $\dFalse$ by \akw{recPropInsert},
when one the fragment produces the empty lists.


}

%

\ifshort{

}

\iflong{The insertion of $\apReq$ into $C$ is quite similar and accomplishes the same aims of 
fast complement-absorption for $\apProp$, as we have already specified.}

\iflong{The  ``fast complement-absorption'' property for $\amaxProps$ is very
easy. When we insert $\anot(\amaxProps(i))$ into a schema that already implies 
$\amaxProps(i)$, we want to have an immediate failure.
Since the complement of $\amaxProps$ is expressed using $\aminProps$, 
we get this effect by checking, whenever we insert $\aminProps$ into a
canonical object type, whether this minimum is compatible with the current value of 
$\amaxProps$.
The case for $\aminProps$ is symmetrical.}

\hide{WAS IFSHORT
In the same way, arrays, numbers, strings, and booleans are represented in a way that ensures that
whenever $\akw{notPush}(K)$ is inserted into a $C$ that already contains $K$ the result reduces immediately
to $\dFalse$, so that the ``fast complement-absorption'' property is always satisfied;
more details can be found in the full \mrevone{paper \cite{full}}{1.10}.}


\iflong{We conclude talking about rapid component access.}

\iflong{
\subsection{Structural all-merge:
rapid component access}

Consider a typical case of object comparison, where, to
save space, we use $S$ and $T$ rather than $\aref(X)$, $\aref(Y)$, and we omit $\atype(\aobj)$,
$\aminProps$, $\amaxProps$ (see also Section \ref{sec:DNFNotObj}):
\[
\begin{array}{l}
	\aall [ \apProp(p'_1,S'_1),\ldots, \apProp(p'_{n'},S'_{n'}), \\
	\hphantom{\aall [ } \apReq(r'_1,T'_1),\ldots, \apReq(r'_{m'},T'_{m'}) ] \\[\NL]
	\subt \aall [ \apProp(p_1,S_1),\ldots, \apProp(p_{n},S_{n}), \\
	\hphantom{\subt \aall [ } \apReq(r_1,T_1),\ldots, \apReq(r_{m},T_{m}) ]
\end{array}
\]

A rule-based system would, for each $i\in 1..n$, look for
a $p'_{f(i)}$ that corresponds to $p_i$, in order to verify that
$S'_{f(i)} \subt S_i$, and similarly for every $j\in 1..m$.
This search phase is $O(n\times n' + m\times m')$ if the LHS is represented as a list,
but is $O(n+m)$ if it is represented as a hash structure. 

We reproduce the same linear efficiency in our approach, despite the fact that we complement the right
hand side.

Consider the following comparison, where we ignore the $\apReq$ component, since it behaves the same
way as the $\apProp$ component.
$$
\begin{array}{lllll}
\aall [ \apProp(p'_1,S'_1),\ldots, \apProp(p'_{n'},S'_{n'})] \\[\NL]
         \subt
           \aall [ \apProp(p_1,S_1),\ldots, \apProp(p_{n},S_{n}) ] 
\end{array}
$$
In our system, it corresponds to the {\CDNF} computation.
$$
\begin{array}{lllll}
\allCS(\aall [\apProp(p'_1,S'_1),\ldots],
       \anot(
           \aall [ \apProp(p_1,S_1),\ldots ] ) \\[\NL]
 \to
\allCS(\aall [\apProp(p'_1,S'_1),\ldots],
           \aany [ \apReq(p_1,\anot(S_1)),\ldots ] )\\[\NL]
\to
\aany[\allCK([\apProp(p'_1,S'_1),\ldots],\apReq(p_1,\anot(S_1))), \ldots  \\[\NL]
\qquad\qquad \allCK([\apProp(p'_1,S'_1),\ldots],\apReq(p_n,\anot(S_n))) ]
\end{array}
$$
We split the RHS, but the LHS is still intact.
Remember that the canonical object is represented as follows:
$$\begin{array}
{llll}
C & = & \{\ \fMap: [f_1,\ldots,f_n],\ \aminProps(m),\ \amaxProps(M)\ \} \\[\NL]
f_i & = & (e_i,X_i,[Y^i_1,\ldots,Y^i_{m_i}])
\end{array}
$$
We implement fMap as a hash table, hashed on the pattern $e_i$,
so that, for every $\apReq(p_i,\anot(S_i))$,
we find the corresponding $p'_{f(i)}$, if it exists, in constant time, as with the rule-based
approach, and we combine $\apReq(p_i,\anot(S_i))$ only with $\apProp(p'_{f(i)},S'_{f(i)})$.
When a pattern $e_i$ that is equal to $p'$ is not found, then we resort to the
complete approach described in the previous section, where we scan the entire fMap to check
for patterns that are included in $p$ or divided by $p$.
Hence, the \emph{rapid component access} principle guarantees that, also in this case,
the Refutational Normalization approach is as efficient as an efficiently implemented rule-based approach,
and it is complete in situations where the rule-based approach is not.
}

\ifshort{
\section{Efficiency}\label{sec:efficiency}

In the full version of the paper~\cite{full},  we show that the Refutational Normalization approach is
(asymptotically)
as fast as the rule-based approach for all judgments $S_1 \subt S_2$ which are 
rule-provable.

There, we consider an algorithm {\RB} that applies the rule-based approach to a problem 
$S_1 \subt S_2$, and an algorithm
{\RN} that applies Refutational Normalization to $\aDNF(\aall[S_1,\anot(S_2)])$,
 and we show that, for any  $S_1 \subt S_2$ which is rule-provable,
{\RN} reduces  $\aDNF(\aall[S_1,\anot(S_2)])$ to $\dFalse$ with a set of steps that
mimics the behavior of {\RB}.
The proof proceeds by cases on the last rule applied by {\RB}, and by induction on the
size of $S_1$.

For rule (\textsc{uninhabited}), $S_1$ is unsatisfiable, and Refutational Normalization and {\RB}  
need the same time to discover this fact. 

For rule (\textsc{anyof-l}), we prove $\aany[S'_1,\ldots, S'_n]\subt S_2$, and both approaches compare 
each $S'_i$ with $S_2$, and verify that all 
comparisons succeed.

For rule (\textsc{anyof-r}), we exploit the hypothesis that $S_1 \subt S_2$ is rule-provable;
in this case, the $\anot(\aany[\ldots])$-optimization
ensures that {\RN} mimics the behavior of {\RB}, as detailed in Section~\ref{sec:notany}.

Finally, we consider the structural case when $S_1$ is an object schema, an array schema,
or an atomic schema.
In this case, the rule-based approach finds, for every assertion~$K^i_2$ in $S_2$, a corresponding
assertion $K^j_1$  in $S_1$ such that $K^i_2$ implies $S_2$. The normalization algorithm
normalizes $\aall[S_1,\anot(K^i_2)]$ for every assertion~$K^i_2$ in~$S_2$, and the combination
of fast complement-absorption and eager reference-evaluation ensures that each 
$\aall[S_1,\anot(K^i_2)]$ is reduced to $\dFalse$ with the same efficiency that 
{\RB}  uses to prove that~$K^j_1$ implies $K^i_2$, as detailed
in Section~\ref{sec:allCK}.

Hence, lazy normalization,
fast complement-absorption,
eager reference-evaluation, and $\anot(\aany[\ldots])$-optimization
together ensure that our approach reaches the same efficiency as {\RB}  in 
situations where
{\RB} can 
prove inclusion.
}

\iflong{
\section{Efficiency}
\label{sec:efficiency}

\iflong{We have seen how the principles of
lazy normalization,
fast comple\-ment-absorption,
eager reference-evaluation,
\hide{rapid component access}
and $\anot(\aany[\ldots])$-optimi\-zation are implemented by 
Refutational Normalization.}
In this section, we show, informally, that the Refutational Normalization approach is
(asymptotically)
as fast as the rule-based approach for all judgments $S_1 \subt S_2$ which are 
rule-provable.

In greater detail, we consider an algorithm {\RB}  that applies the rule-based approach to a problem 
$S_1 \subt S_2$ , and an algorithm
{\RN} that applies Refutational Normalization to $\aDNF(\aall[S_1,\anot(S_2)])$,
returns ``included'' if it rewrites it to $\dFalse$, and invokes a Witness Generator
otherwise.
Consider a pair of schemas $S_1$ and $S_2$ such that $S_1 \subt S_2$ can be 
proved by applying the inclusion rules; we show that, in this case, {\RN} is as efficient
as {\RB} .

For ease of comparison, we consider a set of simplifying 
assumptions\ifshort{, which are explained and justified in the full version of the paper}:
\begin{enumerate}
\item no negation: we assume that $S_1$ and $S_2$ contain neither
    $\anot$ nor $\aone$; 
    \iflong{this assumption is reasonable since $\anot$ that has
    complex object or array operators in its scope is out of the reach of the {\RB}  algorithm,
    while $\anot$ that has only atomic operators in its scope adds nothing interesting to the
    analysis;}
\item normal form: we assume that both $S_1$ and $S_2$ are in DNF, and hence we rename
   them as $D_1$ and $D_2$;
   \iflong{ this assumption is reasonable since both algorithm start by reducing
   $S_1$ in DNF, and {\RB}  even brings $S_2$ in DNF, hence, by eliminating this phase,
   we are actually giving an advantage to the {\RB}  algorithm;}
\item simple objects: 
   \iflong{since we are considering schemas without negation, in an
   object in normal every reqList is generated by JSON schema $\qreq$ operator,
   hence we can assume that every fragment $(e,X,\aY)$ in the representation of the object
   has a singleton c-reference $\Set{x}$ and either has an empty reqList, as in $(e,\Set{x},[])$,
   or a singleton reqList $[\Set{x}]$ as in $(e,\Set{x},[\Set{x}])$;}
      \ifshort{we assume every fragment $f_i=(e_i,X_i,\aY_i)$ in the representation of any 
      object inside\todo{add comma so readers better parse the sentence?}
      $D_1$ or $D_2$
   has a singleton c-reference $\Set{x}$ and either has an empty reqList, as in $(e,\Set{x},[])$,
   or a singleton reqList $[\Set{x}]$ as in $(e,\Set{x},[\Set{x}])$;}
\item stratified form: we assume that $D_1$ and $D_2$ are stratified%
    \iflong{; this simplifying assumption is reasonable since stratification is a linear time operation which does not affect asymptotic analysis}.
\end{enumerate}

\iflong{We will also assume that, whenever the {\RB}  algorithm is able to prove that 
a DNF schema $D$ is unsatisfiable, the {\RN} algorithm reduces $D$ to $\dFalse$
with the same number of operations. This is  reasonable since the {\RB} 
normalization algorithm is a simplified version of  that using by {\RN}.

We now show that  {\RN} normalizes $\allDS(D_1,\anot(D_2),E)$ with at most
the same number of
operations that {\RB}  uses to verify that $D_1\subt D_2$, by cases on the
last rule used in the proof, and by induction on the size of $D_1$.

If $D_1 \subt D_2$ is proved by rule (\textsc{uninhabited}), then, by assumption, {\RN} 
reduces $D_1$ to $\dFalse$ with the same number of operations, and then it
concludes in constant time. }

\ifshort{We now show that  {\RN} normalizes $\allDS(D_1,\anot(D_2),E)$ with at most
the same number of
operations that {\RB}  uses to verify that $D_1\subt D_2$, reasoning by cases on the
last rule used by {\RB}  in its proof of $D_1\subt D_2$, and by induction on the size of $D_1$.

If $D_1 \subt D_2$ is proved by rule (\textsc{uninhabited}), hence {\RB}  has proved
that $D_1$ is unsatisfiable using its simplified algorithm, then we can assume that
{\RN} reduces $D_1$ to $\dFalse$ with the same number of operations, and then it
reduces
$\allDS(\dFalse,\anot(D_2),E)$ to $\dFalse$ in constant time. }

If $D_1 \subt D_2$ is proved by rule (\textsc{anyOf-l}), then 
$D_1 = \aany[C_1,\ldots,C_n]$, and {\RB}  proves that $C_i \subt D_2$ holds for every 
$i\in \SetTo{n}$. 
{\RN} behaves in the same way: the $\aany$ case of $\allDS(D_1,\anot(D_2),E)$ 
invokes $\allCS(C_i,\anot(D_2),E)$ for every $i$, reduces that case to $\dFalse$
with the same number of operations that {\RB}  uses to prove that $C_i \subt D_2$,
and, finally, {\RN} combines all these $\dFalse$ into $\dFalse$.

If $D_1 \subt D_2$ is proved by rule (\textsc{anyOf-r}), then $D_2$ is a disjunction
$\aany[C_1,\ldots,C_n]$, hence $\notPush(D_2)=\aall[\anot(C_1),\ldots,\anot(C_n)]$.
In this case, {\RN}
proceeds in the same way as {\RB}  thanks to the optimization 
$\anot-\aany[\ldots]$ that mirrors the behavior of rule  (\textsc{anyOf-r}).
 In more detail, since inclusion is rule-provable, there exists $C_i$ such that $D_1 \subt C_i$.
The {\RB}  algorithm tests $D_1 \subt C_j$ for every $1\leq j <i$, every such comparison
fails, and, finally, the test $D_1 \subt C_i$ succeeds.
In the same way, {\RN} computes $\allDS(D_1,\anot(C_j),E)$ for every $1\leq j <i$,
using the incomplete $\fastFailAllCS$ algorithm that fails as fast as {\RB} , for every $1\leq j <i$,
and, finally, when it computes $\allDS(D_1,\anot(C_i),E)$, it reduces to $\dFalse$.
By induction, each step uses the same number of operations as {\RB} .

We are left now with structural rules.
Consider rule (\textsc{object}). 

\infrule[object]
{
C_{1}.\atype = \aobj \qquad C_{2}.\atype = \aobj\\
C_{1}.\aminProps \geq C_{2}.\aminProps \qquad C_{1}.\amaxProps \leq C_{2}.\amaxProps\\
\forall (e_2,S_{2}) \in C_{2}.\apProp:\ \exists  (e_1,S_{1}) \in C_{1}.\apProp:\ 
                                                  \ e_1 \supseteq e_2 \ \mbox{\ and\ } \ S_{1} \subtr S_{2} \\
\forall (e,S_{2}) \in C_{2}.\apReq:
\exists\ (e,S_{1}) \in C_{1}.\apReq: \ \ e_1 \subseteq e_2 \ \mbox{\ and\ } \ S_{1} \subtr S_{2} \\
}
{
C_{1} \subtr C_{2}
}

\noindent
Since $ C_{2}.\atype = \aobj$, then $C_{2}$ has shape:
$$
\begin{array}{llllll}
\aall[\  \atype(\aobj),\ 
                        (\apProp(e,\aref(x)),)^*\ (\apReq(e,\aref(x)),)^*\ 
                        \\ 
                       \qquad\qquad 
                       \aminProps(n) (,\ \amaxProps(n\hide{|\Inf}))^? \ ]   \\[\NL]
\end{array}
$$

so that $\notPush(C_2)$ has shape
$$
\begin{array}{llllll}
\aany[\ \anot( \atype(\aobj)),\  
                   \\ \qquad
                       (\apReq(e,\aref(\NV{x})),)^*\ (\apProp(e,\aref(\NV{x})),)^*\ 
                        \\  
                       \qquad\, 
                       \amaxProps(n),\ \aminProps(n+1) \ ]   \\[\NL]
\end{array}
$$

Hence, $\allCS(C_1,C_2,E)$ \iflong{--- case $\aany$ ---}
invokes $\allCK(C_1,K_i,E)$ for every $K_i$ in 
$$
\begin{array}{llllll}
\aK = \SetOpen\!\!\!\!&\anot( \atype(\aobj)),\ 
                       \\ \qquad
                        &(\apReq(e,\aref(\NV{x})),)^*\ (\apProp(e,\aref(\NV{x})),)^*\ 
                        \\ 
                       \qquad 
                       &\amaxProps(n),\ \aminProps(n+1) \ \SetClose   \\[\NL]
\end{array}
$$
We now show that
each $\allCK(C_1,K_i,E)$ call corresponds to one successful test in the 
premise of  (\textsc{object}), so that, by induction, that call requires the same number
of operations as the corresponding test in {\RB} , and returns $\dFalse$,
so that $\allCS(C_1,C_2,E)$ returns $\dFalse$ as well.

\iflong{Hence, consider any $K_i$ in the set $\aK$.

If $K_i=\anot( \atype(\aobj))$, then $\allCK(C_1,K_i,E)$ immediately returns $\dFalse$ because
$C_1$ is a conjunction that includes a $\atype(\aobj)$ argument.

If $K_i=\aminProps(n_2+1)$, this means that $C_2$ contains
the dual clause $\amaxProps(n_2)$, hence, by the second premise
of rule (\textsc{object}), $C_1$ contains $\amaxProps(n_1)$ with $n_1\leq n_2$.
Under these condition, $\amaxProps(n_1)$ and $\aminProps(n_2+1)$
are mutually incompatible, and the implementation of $\allCK(C,\aminProps(m),E)$ 
immediately returns $\dFalse$ when the $\amaxProps$ value of $C$
is incompatible with $\aminProps(m)$.
The same reasoning holds for $K_i=\amaxProps(n)$.
}

\ifshort{
The interesting case is $K_i=\apProp(e_2,\aref(\Set{\NV{x_2}}))$;
the presence of this $K_i$ in $\aK$ means that $C_2$ contains
$\apReq(e_2,\aref(\Set{x_2})$; by the last premise of the successful
(\textsc{object}) rule, there exists $\apReq(e_1,\aref(\Set{x_1})$ in $C_1$ such that
$e_1 \subseteq e_2$ and $\aref(\Set{x_1}) \subt \aref(\Set{x_2})$.
By assumption \emph{simple object}, 
the fragment of $C_1$ corresponding to $\apReq(e_1,\aref(\Set{x_1})$
has shape $f_1=(e_1,\_,[\Set{x_1}])$.
When $\kw{mergeFragProp}$ merges $\apProp(e_2,\aref(\Set{\NV{x_2}}))$ with  
$f_1$, we are in the \emph{included}
case  (line (3) of Alg.~\ref{alg:mergeFragProp}), hence we must refine the reqList $[\Set{x_1}]$ of the 
fragment $f_1$ with $\aref(\Set{\NV{x_2}})$.
Since $\aref(\Set{x_1}) \subt \aref(\Set{x_2})$, by induction, this operation returns 
$\xFalse$ with the same number of operations as
the corresponding operation of {\RB} , hence
$\kw{mergeFragProp}$ returns $[\,]$, hence $\allCK$ returns~$\dFalse$. 
}

\ifshort{The other cases ---
$\anot(\atype(\aobj))$
$\apReq$, $\amaxProps$ and $\aminProps$ --- are analogous or much simpler, and described in the full \mrevone{paper \cite{full}}{1.10}.}

\hide{\iflong{...\todo[]{missing?}
If $K_i=\apReq(e_2,\aref(\Set{\NV{x_2}}))$, this means that $C_2$ contains
the dual clause $\apProp(e_2,\aref({x_2})$;\todo[]{paren missing?} by the third premise of the successful
(\textsc{object}) rule, there exists $\apProp(e_1,\aref({x_1})$\todo[]{parenthesis?} in $C_1$ such that
$e_1 \supset e_2$ and $\aref({x_1}) \subt \aref({x_2})$;
Since the canonical object $C_1$ enjoys the \emph{partitioning}
property, this means that all other fragments of~$C_1$, being disjoint
from $e_1$, are also disjoint from $e_2$.
Hence, when we call $\allCK(C_1,K_i,E)$, we are in the ``included'' case, where
we combine $e_2$ with a fragment $e_1$ that is included\todo[]{sentence ?} 

By induction, we know that $\allXX(\Set{x_1},\Set{\NV{x_2}})$ returns
$\dFalse$, and does that with the same number of operations that are used by the {\RB}  algorithm
to perform the corresponding test. TO BE COMPLETED \todo{unfinished}
}}

}


%

\section{Experimental Evaluation}
\label{sec:expeval}

\subsection{Research Hypotheses}
\label{sec:hypotheses}

The experimental evaluation is guided by these hypotheses:

{\textbf{H1}: Completeness:}\ \  Our approach is at least as complete as the witness-generation approach. Here ``completeness'' refers to the set of inclusion problems the algorithm can solve.

{\textbf{H2}: Efficiency:}\ \  Our approach is at least as 
efficient as the rule-based approach.
	
{\textbf{H3}: \emph{Significant} increase in coverage:}\ \  The combined effect of completeness and efficiency leads to a significant increase in the ability to analyze large and complex schemas.
\hide{
\begin{itemize}
	\item \textbf{H1}: Completeness: Our approach is at least as complete as the witness-generation approach.
    Here ``completeness'' refers to the set of inclusion problems
	that the algorithm is able to solve without a time limit.
	\item \textbf{H2}: Efficiency: Our approach is at least as 
	efficient as the rule-based approach.
	\item \textbf{H3}: \emph{Significant} increase in coverage: 
	The combined effect of completeness and efficiency leads to a significant increase in the
	ability to analyze large and complex schemas.

\end{itemize}}


\subsection{Implementation and experimental setup}
\label{sec:setup}
\iflong{We implemented our refutational normalization algorithm for JSON Schema Draft-06 in Scala~3.3. Our experiments were run on a server with 2$\times$24-core Intel Xeon Gold 6248R 3.0GHz, 384GB RAM, operating under Debian 12 and OpenJDK 21. We assigned 32GB of heap space to the JVM. The dispatcher scripts are implemented in Bash and Python 3.14.
Each schema is processed by a single thread, and all reported times are	measured for a single run. We enforce a 10-minutes timeout per schema.
The experimental setup for existing tools is based on the reproduction package provided by Attouche et al.~\cite{repro_package}. The existing tools run on the same machine, using the same timeout and JVM memory configuration.
}
\ifshort{
Our prototype is implemented in Scala~3.3. Experiments ran on a server with
2$\times$24-core Intel Xeon Gold 6248R 3.0GHz and 384GB RAM, with 32GB of heap space assigned to the JVM. Each inclusion test runs in a single thread, with a 10-minute timeout per test.
The detailed experimental setup is described in the full \mrevone{paper \cite{full}}{1.10}.
}


\subsection{Tools for Comparative Experiments}
\label{sec:tools}
We compare against the following tools: 

\emph{Rule-based containment checker ({\RB}).}
Developed in Python, proposed by Habib et al.~\cite{DBLP:conf/issta/HabibSHP21}
(vers.~0.0.8),
as described in Section~2, and originally called
\emph{jsonsubschema}. \iflong{It only supports Draft-04 schemas and has
restrictions w.r.t.\ negation and recursion, which constrains its applicability
in more expressive inclusion-checking scenarios.}
	
\emph{Witness Generator ({\WGBI}).} A JSON Schema witness generation tool
developed in Java, proposed by Attouche et al.~\cite{10.1145/3799416}, used to
check inclusion as described in Section~4. \iflong{Although it was shown to be
more complete than {\RB} in most cases, it usually shows longer run times.}

\emph{Refutational Witness Generator ({\RN}).} Our Scala prototype, it applies
the refutational normalization algorithm and invokes a recursive witness
generation procedure on the produced DNF when it differs from $\dFalse$.
\mrevone{It supports Classical JSON Schema with the only exception of
{\quniqIts}, a limitation shared by {\WGBI}.}{1.4}

\iflong{
All witnesses generated by {\WGBI} and {\RN} are validated
using two independent reliable\footnote{According to the Bowtie report (\url{https://bowtie.report}, retrieved 12 February 2026), both tools have perfect coverage of the official JSON Schema Draft-06 test suite.} validators.
The validators are:
	\begin{itemize}
		\item Networknt JSON Schema Validator~\cite{jsvalidator1}, a Java validator 
		\item jsonschema~\cite{jsvalidator_python}, a Python library
	\end{itemize}
}

\subsection{Schema Collections}
\label{sec:schemas}
\begin{table}[tb]
	\centering
	\caption{Characteristics of schema collections. }
	\label{tab:collections}

    \renewcommand{\arraystretch}{0.9}
    \setlength{\aboverulesep}{0.3ex}
    \setlength{\belowrulesep}{0.3ex}
    \setlength\tabcolsep{3.7pt}
    \footnotesize
    
	\begin{tabular}{lrrrrrrr}
		\toprule
		\textbf{Collection} &
		\textbf{\#Total} &
		{\bf \#$\not\subseteq$} &
		{\bf \#$\subseteq$ } &
		\textbf{\#Unk.} &
		\textbf{Avg Size} &
		\textbf{Max Size} \\
		\toprule
		MergeAllOf     & 174 & 7 & 167 & 0 & 0.6 KB & 2.1 KB \\
		Synthesized & 1{,}331 &  450 & 881 & 0 & 0.5 KB & 2.9 KB \\
		Handwritten SC &   282   &   120 & 162 & 0 & 0.9 KB & 3.9 KB \\
		{\RB}-testset & 300 & 192 & 108 & 0 & 48.3 KB &  949.8 KB \\
		\midrule
		SchemaStore vers. & 1{,}056 & 153 & 789 & 114 & 28.8 KB & 938.4 KB \\
		oneOf as anyOf        & 1{,}822    &   156 & 1{,}185 & 481 & 32.6 KB & 587.8 KB \\     
		uneval as additional  & 608     &   181 & 302 & 125 & 82.4 KB & 1{,}174.2 KB \\
		additional as uneval  & 5{,}698    &   140 & 3{,}572 & 1{,}986 & 23.6 KB & 1{,}049.6 KB \\
		\midrule\midrule
		\multicolumn{2}{l}{\textbf{Cut-off ($>$ 25 KB)}} & & & & & \\
		oneOf as anyOf (L)        & 702    &   45 & 339 & 318 & 71.0 KB & 587.8 KB \\ 
		uneval as additional (L) & 176    &   78 & 11 & 87 & 273.5 KB & 1{,}174.2 KB \\
		additional as uneval (L) & 1{,}201    &   38 & 81 & 1{,}082  & 85.3 KB & 1{,}049.6 KB \\
		\bottomrule
	\end{tabular}

\end{table}
Our collections of schema inclusion tests model
three use cases: (1)~schema evolution, where we compare
different versions of (semantically related) schemas,
(2)~tool checking, where we verify the correctness of a tool, typically one that performs schema rewriting, and
(3)~schema analysis, where we analyze schemas to check for properties such as whether certain keywords are interchangeable.

Table~\ref{tab:collections} describes each collection, stating the total number of inclusion tests, broken down into included, non-included\mrevone{, and unknown}{1.5} cases, as well as the mean and maximum size of the tests.
\mrevone{The collections ``Synthesized'', ``Handwritten SC'' and ``RB-testset'' have pre-defined ground truths. For the others, we rely on the result of {\WGBI}, which has been tested extensively, 
defaulting to ``unknown'' when {\WGBI} does not produce a result, 
unless {\RN} produces a valid witness (checked by an independent validator), in which case we also classify tests as non-included.}{1.5}
%
Five collections were already used in evaluating {\WGBI}~\cite{10.1145/3799416}: 
``MergeAllOf'', ``Synthesized'', ``Handwritten SC'', ``{\RB}-testset''\footnote{The ``{\RB}-testset'' was referred to as ``CC-testset'' by Attouche et al.\ in~\cite{DBLP:journals/pvldb/AttoucheBCGSS22,10.1145/3799416}.}, and ``SchemaStore versions''.

The ``SchemaStore versions'' dataset compares successive sche\-ma versions, as retrieved from the SchemaStore repository. It is arguably the most important collection,
as it covers the fundamental use case ``schema evolution'' with real-world non-trivial schemas.
The ``MergeAllOf'' and ``{\RB}-testset'' 
collections contain test cases for the MergeAllOf and {\RB} tools, respectively, and cover the use case ``tool checking''. ``Handwritten SC'' and ``Synthesized'', produ\-ced in the context of the {\WGBI}
project\iflong{ to test correctness and completeness of inclusion-checking tools}, cover the same use case.

We introduce three new datasets of category ``schema analysis'', which were not 
considered in previous work because they were too hard to be processed.
These datasets are derived from real-world GitHub schemas to analyze specific usage patterns of keywords:

\emph{``oneOf as anyOf''}, derived from the GitHub  dataset by 
Attouche et al.~\cite{10.1145/3799416}: Starting from a schema $S$ 
with {\qone}, we create a schema $S^{one\to any}$ by replacing each 
occurrence of {\qone}
with {\qany}, and test the equivalence of~$S$ and  $S^{one\to any}$.
This has practical interest, as {\qone} is often used in situations where it 
is equivalent to {\qany}~\cite{DBLP:conf/er/BaaziziCGSS21}, but {\qany} is much easier to validate and analyze.
\iflong{
	
	\emph{``uneval as additional''}: To create this dataset, we used the GitHub 
	Code Search API to obtain a schema $S$ containing the Modern JSON 
	Schema keywords {\qunProps} or {\qunIts}. 
	We transform $S$ into an equivalent Classical schema $S^{M\to C}$, using the
	approach proposed by Attouche et al.~\cite{DBLP:journals/tcs/AttoucheBCGKSS26}.
	We then check whether $S^{M\to C}$ is equivalent to the schema 
	$S^{un\to add}$ obtained by just
	replacing all {\qunStar} keywords with {\qaddStar} keywords, 
	which is a much
	easier transformation, that does not, in general, preserve the schema meaning. 
	When $S^{M\to C}$ is equivalent to $S^{un\to add}$, we can affirm that the user 
	used the complex modern operators in a situation where they were equivalent 
	to the corresponding, simpler, Classical operators.
	
	\emph{``additional as uneval''}: This dataset is also derived from the 
	Git\-Hub  dataset
	by Attouche et al.~\cite{10.1145/3799416}. 
	Given a Classical schema $S$, containing {\qaddStar}, we replace each
	 {\qaddStar} keyword with the corresponding {\qunStar} keyword, to
	obtain a schema $S^{add\to un}$. 
	We then use the approach by Attouche et al.~\cite{DBLP:journals/tcs/AttoucheBCGKSS26}, 
	to transform $S^{add\to un}$ into an equivalent Classical schema 
	$S^{add\to un;M\to C}$, and check the inclusion  between $S$ and $S^{add\to un;M\to C}$.
	We use this dataset to check the hypothesis that, in most practical cases, the 
	Classical-to-Modern substitution {\qaddStar} $\to$ {\qunStar} has no 
	effect on the schema.}
\ifshort{	
	
	\emph{``uneval as additional''} and \emph{``additional as uneval''},
	obtained by substituting Modern JSON 
	Schema keywords {\qunIts} and {\qunProps}
	with their counterparts in Classical JSON Schema,  
	{\qaddIts} and {\qaddProps}, and vice versa
	(details in the full \mrevone{paper \cite{full}}{1.10}).
	This allows us to assess the practical relevance of semantic differences 
	between the operators.
}

These three datasets correspond to open research problems regarding the use of those
JSON Schema keywords. 
They contain many small schemas whose inclusion is easy to prove,
but also a good \mrevtwo{number of sizeable schemas}{2.8}, 
where we expected a significant 
difference over the state of the art.
To verify this hypothesis, for each of these collections, we also analyze the subset that contains only inclusion tests with a size larger than 25~KB, termed ``cut-off'' and marked ``(L)''.
This threshold removes trivial examples while leaving a good number of tests, as shown in Table~\ref{tab:collections}.



\subsection{Results and Analysis}
\label{sec:charts}




\iflong{
In our experiments, we compare {\RB} and {\WGBI} against our {\RN}.

In case of non-inclusion, the generation-involving approaches produce a witness for the non-inclusion test $S_1\setminus S_2 =$ {\{\qall: [$S_1$, \{{\qnot}: $S_2$\}]\}}. We check the validity of the witness using an external validator. Although the Java validator used in our experiments is generally very reliable, we observe that it produces false negatives in a few recurring cases. In these cases, we manually inspect the result and invoke the Python validator to confirm our evaluation. Both our manual assessment and the result of the Python validator confirm the validity of the witness in all cases.}

\subsubsection{Completeness and Coverage}

Completeness (H1) is measured by the absence of 
errors: \mrevone{We define runtime errors as all cases where a tool fails to produce a definite result (i.e., ``included'' or ``non-included'') but does not time out. Logical errors occur when there is a definite result that contradicts the known ground truth,
so that we do not impute logical errors to a tool when the ground truth is ``unknown''.}{1.5}
Fig.~\ref{fig:failure_rate} shows the failure rates of the three tools, 
categorized into logical errors, timeouts, and runtime errors. We distinguish two groups of collections: 
artificial and real-world sche\-mas, allowing us to evaluate the behavior of the tools across both controlled synthetic benchmarks and practical real-world inputs. 

\ifshort{
\begin{figure}[t]
	\centering
	\marginnote{\bcolorone{\Small chart updated}}
	{\setlength{\fboxsep}{0pt}%
		\setlength{\fboxrule}{1pt}%
	\fcolorbox{red}{white}{
	\includegraphics[width=\columnwidth]{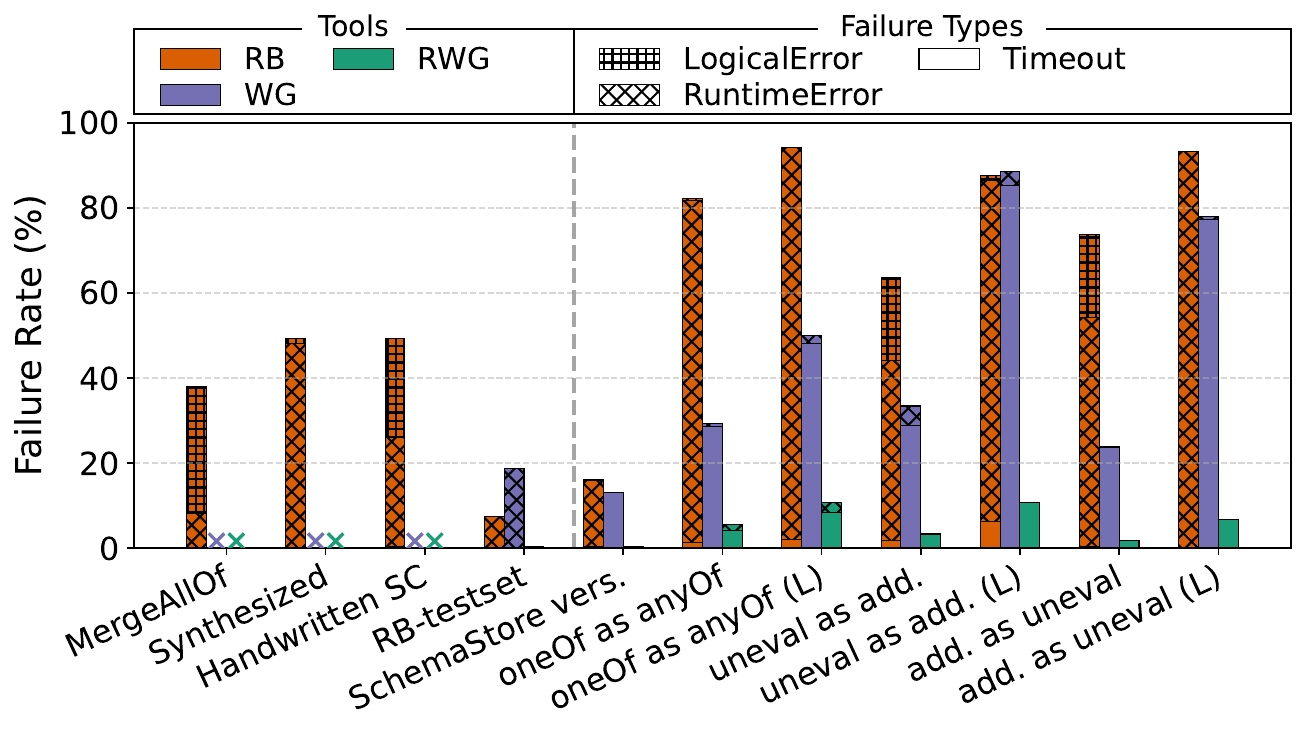}
	}}
	\caption{Failure rates of tools across datasets (lower is better). Tools distinguished by color, failure types by pattern. ``$\mathsf{x}$'' indicates zero-height bars (i.e., tools without failures).}
	\label{fig:failure_rate}
\end{figure}
}
\iflong{
\begin{figure}[t]
	\centering
	\includegraphics[width=\columnwidth]{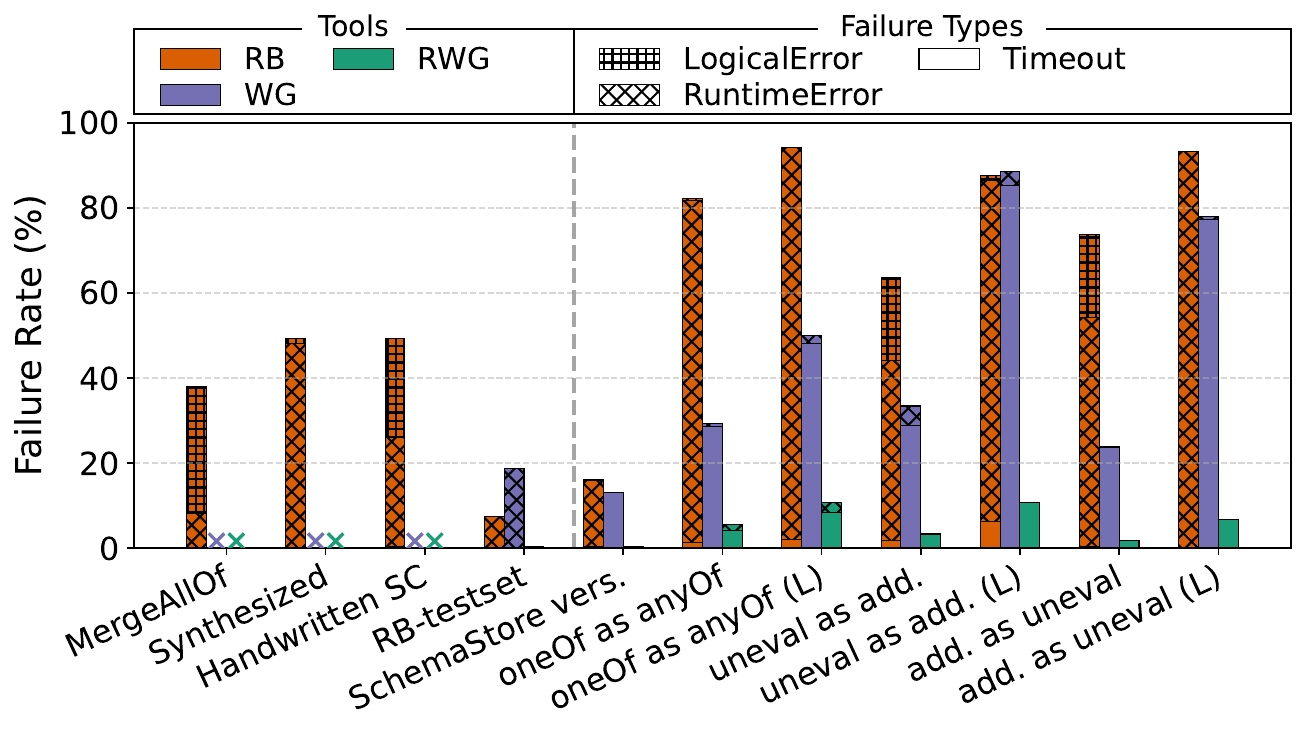}
	\caption{Failure rates of tools across datasets (lower is better). Tools distinguished by color, failure types by pattern. ``$\mathsf{x}$'' indicates zero-height bars (i.e., tools without failures).}
	\label{fig:failure_rate}
\end{figure}
\begin{figure}[t]
	\centering
	\includegraphics[width=\columnwidth]{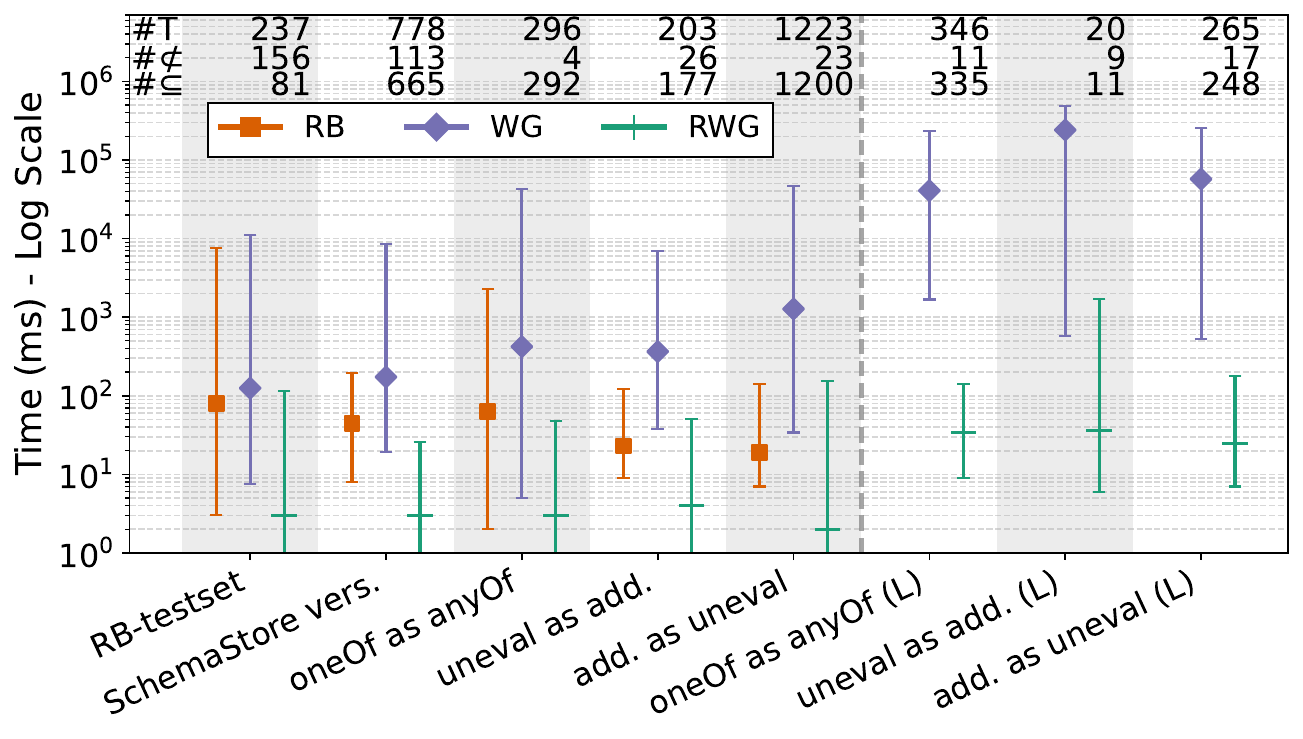}
	\caption{Run time comparison on intersection of successfully processed inclusion tests. \#T indicates the number of inclusion tests in the intersection, \#$\subseteq$/\#$\not\subseteq$ shows the number of included/non-included tests.
		The markers on the vertical bars indicate the $5^{th}$, $50^{th}$, $95^{th}$
		run time percentiles.}
	\label{fig:runtime}
\end{figure}
	
\begin{table*}[ht]
	\centering
	\caption{Experimental results. ``Error'' is divided into logical errors (Log.) and runtime errors (Run.). Ind. = Individual Runtime (calculated on schemas for which the specific tool succeeded); Int. = Intersection Runtime (calculated on schemas for which every tool succeeded). Intersection runtimes for cut-off datasets exclude {\RB}.}
	\label{tab:performance_by_collection}
	\begin{tabular}{llrrrrrrrrrrr}
		\toprule
		\textbf{Collection} & \textbf{Tool} & \textbf{Success} & \textbf{Timeout} & \multicolumn{2}{c}{\textbf{Error}} & \multicolumn{2}{c}{\textbf{Avg. Time (ms)}} & \multicolumn{2}{c}{\textbf{Med. Time (ms)}} & \multicolumn{2}{c}{\textbf{95\%-tile Time (ms)}} \\
		\cmidrule(lr){5-6} \cmidrule(lr){7-8} \cmidrule(lr){9-10} \cmidrule(lr){11-12}
		& & & & \small{Log.} & \small{Run.} & \small{Ind.} & \small{Int.} & \small{Ind.} & \small{Int.} & \small{Ind.} & \small{Int.} \\
		\midrule
		MergeAllOf 
		& {\RN} & 100.00\% & 0.00\% & 0.00\% & 0.00\% & 8 & 4 & 4 & 4 & 13 & 10 \\
		& {\WGBI} & 100.00\% & 0.00\% & 0.00\% & 0.00\% & 19 & 9 & 8 & 7 & 45 & 24 \\
		& {\RB} & 45.40\% & 0.00\% & 12.07\% & 42.53\% & 179 & 179 & 21 & 21 & 55 & 55 \\
		\midrule
		Synthesized & {\RN} & 100.00\% & 0.00\% & 0.00\% & 0.00\% & 11 & 13 & 6 & 13 & 26 & 26 \\
		& {\WGBI} & 100.00\% & 0.00\% & 0.00\% & 0.00\% & 6 & 2 & 2 & 2 & 20 & 6 \\
		& {\RB} & 35.91\% & 0.00\% & 1.13\% & 59.73\% & 30 & 29 & 5 & 5 & 138 & 138 \\
		\midrule
		Handwritten SC & {\RN} & 100.00\% & 0.00\% & 0.00\% & 0.00\% & 8 & 9 & 3 & 3 & 36 & 34 \\
		& {\WGBI} & 99.29\% & 0.00\% & 0.71\% & 0.00\% & 22 & 23 & 15 & 16 & 51 & 38 \\
		& {\RB} & 40.43\% & 0.00\% & 21.28\% & 38.30\% & 983 & 983 & 48 & 48 & 4{,}172 & 4{,}172 \\
			\midrule
		{\RB}-testset & {\RN} & 98.67\% & 0.33\% & 0.00\% & 1.00\% & 7{,}566 & 72 & 4 & 3 & 505 & 116 \\
		& {\WGBI} & 81.33\% & 0.00\% & 0.00\% & 18.67\% & 2{,}002 & 2{,}023 & 131 & 125 & 10{,}939 & 11{,}107 \\
		& {\RB} & 92.67\% & 0.00\% & 0.00\% & 7.33\% & 987 & 1{,}064 & 74 & 73 & 7{,}194 & 7{,}360 \\
		\midrule
		SchemaStore & {\RN} & 99.72\% & 0.28\% & 0.00\% & 0.00\% & 894 & 19 & 4 & 3 & 112 & 26 \\
		vers. & {\WGBI} & 86.93\% & 13.07\% & 0.00\% & 0.00\% & 5{,}657 & 4{,}593 & 198 & 174 & 13{,}238 & 5{,}613 \\
		& {\RB} & 79.17\% & 0.00\% & 0.00\% & 20.83\% & 22{,}156 & 3{,}421 & 43 & 38 & 2{,}898 & 309 \\
		\midrule
		oneOf as anyOf & {\RN} & 94.29\% & 4.17\% & 0.00\% & 1.54\% & 2{,}152 & 28 & 19 & 3 & 2{,}214 & 49 \\
		& {\WGBI} & 70.64\% & 28.49\% & 0.00\% & 0.88\% & 31{,}716 & 8{,}630 & 2{,}958 & 422 & 192{,}744 & 45{,}906 \\
		& {\RB} & 16.63\% & 1.21\% & 0.60\% & 81.56\% & 1{,}635 & 1{,}358 & 71 & 64 & 2{,}409 & 1{,}095 \\
		\midrule
		uneval as & {\RN} & 96.05\% & 3.45\% & 0.00\% & 0.49\% & 4{,}569 & 8 & 6 & 3 & 27{,}685 & 22 \\
		additional & {\WGBI} & 66.12\% & 28.45\% & 0.82\% & 4.61 & 18{,}733 & 13{,}484 & 446 & 1{,}372 & 77{,}359 & 32{,}276 \\
		& {\RB} & 9.70\% & 1.81\% & 27.80\% & 60.69\% & 57 & 33 & 19 & 18 & 169 & 80 \\
		\midrule
		additional as & {\RN} & 98.16\% & 1.84\% & 0.00\% & 0.00\% & 235 & 35 & 6 & 3 & 254 & 153 \\
		uneval & {\WGBI} & 64.13\% & 35.33\% & 0.33\% & 0.21\% & 4{,}771 & 2{,}166 & 1{,}614 & 781 & 20{,}496 & 9{,}586 \\
		& {\RB} & 22.39\% & 0.46\% & 19.95\% & 57.20\% & 416 & 45 & 22 & 19 & 312 & 135 \\
		\midrule\midrule
		\multicolumn{2}{l}{\textbf{Cut-off ($>$ 25 KB)}} &&&&&&&&&&\\
		oneOf as anyOf (L) & {\RN} & 89.32\% & 8.40\% & 0.00\% & 2.28\% & 4{,}920 & 168 & 43 & 34 & 6{,}897 & 141 \\
		& {\WGBI} & 50.00\% & 48.01\% & 0.00\% & 1.99\% & 68{,}280 & 68{,}144 & 40{,}823 & 40{,}823 & 231{,}942 & 232{,}832 \\
		& {\RB} & 5.13\% & 1.99\% & 0.57\% & 92.31\% & 3{,}172 & -- & 364 & -- & 21{,}075 & -- \\
		\midrule
		additional as & {\RN} & 92.84\% & 7.16\% & 0.00\% & 0.00\% & 874 & 58 & 80 & 26 & 4{,}576 & 242 \\
		uneval (L)& {\WGBI} & 7.58\% & 91.76\% & 0.25\% & 0.42\% & 11{,}261 & 11{,}261 & 8{,}214 & 8{,}214 & 26{,}231 & 26{,}231 \\
		& {\RB} & 5.91\% & 1.75\% & 3.91\% & 88.43\% & 6{,}356 & -- & 268 & -- & 30{,}336 & -- \\
		\midrule
		uneval as & {\RN} & 88.07\% & 10.80\% & 0.00\% & 1.14\% & 17{,}195 & 171 & 990 & 35 & 66{,}967 & 1{,}235 \\
		additional (L) & {\WGBI} & 13.07\% & 83.52\% & 0.00\% & 3.41\% & 257{,}095 & 257{,}095 & 273{,}508 & 273{,}508 & 500{,}758 & 500{,}758 \\
		& {\RB} & 3.98\% & 6.25\% & 5.11\% & 84.66\% & 293 & -- & 159 & -- & 798 & -- \\
		\bottomrule
	\end{tabular}
\end{table*}

Table~\ref{tab:performance_by_collection} provides a more detailed breakdown of these results, including success rates, timeouts, and error types for each dataset. Overall, {\RN} consistently outperforms the other tools, achieving 100\% success on 3 out of 8 datasets and at least 94\% on the remaining ones, while maintaining a low timeout rate (around 4\%). These observations further confirm the results illustrated in Fig.~\ref{fig:failure_rate}.
}

{\RN} has no logical errors and almost no runtime errors, similar to or better than {\WGBI};
for both tools, the vast majority of failures are timeouts. This result is perfectly in line with hypothesis~H1. 


For coverage (H3), {\RN} significantly outperforms existing tools in all cases where both of them had issues.
The difference is already visible for ``{\RB}-testset'' and ``SchemaStore versions'':
 pre-existing tools show significant failure rates, while {\RN} has close to 100\% coverage.
The differences are even more pronounced on the particularly challenging cut-off datasets\iflong{. As shown in Table~\ref{tab:performance_by_collection}, {\RN} still achieves success rates close to 90\% on these datasets}, where {\WGBI} and {\RB} fall significantly behind. {\RB} has failure rates around 90\% across all cut-off datasets, 
mainly due to runtime errors. {\WGBI} performs moderately better on the ``oneOf-as-anyOf'' cut-off dataset, with a failure rate of roughly 50\%, but its performance drops sharply on the other cut-off datasets, with failure rates around 80\%, largely due to timeouts.


These results strongly confirm hypothesis H3:
{\RN} significant\-ly increases coverage, enabling complex analyses on non-trivial sche\-mas that were de-facto infeasible before. \iflong{Unlike {\RB}, {\RN} does not suffer from systematic runtime failures, and unlike {\WGBI}, it maintains strong scalability. }

\iflong{
\begin{table*}[ht]
	\centering	\caption{Experimental results of {\RN} for included cases. Distinguishing cases where the algorithm arrived at the correct result by only using Refutational Normalization (``RN only''), and cases where the algorithm produced the correct result after invoking witness generation (``w/ WitGen'').}
	\label{tab:rn_unsat_results}
	\begin{tabular}{l r r r r r r r}
		\toprule
		\textbf{Collection ($\subseteq$)} & \textbf{\#Schemas} & \textbf{Success} & \textbf{Success} & \textbf{Success} & \textbf{Timeout} & \textbf{Logical} & \textbf{Runtime} \\
		\ & & \textbf{(overall)} & \textbf{(RN only)} & \textbf{(w/ WitGen)} & & \textbf{Error} & \textbf{Error} \\
		\midrule
		MergeAllOf & 167 & 100.0\% & 100.0\% & 0.0\% & 0.0\% & 0.0\% & 0.0\% \\
		Synthesized & 881 & 100.0\% & 99.7\% & 0.3\% & 0.0\% & 0.0\% & 0.0\% \\
		Handwritten SC & 162 & 100.0\% & 86.4\% & 13.6\% & 0.0\% & 0.0\% & 0.0\% \\
		RB-testset & 106 & 100.0\% & 98.1\% & 1.9\% & 0.0\% & 0.0\% & 0.0\% \\
		SchemaStore vers. & 903 & 100.0\% & 100.0\% & 0.0\% & 0.0\% & 0.0\% & 0.0\% \\
		oneOf as anyOf & 1666 & 95.5\% & 91.6\% & 3.9\% & 4.3\% & 0.0\% & 0.2\% \\
		additional as uneval & 5558 & 98.1\% & 90.9\% & 7.3\% & 1.9\% & 0.0\% & 0.0\% \\
		uneval as additional & 427 & 95.6\% & 95.1\% & 0.5\% & 4.0\% & 0.0\% & 0.5\% \\
		\bottomrule
	\end{tabular}
\end{table*}
To further analyze the effectiveness of our Refutational Normalization approach, we investigate the percentage of cases that were successfully solved by our algorithm without the need to invoke witness generation. For this analysis, we only consider the included subsets of the datasets, since our approach always invokes witness generation for non-included cases. Table~\ref{tab:rn_unsat_results} shows the overall success rate of {\RN}, the percentage of cases successfully processed without invoking witness generation, and the rate of cases solved after invoking witness generation. Further, the table shows the overall timeouts, logical errors, and runtime errors.

Across all collections, except Handwritten SC, {\RN} successfully processes at least 90\% of the cases through Refutational Normalization alone, invoking witness generation only for recursive schemas. Handwritten SC is the only dataset where witness generation is invoked in non-recursive schemas, as these contain string patterns that our normalization does not cover. 

The fact that the only non-recursive schemas requiring witness generation in this experiment are highly synthetic, highlights the effectiveness of our approach in practical settings. 

Overall, these results indicate that the approach is effective for inclusion checking on its own.
}


\subsubsection{Run time} 
To study hypothesis H2, we compare the run times in Figure~\ref{fig:runtime}. 
We consider only test cases that are successfully analyzed by all tools. For each dataset, we report the total number of test cases in that intersection, broken down into included and non-included cases.\footnote{\crevone{Cases with unknown ground truth are, by definition, not part of this intersection, since {\WGBI} did not successfully analyze them.}} 
We exclude the ``MergeAllOf'', ``Synthesized'', and ``Handwritten SC'' datasets, 
since they contain only very small schemas, with negligible run time differences. For the cut-off data\-sets, we limit the comparison to {\RN} and {\WGBI}, since the overlap with {\RB} is too small to support a meaningful analysis. 

All tools have median run times 
in the 1~$msec$–1~$sec$ range, while the timeout threshold is set at 600~$sec$. Hence, the mean run time is not representative of typical performance, being disproportionately influenced by a small number of schemas that approach the timeout limit.
For this reason, we report 
the run time only at the $5^{th}$, the $50^{th}$ (median), and the $95^{th}$ percentile.
The median indicates a ``typical case'', and the $95^{th}$ percentile  a ``difficult case''. 
\iflong{We consider these measures more informative in practice than the high\-ly skewed mean run time. More detailed statistics including average, median, and $95^{th}$ percentile values for both individual and intersection run times are reported in Table~\ref{tab:performance_by_collection}. }

Figure~\ref{fig:runtime}
shows that {\WGBI} is mostly slower than {\RB} and {\RN}, 
 {\RN} being fastest.\footnote{\crevfour{There were only four schema pairs where {\WGBI} was significantly faster than {\RN}, i.e., with an absolute advantage greater than or equal to 4,000 ms.}}
In particular, {\RN} exhibits a median run time  at least
one order of magnitude lower than {\RB}, and likewise for the~$5^{th}$ and the $95^{th}$ percentiles, in the majority of datasets.\iflong{ These gains are also reflected in the aggregated statistics reported in Table~\ref{tab:performance_by_collection}, confirming that the advantage of {\RN} holds not only for typical cases but also for more challenging instances.}\textsc{}

The comparison between {\RB} (implemented in Python) and {\RN} (implemented in Scala) must be interpreted with caution. The tools rely on different runtime infrastructures, which may significantly influence absolute run time measurements. \hide{Therefore, the observed differences,in some cases up to three orders of magnitude, cannot be interpreted as a strict measurement of language-level or implementation-level efficiency alone, but rather as an indication of the overall performance of the complete systems.}
However, we believe that the size and consistency of the distance between the two tools 
aligns with hypothesis H2, namely: Our approach is at least as 
efficient as the rule-based approach.

\iflong{



}

\ifshort{
	\begin{figure}[bt]
		\centering
		\marginnote{\bcolorone{\Small chart updated}}
		{\setlength{\fboxsep}{0pt}%
			\setlength{\fboxrule}{1pt}%
			\fcolorbox{red}{white}{
		\includegraphics[width=\columnwidth]{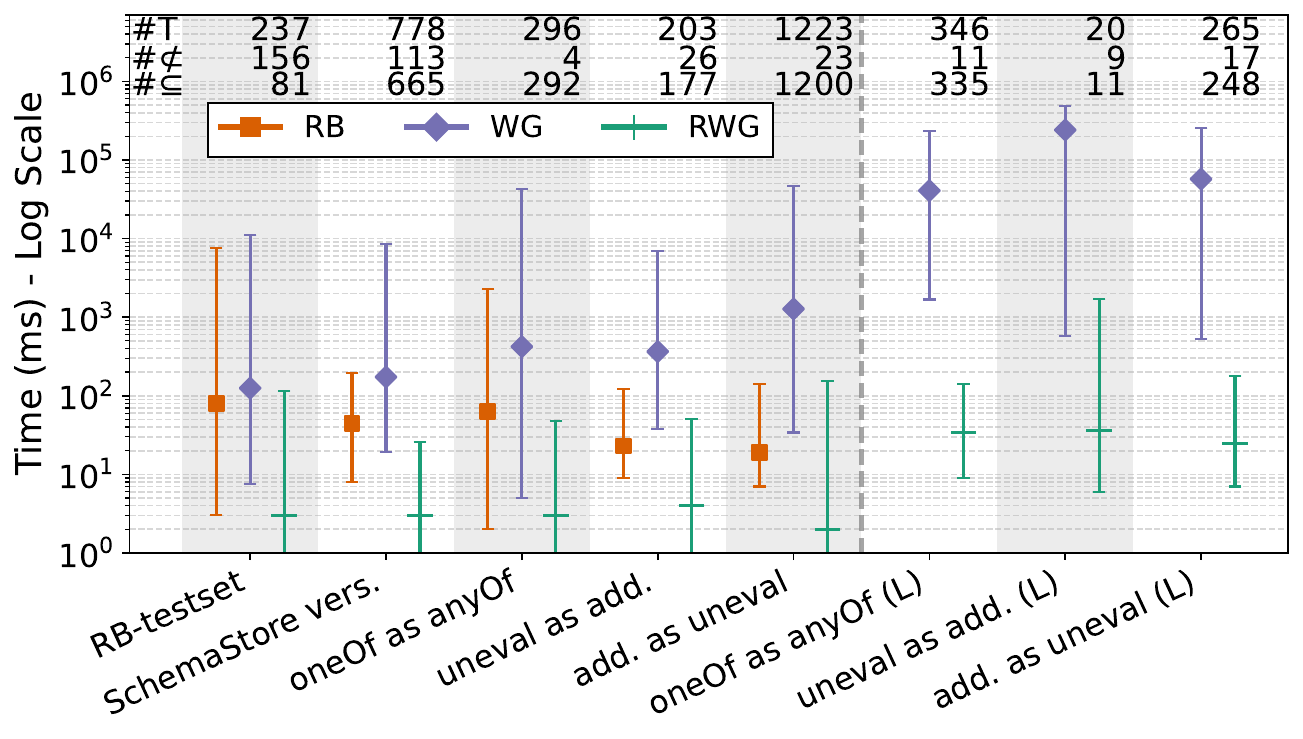}
		}}
		\caption{Run time on intersection of successfully processed inclusion tests. Showing the number of tests in the intersection (\#T) and of included/non-included tests (\#$\subseteq$/\#$\not\subseteq$).
			Markers indicate the $5^{th}$, $50^{th}$, $95^{th}$
			run time percentiles.}
		\label{fig:runtime}
	\end{figure}
}


\mrevfour{\subsubsection{Limitations} Despite these good results, {\RN} may still exhibit long run times, possibly leading to timeouts, particularly on schemas with $\geq 5$ branches in {\qall} and/or {\qone} operators, or on schemas with very complex patterns as arguments of {\qpatt} operators.}{4.2}
\mrevfour{Yet, the former scenario is where {\RN} pays off most: on schemas with $\geq 5$ branches in {\qall} and/or {\qone} operators (462 files, avg size 143KB), it clearly outperforms {\WGBI} and~{\RB} (66\% success rate in {\RN} vs.\ 7\% in {\WGBI} and 4\% in {\RB}).
}{4.6,4.8,M.4}

\mrevfour{\subsubsection{Impact of Optimizations} A preliminary ablation study indicates that fast complement-absorption through eager reference-normalization is  the most effective aspect of refutational normalization, immediately followed by lazy normalization;
whereas
the $\anot(\aany[\ldots])$-optimization, although crucial to prove that {\RN} is as fast as {\RB} (Section~\ref{sec:efficiency}), is much less relevant in our datasets.}{4.4,M.4}

\iflong{
\subsubsection{Summary} Our experimental results show that our new approach performs at least as well as the generation-based approach and is able to handle inclusion tests that previous approaches cannot process, clearly supporting hypotheses H1 and H3. Further, we show that the run time of our new approach is consistently better than the run time of existing tools. Although differences in execution environments restrict the comparison between {\RB} and the other tools, we consider our results to be a strong indication that hypothesis H2 holds.
}

\ifshort{
\subsubsection{Summary} Our experimental results support the claim that our approach reconciles
the completeness of the {\WGBI} approach with the efficiency of the {\RB} approach. This makes it possible to address use cases that were previously intractable.
}

\section{Conclusions}\label{sec:concl}

\mrevone{Inclusion for JSON Schema is EXPTIME-complete, and the two known approaches suffer from
complementary problems: the rule-based one is fast but incomplete, while witness 
generation is complete, but, under any reasonable timeout, incomplete in practice.}{1.1, 1.14, 4.1}\bcolorone

We presented a solution for this problem:
we redesigned the normalization core of the complete algorithm to make it as fast as the rule-based algorithm, without losing completeness.
Our experiments show that the result  analyzes 
real-world schemas whose size or complexity put them out of reach of previous tools.

The design is based on the observation that the normalization of $\qall : [ S_1, \{ \qnot : S_2 \} ]$,
seemingly unrelated to the structural comparison of $S_1$ with $S_2$,
can be implemented in a way that retraces that comparison, 
hence combining the completeness of the former with the 
efficiency of the latter.
This was our key idea, and it may reach beyond JSON Schema inclusion.
\ecolor

\hide{
\begin{acks}
 This work was supported by the [...] Research Fund of [...] (Number [...]). Additional funding was provided by [...] and [...]. We also thank [...] for contributing [...].
\end{acks}
}


\bibliographystyle{ACM-Reference-Format}
\bibliography{references}

\end{document}